\documentclass[aps,prd,twocolumn,floatfix,showpacs,superscriptaddress,nofootinbib]{revtex4-1}  
                
\usepackage{graphicx}  
\usepackage{dcolumn}   
\usepackage{bm}        
\usepackage{amssymb}   
\usepackage{epsfig}  
\usepackage{latexsym}
\usepackage{amssymb} 
\usepackage[english]{babel}
\usepackage{amsmath}
\usepackage{enumerate}

\newcommand{\mnras}{Mon. Not. R. Astron. Soc.}

\newcommand{\jcap}{Journal of Cosmology and Astroparticle Physics}
\newcommand{\ssr}{Space Science Reviews}
\newcommand{\aapr}{Astron. and Astrophys. Review}

\newcommand{\jqsrt}{J. Quant. Spec. Rad. Trans.}
    
\newcommand{\beq}{\begin{equation}}
\newcommand{\eeq}{\end{equation}}
\newcommand{\bga}{\begin{gathered}}
\newcommand{\ega}{\end{gathered}}
\newcommand{\eq}[1]{{Eq.~(#1)}}
\begin{document}

\widetext    
\title{A new probe of magnetic fields in the pre--reionization epoch: II. Detectability}
\author{Vera Gluscevic}
\author{Tejaswi Venumadhav}
\affiliation{Institute for Advanced Study, Einstein Drive, Princeton, NJ 08540, USA} 
\author{Xiao Fang}
\author{Christopher Hirata}
\affiliation{Center for Cosmology and Astroparticle Physics, The Ohio State University, 191 West Woodruff Lane, Columbus, Ohio 43210, USA}
\author{Antonija Oklop\v ci\' c}
\author{Abhilash Mishra} 
\affiliation{California Institute of Technology, Mail Code 350-17, Pasadena, CA 91125, USA}
\date{\today}

\begin{abstract} 
In the first paper of this series, we proposed a novel method to probe large--scale intergalactic magnetic fields during the cosmic Dark Ages, using 21--cm tomography. This method relies on the effect of spin alignment of hydrogen atoms in a cosmological setting, and on the effect of magnetic precession of the atoms on the statistics of the 21--cm brightness--temperature fluctuations. In this paper, we forecast the sensitivity of future tomographic surveys to detecting magnetic fields using this method. For this purpose, we develop a minimum--variance estimator formalism to capture the characteristic anisotropy signal using the two--point statistics of the brightness--temperature fluctuations. We find that, depending on the reionization history, and subject to the control of systematics from foreground subtraction, an array of dipole antennas in a compact--grid configuration with a collecting area slightly exceeding one square kilometer can achieve a $1\sigma$ detection of $\sim$$10^{-21}$ Gauss comoving (scaled to present--day value) within three years of observation. Using this method, tomographic 21--cm surveys could thus probe ten orders of magnitude below current CMB constraints on primordial magnetic fields, and provide exquisite sensitivity to large--scale magnetic fields \textit{in situ} at high redshift. 
\end{abstract} 
        
\pacs{} 
\maketitle   
\section{Introduction}
\label{sec:intro}

Magnetic fields are ubiquitous in the universe on all observed scales \cite{2013A&ARv..21...62D,Vallee04,Neronov10,2005LNP...664...89W,2012SSRv..166..215B}. However, the question of origins of the magnetic fields in galaxies and on large scales is as of yet unresolved. Various forms of dynamo mechanism have been proposed to maintain and amplify them \cite{2013PhRvE..87e3110P}, but they typically require the presence of seed fields \cite{2013A&ARv..21...62D}. Such seed fields may be produced during structure formation through the Biermann battery process or similar mechanisms \cite{Naoz13,2013PhRvL.111e1303N}, or may otherwise be relics from the early universe \cite{2013A&ARv..21...62D,2012SSRv..166...37W,2014JCAP...05..040K}. Observations of large--scale low--strength magnetic fields in the high--redshift intergalactic medium (IGM) could thus probe the origins of present--day magnetic fields and potentially open up an entirely new window into the physics of the early universe.

Many observational probes have been previously proposed and used to search for large--scale magnetic fields locally and at high redshifts (e.~g.~\cite{Yamazaki10,Blasi99,Tavecchio10,Dolag11,2005LNP...664...89W,2014JCAP...01..009K,2013ApJ...770...47K,2014PhRvD..89j3522S,2006MNRAS.372.1060T,2009ApJ...692..236S}). Amongst the most sensitive tracers of cosmological magnetic fields is the cumulative effect of Faraday rotation in the cosmic--microwave--background (CMB) polarization maps, which currently places an upper limit of $\sim$$10^{-10}$ Gauss (in comoving units) using data from the Planck satellite \cite{2015arXiv150201594P}. In Paper I of this series \cite{2014arXiv1410.2250V}, we proposed a novel method to detect and measure extremely weak cosmological magnetic fields during the pre--reionization epoch (the cosmic Dark Ages). This method relies on data from upcoming and future 21--cm tomography surveys \cite{1997ApJ...475..429M,2004PhRvL..92u1301L}, many of which have pathfinder experiments currently running \cite{2012arXiv1201.1700G,2011AAS...21813206B,2014ApJ...788..106P,2008arXiv0802.1727C,Vanderlinde14,2015AAS...22532803D}, with the next--stage experiments planned for the coming decade \cite{2008arXiv0802.1727C,2015AAS...22532803D}. 

In Paper I, we calculated the effect of a magnetic field on the observed 21--cm brightness--temperature fluctuations, and in this Paper, we focus on evaluating the sensitivity of future 21--cm experiments to measuring this effect. As we pointed out in Paper I, the 21--cm signal from the cosmic Dark Ages has an intrinsic sensitivity to capturing the effect of the magnetic fields in the IGM that are more than \textit{ten orders of magnitude smaller than the current upper limits on primordial magnetic fields from the CMB}. In the following, we demonstrate that a square--kilometer array of dipole antennas in a compact grid can reach the sensitivity necessary to detect large--scale magnetic fields that are on the order of $10^{-21}$ Gauss comoving (scaled to present day, assuming adiabatic evolution of the field due to Hubble expansion). 

The rest of this Paper is organized as follows. In \S\ref{sec:method}, we summarize the main results of Paper I. In \S\ref{sec:basics}, we define our notation and review the basics of the 21--cm signal and its measurement. In \S\ref{sec:estimators}, we derive minimum--variance estimators for uniform and stochastic magnetic fields. In \S\ref{sec:fisher}, we set up the Fisher formalism necessary to forecast sensitivity of future surveys. In \S\ref{sec:results}, we present our sensitivity forecasts. In \S\ref{sec:conclusions} we summarize and discuss the implications of our results. Supporting materials are presented in the appendices.
\section{Summary of the Method}
\label{sec:method}
\begin{figure}
\centering
\includegraphics[width=.45\textwidth,keepaspectratio=true]{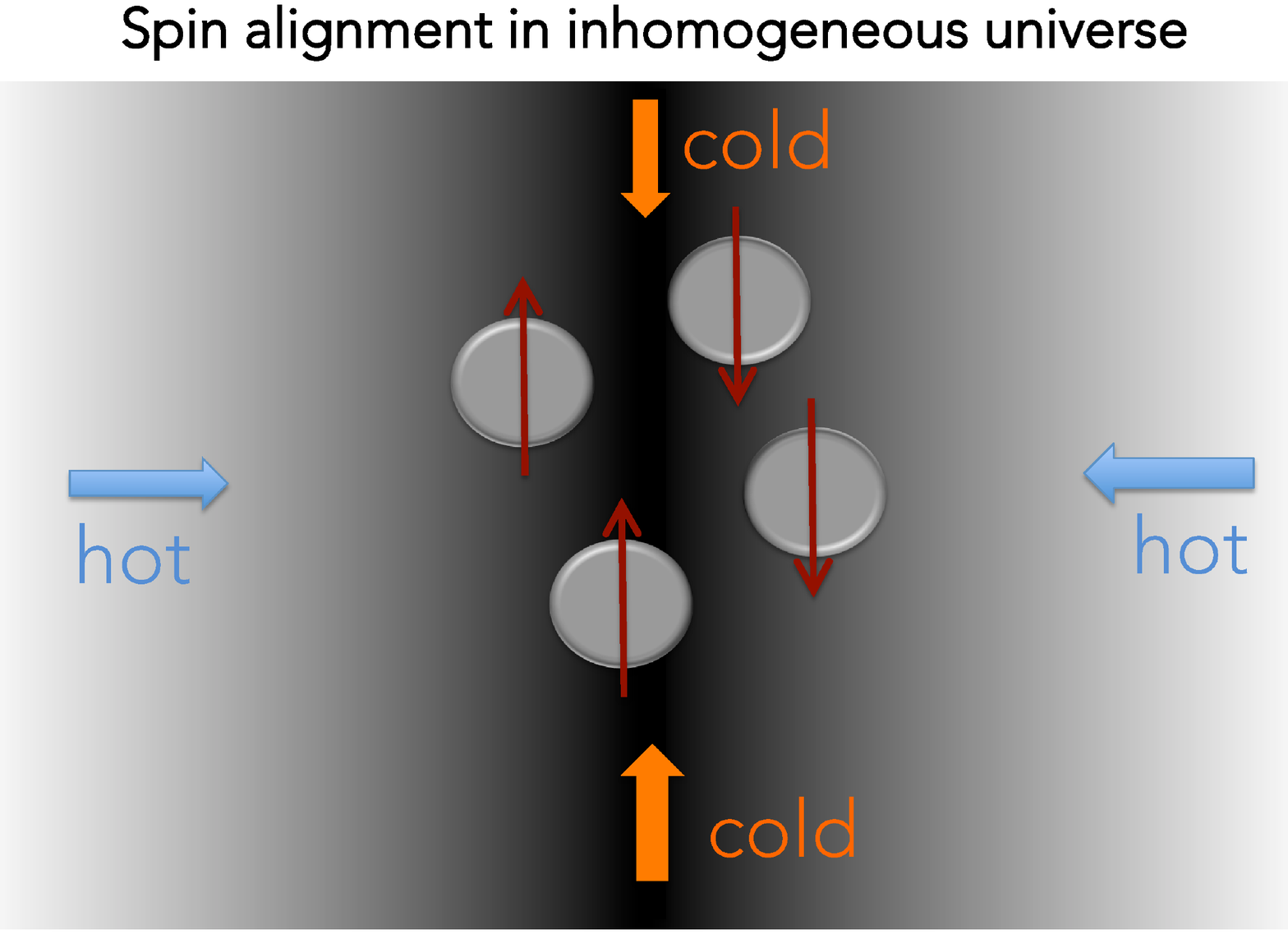}
\includegraphics[width=.45\textwidth,keepaspectratio=true]{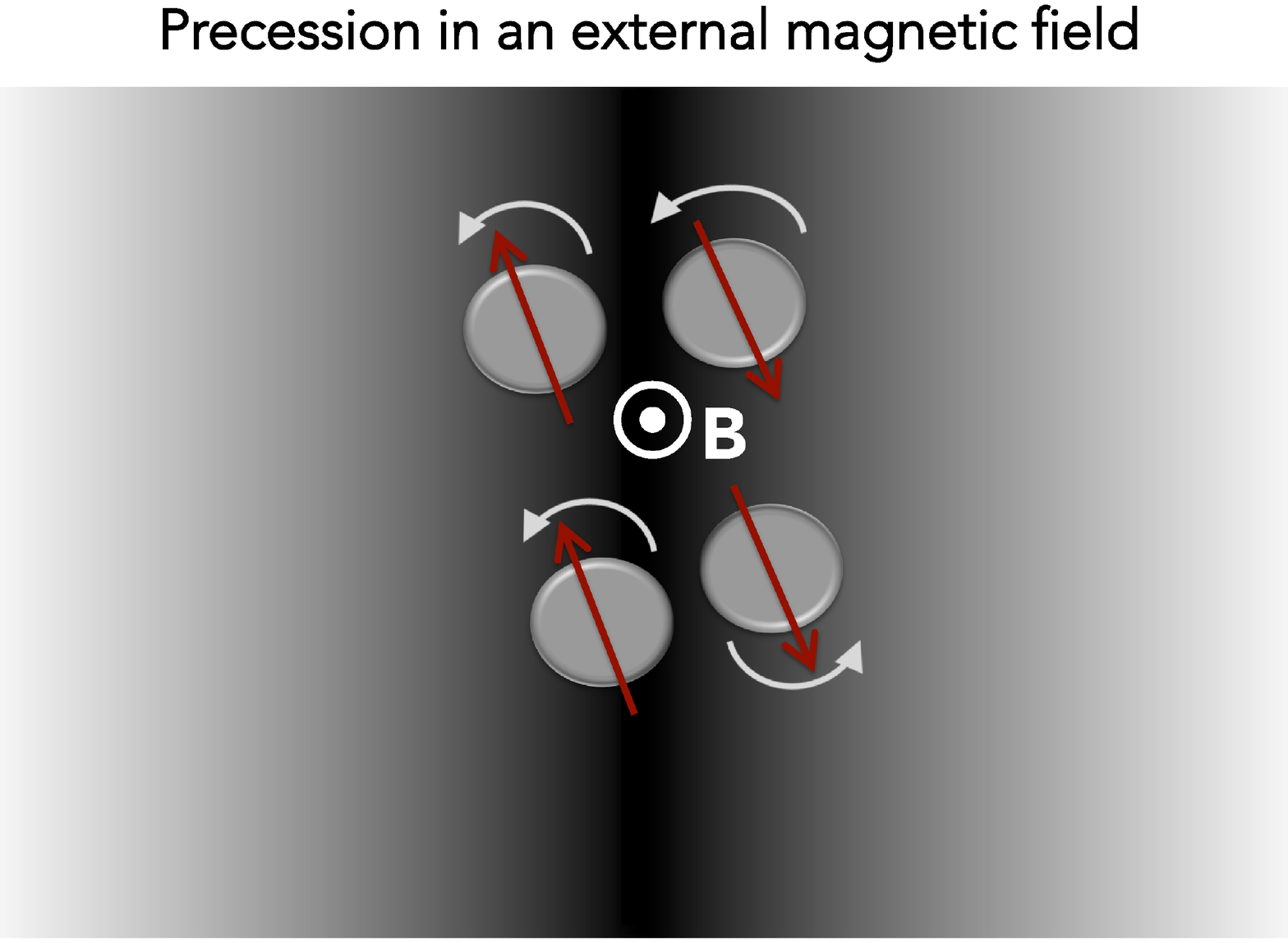}
\caption{Illustration of the effect of a magnetic field on hydrogen atoms in the excited state of 21--cm transition in cosmological setting. In the classical picture, magnetic moments of the atoms (depicted as red arrows) are aligned with density gradients (see upper panel; the gradient is depicted with the background shading), unless they precess about the direction of ambient magnetic field (pointing out of the page on the lower panel). When the precessing atoms decay back into the ground state, the emitted quadrupole (aligned with the direction of the magnetic moments) is misaligned with the incident quadrupole. This offset can be observed as a statistical anisotropy in 21--cm brightness--temperature signal, and used to trace cosmological magnetic fields.\label{fig:precession}}
\end{figure}
Magnetic moments of hydrogen atoms in the excited (triplet) state of the 21--cm line transition tend to align with the incident quadrupole of the 21--cm radiation from the surrounding medium. This effect of ``ground--state alignment'' \cite{Yan08,Yan12} arises in a cosmological setting due to velocity--field gradients. In the presence of an external magnetic field, the emitted 21--cm quadrupole is misaligned with the incident quadrupole, due to atomic precession; this is illustrated in Fig.~\ref{fig:precession}. The resulting emission anisotropy can be used to trace magnetic fields at high redshifts.

\begin{figure}
\centering
\includegraphics[width=.4\textwidth,keepaspectratio=true]{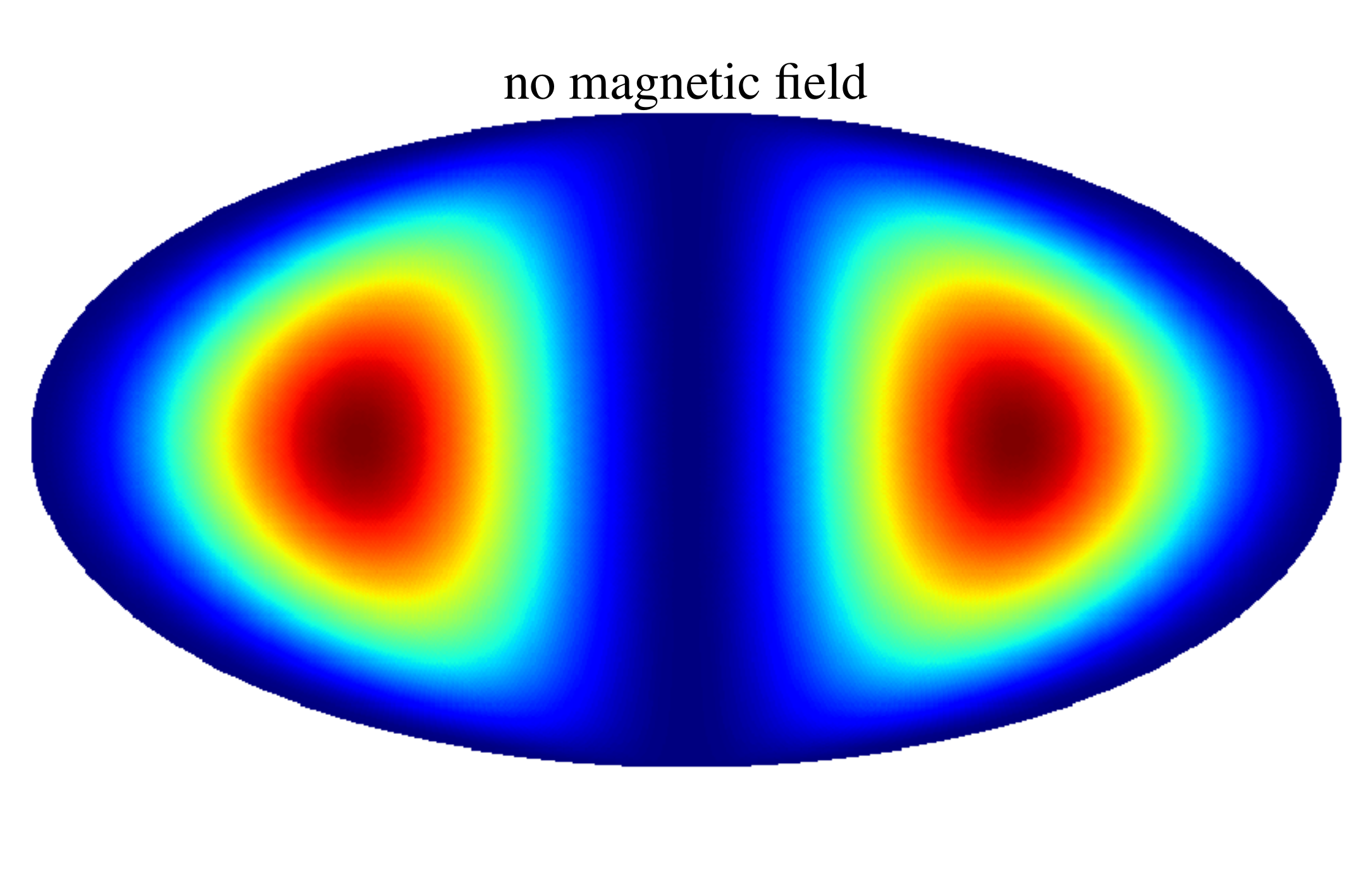}
\includegraphics[width=.4\textwidth,keepaspectratio=true]{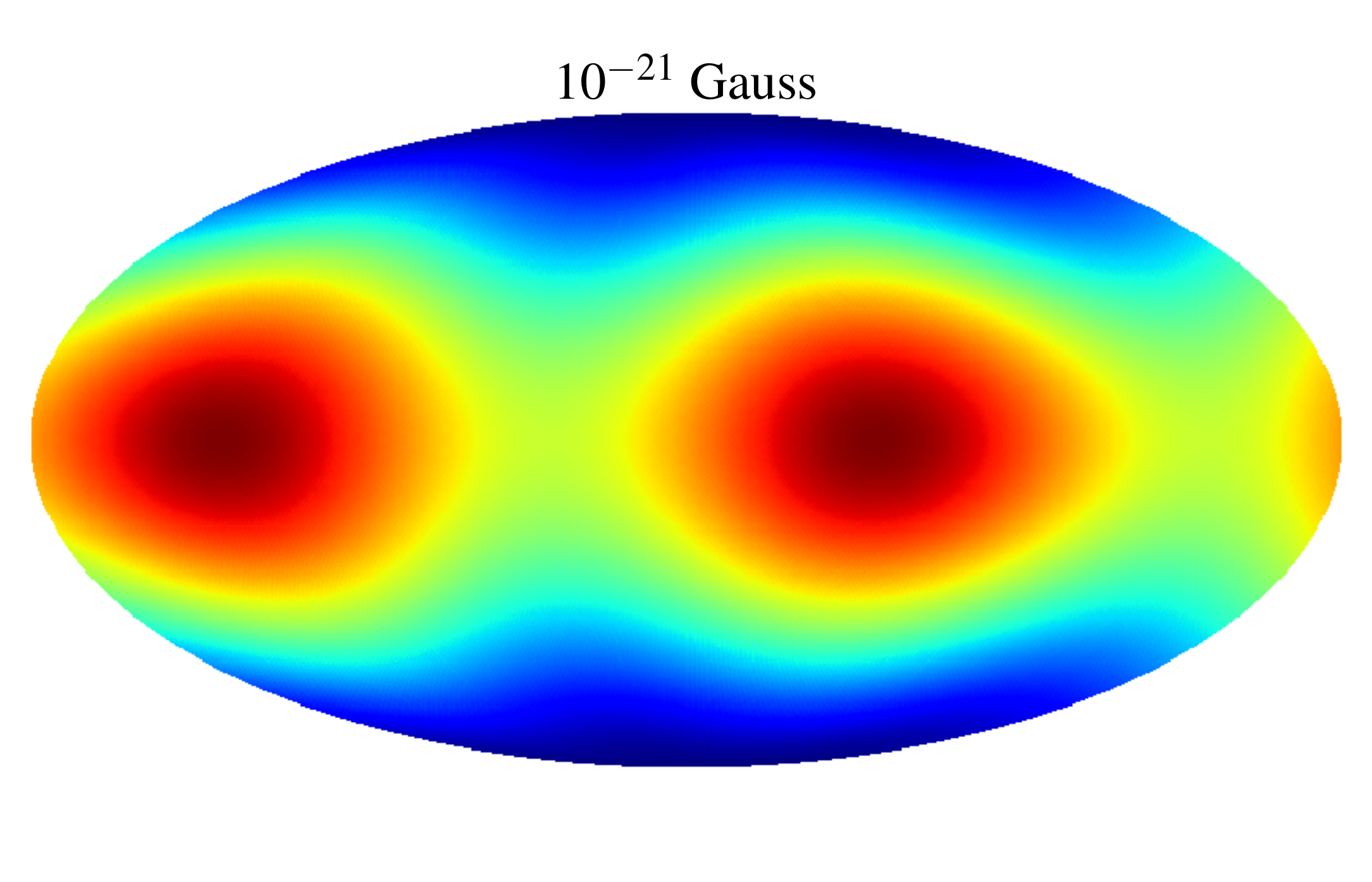}
\includegraphics[width=.4\textwidth,keepaspectratio=true]{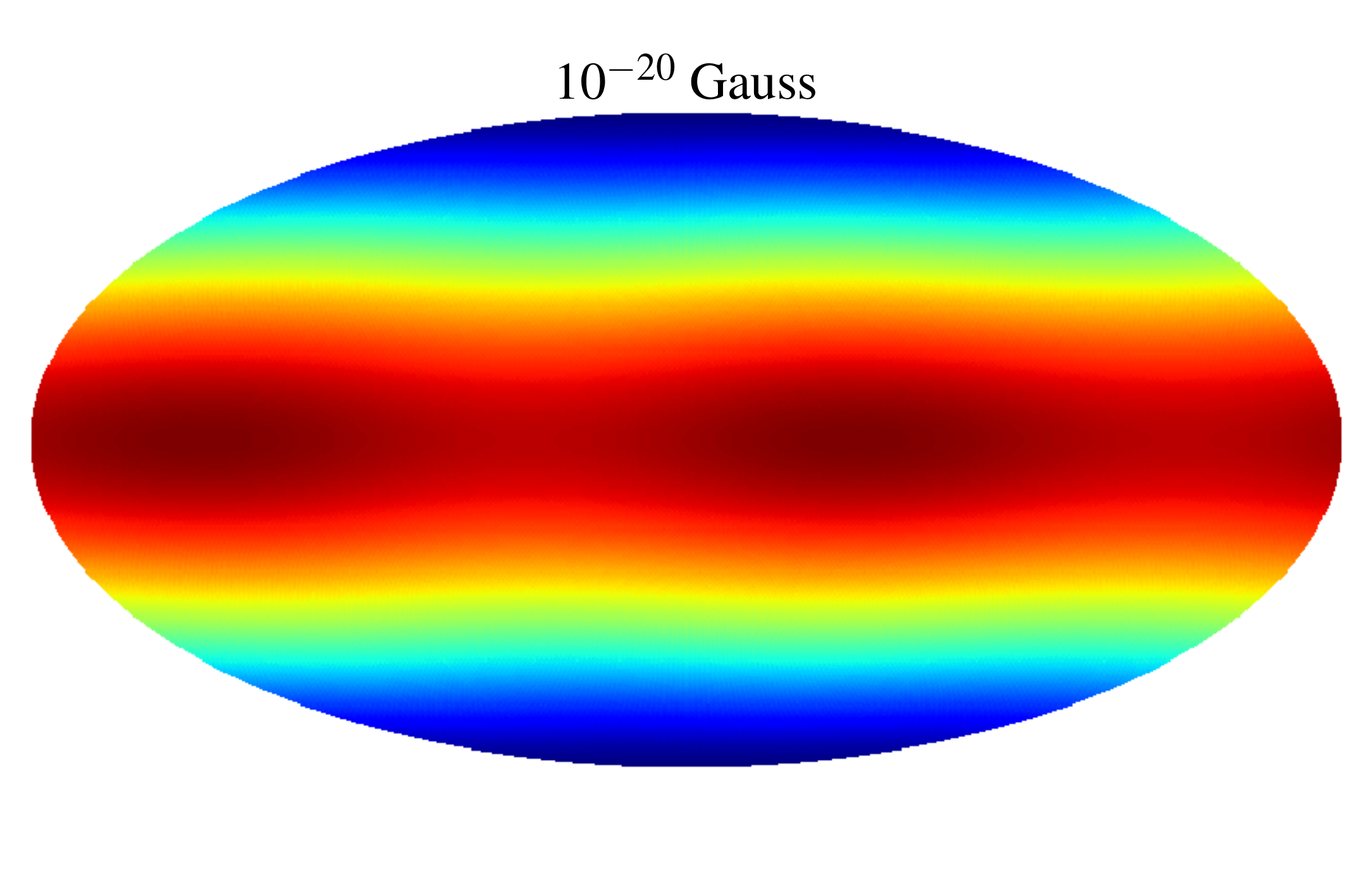}
\includegraphics[width=.4\textwidth,keepaspectratio=true]{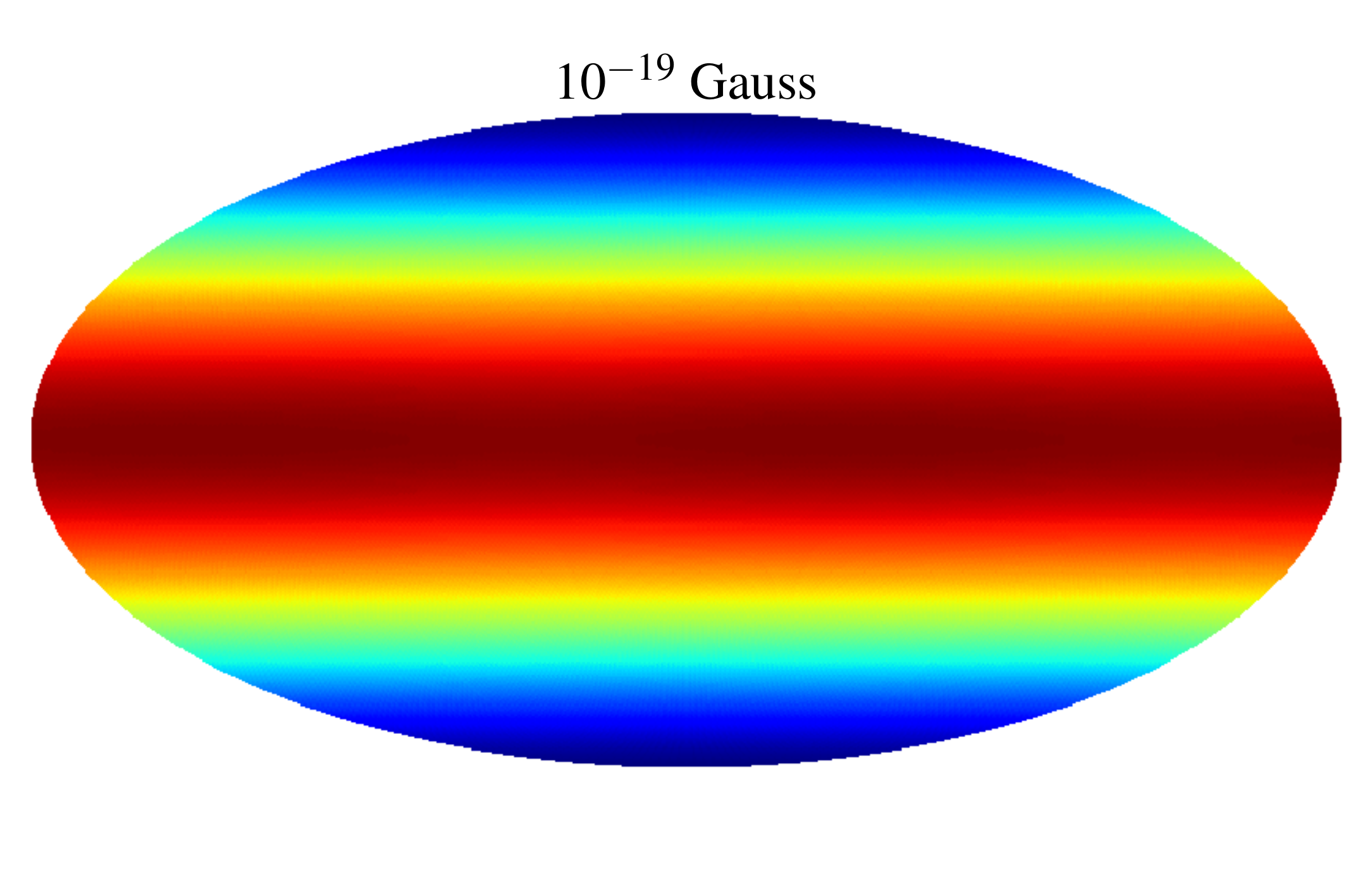}
\caption{Illustration of the quadrupolar pattern of 21--cm emission from the last ($\vec B$--dependent) term of \eq{\ref{eq:tbsoln}} in the frame of the emitting atoms, shown in Molleweide projection, where the intensity increases from blue to red shades. This illustration in all panels shows the case where $\vec k$ matches the direction of the hot spots in the top panel, and is perpendicular to the direction of the magnetic field (along the vertical axis in all panels). Every pixel in the maps corresponds to a unique direction $\bf{\widehat n}$ in Eq.~(\ref{eq:tbsoln}). Lower panels correspond to increasingly stronger magnetic field (strength denoted on each panel in comoving units, for $z=21$), with the bottom panel corresponding to the saturated case. Notice how the type of quadrupole in the top panel (weak--field regime) is distinct from that in the bottom panel (saturated regime). \label{fig:hp}}
\end{figure}

The main result of Paper I was derivation of the 21--cm brightness--temperature fluctuation\footnote{Standard notation, used in other literature and in Paper I of this series, for this quantity is $\delta T_b$; however, we use $T$ here to simplify our expressions.} $T_{\rm }$, including the effects of magnetic precession, as a function of the line--of--sight direction ${\bf{\widehat n}}$, 
\beq
\bga
   T_{\rm }({\bf{\widehat n}}, {{\vec k}}) = \left( 1 - \frac{T_\gamma}{T_{\rm s}} \right) x_{1{\rm s}} \left( \frac{1+z}{10} \right)^{1/2} \\
  \times \biggl[ 26.4 \ {\rm mK} \Bigl\{ 1 + \left(1 + ({\bf{\widehat k}} \cdot {\bf{\widehat n}})^2 \right)\delta(\vec k) \Bigr\}  
- 0.128 \ {\rm mK} \left( \frac{T_\gamma}{T_{\rm s}} \right)\\ \times x_{1{\rm s}} \left( \frac{1+z}{10} \right)^{1/2}  
 \Bigl\{ 1 + 2 \left(1 + ({\bf{\widehat k}} \cdot {\bf{\widehat n}})^2 \right)\delta(\vec k) \\
- \frac{ \delta(\vec k) }{15} \sum_m \frac{4\pi}{5} \frac{Y_{2 m}({\bf{\widehat k}}) \left[ Y_{2 m} ({\bf{\widehat n}}) \right]^* }{1 + x_{ \alpha, (2) } + x_{{\rm c}, (2)} - i m x_{\rm B}} \Bigr\} \biggr] \mbox{,} 
\ega
\label{eq:tbsoln}
\eeq
where the magnetic field is along the $z$ axis in the rest frame of the emitting atoms (in which the spin--zero spherical harmonics $Y_{2 m}$ are defined in the usual way); $\delta(\vec k)$ is a density--fluctuation Fourier mode corresponding to the wave vector $\vec k$ whose direction is along the unit vector $\bf{\widehat k}$; $x_{\alpha, (2)}$, $x_{{\rm c}, (2)}$, and $x_{\rm B}$ parametrize the rates of depolarization of the ground state by optical pumping and atomic collisions, and the rate of magnetic precession (relative to radiative depolarization), respectively (defined in detail in Paper I), and are all functions of redshift $z$; $T_{\rm s}$ and $T_\gamma$ are the spin temperature and the CMB temperature at redshift $z$, respectively. Fig.~\ref{fig:hp} illustrates the effect of the magnetic field on the brightness temperature emission pattern in the frame of the emitting atoms; shown are the quadrupole patterns corresponding to the last term of \eq{\ref{eq:tbsoln}}, for various strengths of the magnetic field. Notice that there is a saturation limit for the field strength---for a strong field, the precession is much faster than the decay of the excited state of the forbidden transition, and the emission pattern asymptotes to the one shown in the bottom panel of Fig.~\ref{fig:hp}. Above this limit, the signal cannot be used to reconstruct the strength of the field. However, in this ``saturated regime'', it is still possible to distinguish the presence of a strong magnetic field from the case of no magnetic field, as we discuss in detail in \S\ref{sec:fisher}.

The effect of quadrupole misalignment arises at second order in optical depth (it is a result of a two--scattering process), and is thus a small correction to the total brightness--temperature fluctuation. However, owing to the long lifetime of the excited state of the forbidden transition (during which even an extremely slow precession can have a large cumulative effect on the direction of the quadrupole, at second order), the misalignment is exquisitely sensitive to magnetic fields in the IGM at redshifts prior to cosmic reionization. As we showed in Paper I, a minuscule magnetic field of  $10^{-21}$ Gauss (in comoving units) produces order--one changes in the direction of the quadrupole. This implies that a high--precision measurement of the 21--cm brightness--temperature two--point correlation function intrinsically has that level of sensitivity to magnetic fields prior to the epoch of reionization (when most of the IGM is still neutral). We now proceed to develop a formalism to search for magnetic fields at high redshifts using this effect, and to forecast the sensitivity of future 21--cm experiments. 
\vspace{-12pt}
\section{Basics}
\label{sec:basics}

Before focusing on the estimator formalism (presented in the following Section), we review the basics of 21--cm brightness--temperature fluctuation measurements. In \S\ref{subsec:def}, we set up our notation and review definitions of quantities describing sensitivity of interferometric radio arrays; in \S\ref{subsec:noise}, we focus on the derivation of the noise power spectrum; and in \S\ref{subsec:uv}, we discuss the effects of the array configuration and its relation to coverage of modes in the $uv$ plane. 

\subsection{Definitions}
\label{subsec:def}

The redshifted 21--cm signal can be represented with specific intensity at a location in physical space $I(\vec{r})$ or in Fourier space $\widetilde{I}(\vec{k})$. In sky coordinates (centered on an emitting patch of the sky), these functions become $\mathcal{I}(\theta_x, \theta_y, \theta_\nu)$ and $\widetilde{\mathcal{I}}(u,v,\eta)$, respectively. Here, vector $\vec{k}$ (in the units of comoving Mpc$^{-1}$) is a Fourier dual of $\vec{r}$ (comoving Mpc), and likewise, $\theta_x$ (rad), $\theta_y$ (rad), and $\theta_\nu$ (Hz) are duals of the coordinates $u$ (rad$^{-1}$), $v$ (rad$^{-1}$), and $\eta$ (seconds), respectively.  Notice that $\theta_x$ and $\theta_y$ represent the angular extent of the patch in the sky, while $\theta_\nu$ represents its extent in frequency space. The two sets of coordinates are related through linear transformations in the following way
\begin{equation}
\begin{gathered}
\theta_x = \frac{r_x}{\chi(z)}, \hspace{0.5in} u = \frac{k_x\chi(z)}{2\pi},\\
\theta_y = \frac{r_y}{\chi(z)}, \hspace{0.5in} v = \frac{k_y\chi(z)}{2\pi},\\
\theta_\nu = \frac{H(z)\nu_{21}}{c(1+z)^2} r_z, \hspace{0.5in} \eta = \frac{c(1+z)^2}{2\pi H(z)\nu_{21}}k_z,
\end{gathered}
\label{eq:fourier_duals}
\end{equation} 
where $\nu_{21}=1420.4$ MHz is the frequency corresponding to the 21--cm line in the rest frame of the emitting atoms; $H(z)$ is the Hubble parameter; and $\chi(z)$ is the comoving distance to redshift $z$ which marks the middle of the observed data cube where $r_z$ and $\theta_\nu$ intervals are evaluated. Note that $2\pi\theta_iu = r_ik_i$, for $i\in\{x,y\}$. The convention we use for the Fourier transform is 
\beq
\bga
I(\vec{r}) = \frac{1}{(2\pi)^3}\int\widetilde{I}(\vec{k})e^{i\vec{k} \cdot \vec{r}}d\vec{k},\\
\widetilde{I}(\vec{k}) = \int{I}(\vec{r})e^{-i\vec{k} \cdot \vec{r}}d\vec{r},
\ega
\label{eq:tildeI_I}
\eeq
where Fourier--space functions are denoted with tilde. Similarly,
\beq
\bga
\mathcal{I}(\theta_x,\theta_y,\theta_\nu) = \int\widetilde{\mathcal{I}}(u,v,\eta)e^{2\pi i(u\theta_x + v\theta_y+\eta \theta_\nu)}dudvd\eta,\\
\widetilde{\mathcal{I}}(u,v,\eta) = \int{\mathcal{I}}(\theta_x,\theta_y,\theta_\nu)e^{-2\pi i(u\theta_x + v\theta_y+\eta\theta_\nu)}d\theta_xd\theta_yd\theta_\nu.
\ega
\label{eq:mathcal_tilde_I}
\eeq
From Eqs.~(\ref{eq:fourier_duals})--(\ref{eq:mathcal_tilde_I}), the following relation is satisfied
\beq
\widetilde{I}(\vec{k}) = \frac{c(1+z)^2\chi(z)^2}{H(z)\nu_{21}}\widetilde{\mathcal{I}}(u,v,\eta),
\label{eq_tilde_I_vs_Ik_scaling}
\eeq
where the proportionality factor contains the transformation Jacobian $\frac{dr_xdr_ydr_z}{d\theta_xd\theta_yd\theta_\nu}$. Finally, the relationship between the specific intensity in the $uv$--plane and the visibility function $\mathcal V(u,v,\theta_\nu)$ is given by the Fourier transform of the frequency coordinate,
\beq
\bga
 \mathcal{V}(u,v,\theta_\nu)= \int \mathcal{\widetilde{I}}(u,v,\eta)e^{2\pi i \theta_\nu\eta}d\eta,\\
\mathcal{\widetilde{I}}(u,v,\eta) = \int \mathcal{V}(u,v,\theta_\nu)e^{-2\pi i \theta_\nu\eta}d\theta_\nu.
\ega
\label{eq:visibility}
\eeq
Here, $\theta_{\nu,\text{max}}-\theta_{\nu,\text{min}}=\Delta\nu$ is the bandwidth of the observed data cube, centered on $z$ (see also Appendix \ref{app:Vrms}).
\subsection{Power spectra and noise}
\label{subsec:noise}

In this Section, we derive the noise power spectrum for the brightness--temperature fluctuation measurement. We start by defining a brightness--temperature power spectrum as
\beq
\langle \widetilde{I}(\vec{k})\widetilde{I}^*(\vec{k}')\rangle \equiv (2\pi)^3P_{\widetilde{I}}\delta_D(\vec{k}-\vec{k}'),
\label{eq_tildeI_power}
\eeq
where $\delta_D$ is Dirac delta function. The observable quantity of the interferometric arrays is the visibility function---a complex Gaussian variable with a zero mean and the following variance (see detailed derivation in Appendix \ref{app:Vrms}) 
\beq\bga
\langle \mathcal{V}({u},v,\theta_\nu)\mathcal{V}({u'},v',\theta_\nu')^*\rangle \\
= \frac{1}{\Omega_\text{beam}}\left(\frac{2k_BT_\text{sky}}{A_e\sqrt{\Delta\nu t_1}}\right)^2 \delta_D({u}-{u}')\delta_D({v}-{v}')\delta_{\theta_\nu\theta_{\nu}'},
\ega
\label{eq_Vrms}
\eeq 
where $T_\text{sky}$ is the sky temperature (which, in principle, includes both the foreground signal from the Galaxy, and the instrument noise, where we assume the latter to be subdominant in the following); $t_1$ is the total time a single baseline observes element $(u,v)$ in the $uv$ plane; $A_e$ is the collecting area of a single dish; $k_B$ is the Boltzmann constant; $\Delta\nu$ is the bandwidth of a single observation centered on $z$; and the last $\delta$ in this expression denotes the Kronecker delta.

Combining Eqs.~(\ref{eq:visibility}) and (\ref{eq_Vrms}), and taking the ensemble average,
\beq
\bga
\langle\widetilde{\mathcal{I}}(u,v,\eta) \widetilde{\mathcal{I}}^*(u',v',\eta')\rangle \\
 = \frac{1}{t_1\Omega_\text{beam}}\left(\frac{2k_BT_\text{sky}}{A_e}\right)^2 \delta_D({u}-{u}')\delta_D({v}-{v}')\delta_D(\eta-\eta'),
\ega
\label{eq:mathcal_power_Vrms}
\eeq 
where we used the standard definition
\beq
\int e^{2\pi i \theta_\nu(\eta-\eta')}d\theta_\nu =\delta_D(\eta-\eta').
\eeq
Taking into account the relation of \eq{\ref{eq_tilde_I_vs_Ik_scaling}}, using \eq{\ref{eq_tildeI_power}}, and keeping in mind the scaling property of the delta function, we arrive at
\beq
P_1^N(\vec k) = \frac{c(1+z)^2\chi^2(z)}{\Omega_\text{beam}t_1H(z)\nu_{21}}\left(\frac{2k_BT_\text{sky}}{A_e}\right)^2 ,
\label{eq:Pnoise_1mode}
\eeq
for the noise power per $\vec k$ mode, per baseline.

In the last step, we wish to get from \eq{\ref{eq:Pnoise_1mode}} to the expression for the noise power spectrum that corresponds to observation with all available baselines. To do that, we need to incorporate information about the array configuration and its coverage of the $uv$ plane. In other words, we need to divide the expression in \eq{\ref{eq:Pnoise_1mode}} by the number density of baselines $n_\text{base}(\vec k)$ that observe a given mode $\vec k$ at a given time (for a discussion of the $uv$ coverage, see the following Section). The final result for the noise power spectrum per mode $\vec k$ in intensity units is 
\beq
P^N(\vec k) = \frac{c(1+z)^2\chi^2(z)}{\Omega_\text{beam}t_1H(z)\nu_{21}}\frac{\left(2k_BT_\text{sky}\right)^2}{A_e^2n_\text{base}(\vec k)},
\label{eq:Pnoise_Jy}
\eeq
and in temperature units
\beq
P^N(\vec k) =\frac{\lambda^4c(1+z)^2\chi^2(z)}{\Omega_\text{beam}t_1H(z)\nu_{21}}\frac{T_\text{sky}^2}{A_e^2n_\text{base}(\vec k)},
\label{eq:Pnoise_K}
\eeq
where $\lambda=c/\nu_{21}(1+z)$.

\subsection{The UV coverage}
\label{subsec:uv}

The total number density $n_\text{base}(\vec k)$ of baselines that can observe mode $\vec k$ is related to the (unitless) number density $n(u,v)$ of baselines per $dudv$ element as
\beq
n_\text{base}(\vec k) = \frac{n(u,v)}{\Omega_\text{beam}},
\label{eq:nuv_nk}
\eeq
where $\frac{1}{\Omega_\text{beam}}$ represents an element in the $uv$ plane. The number density integrates to the total number of baselines $N_\text{base}$,
\beq
N_\text{base}=\frac{1}{2}N_\text{ant}(N_\text{ant}+1) = \int_\text{half} n(u,v)dudv,
\label{eq:nk}
\eeq
where $N_\text{ant}$ is the number of antennas in the array, and the integration is done on one half of the $uv$ plane\footnote{This is because the visibility has the following property $V(u,v,\theta_\nu)=V^*(-u,-v,\theta_\nu)$, and only a half of the plane contains independent samples.}. We assume that the array consists of many antennas, so that time dependence of $n(u,v)$ is negligible; if this is not the case, time average of this quantity should be computed to account for Earth's rotation.

In this work, we focus on a specific array configuration that is of particular interest to cosmology---a compact grid of dipole antennas, with a total collecting area of $(\Delta L)^2$, and a maximum baseline length\footnote{Note that for a square with area $(\Delta L)^2$ tiled in dipoles, there is a very small number of baselines longer than $\Delta L$, but we neglect this for simplicity.} of $\Delta L$. In this setup, the beam solid angle is 1 sr, the effective area of a single dipole is $A_e = \lambda^2$, and the effective number of antennas is $N_\text{ant} = \frac{(\Delta L)^2}{\lambda^2}$. For such a configuration, the number density of baselines entering the calculation of the noise power spectrum reads
\beq
n(u,v) = (\frac{\Delta L}{\lambda} - u)(\frac{\Delta L}{\lambda} - v).
\label{eq:nuv_fftt}
\eeq
The relation between $\vec k=(k,\theta_k,\phi_k)$ and $(u,v)$ is
\beq
\bga
u_\perp \equiv \frac{\chi(z)}{2\pi}k\sin\theta_k,\\
u = u_\perp \cos\phi_k,\\
v = u_\perp \sin\phi_k,
\ega
\label{eq:k_uv}
\eeq
where the subscript $\perp$ denotes components perpendicular to the line--of--sight direction ${\bf{\widehat n}}$, which, in this case, is along the $z$ axis. From this, the corresponding number of baselines observing a given $\vec k$ is
\beq
\bga
n_\text{base}(\vec k) = (\frac{\Delta L}{\lambda} - \frac{\chi(z)}{2\pi}k\sin\theta_k\cos\phi_k)\\\times (\frac{\Delta L}{\lambda} - \frac{\chi(z)}{2\pi}k\sin\theta_k\sin\phi_k).
\ega
\label{eq:nk_fftt}
\eeq

As a last note, when computing numerical results in \S\ref{sec:results}, we substitute the $\phi_k$--averaged version of the above quantity (averaged between $0$ and $\pi/2$ only, due to the four--fold symmetry of the experimental setup of a square of dipoles) when computing the noise power, in order to account for the rotation of the baselines with respect to the modes in the sky. This average number density reads
\beq
\bga
\langle n_\text{base}(\vec k) \rangle_{\phi_k}= \left(\frac{\Delta L}{\lambda}\right)^2 -\frac{4}{\pi}\frac{\Delta L}{\lambda}\frac{\chi(z)}{2\pi}k\sin\theta_k \\+ \frac{1}{\pi}\left(\frac{\chi(z)}{2\pi}k\sin\theta_k\right)^2,
\ega
\label{eq:nk_fftt_mean}
\eeq
assuming a given mode $k$ is observable by the array, such that its value is between $2\pi L_\text{min}/(\lambda(z)\chi(z)\sin\theta_k)$ and $2\pi L_\text{max}/(\lambda(z)\chi(z)\sin\theta_k)$, where $L_\text{min}$ and $L_\text{max}$ are the maximum and minimum baseline lengths, respectively. If this condition is not satisfied, $\langle n_\text{base}(\vec k) \rangle_{\phi_k}=0$.
\section{Quadratic estimator formalism}
\label{sec:estimators}

We now derive an unbiased minimum--variance quadratic estimator for a magnetic field $\vec B$ present in the IGM prior to the epoch of reionization. This formalism is applicable to tomographic data from 21--cm surveys, and is similar to that used in CMB lensing analyses \cite{2003PhRvD..67h3002O}, for example. We assume that the magnetic field only evolves adiabatically, due to Hubble expansion, 
\beq
B(z) = B_0(1+z)^2,
\label{eq:B0}
\eeq
where $B_0$ is its present--day value (the value of the field in comoving units). The corresponding estimator is denoted with a hat sign, $\widehat B_0$. 

We start by noting that the observed brightness--temperature fluctuations $T(\vec k)$ contain contributions from the noise fluctuation $T^N(\vec k)$ (from the instrumental noise plus Galactic foreground emission\footnote{Note that this term adds variance to the visibilities due to foregrounds, but we assume the bias in the visibilities is removed via foreground cleaning.}) and the signal $T^S(\vec k)$, 
\beq
\bga
T(\vec k) = T^N(\vec k) + T^S(\vec k),
\ega
\label{eq:Ttot}
\eeq
where $T^S(\vec k)$ can get contribution from both the magnetic--field effects and the (null--case) cosmological 21--cm signal, $T^S_0(\vec k)$. The signal temperature fluctuation is proportional to the density fluctuation $\delta$, with transfer function $G({\bf{\widehat k}})$ as the proportionality factor, 
\beq
\bga
G({\bf{\widehat k}}) \equiv \frac{\partial T}{\partial\delta}({\bf{\widehat k}},\delta=0),
\ega
\eeq
and
\beq
\bga
T^S(\vec k) = G({\bf{\widehat k}})\delta(k),\\
T^S_0(\vec k) = G_0({\bf{\widehat k}})\delta(k),
\ega
\label{eq:def_G}
\eeq
where ${\bf{\widehat k}}=(\theta_k,\phi_k)$ is a unit vector in the direction of $\vec k$. Note that we use the subscript ``0'' to denote when the transfer function $G$, the temperature fluctuation $T$, their derivatives, or the power spectrum $P$, are evaluated at $B_0=0$. Furthermore, we omit explicit dependence of $G$ on redshift and on cosmological parameters, and consider it implied. Finally, note that $G$ is a function of the direction vector ${\bf{\widehat k}}$, while the power spectrum $P_\delta$ is a function of the magnitude $k$, in an isotropic universe. The expression for the transfer function is obtained from Eq.~(\ref{eq:tbsoln}),
\beq
\bga
G({\bf{\widehat k}})=\left( 1 - \frac{T_\gamma}{T_{\rm s}} \right) x_{1{\rm s}} \left( \frac{1+z}{10} \right)^{1/2} \\
\times \biggl[ 26.4 \ {\rm mK}  \left(1 + ({\bf{\widehat k}} \cdot {\bf{\widehat n}})^2 \right)  
- 0.128 \ {\rm mK} \left( \frac{T_\gamma}{T_{\rm s}} \right)\\
\times x_{1{\rm s}} \left( \frac{1+z}{10} \right)^{1/2}  
 \Bigl\{ 2 \left(1 + ({\bf{\widehat k}} \cdot {\bf{\widehat n}})^2 \right) \\
- \sum_m \frac{4\pi}{75} \frac{Y_{2 m}({\bf{\widehat k}}) \left[ Y_{2 m} ({\bf{\widehat n}}) \right]^* }{1 + x_{ \alpha, (2) } + x_{{\rm c}, (2)} - i m x_{\rm B}} \Bigr\} \biggr] ,
\label{eq:G_def}
\ega
\eeq
for a reference frame where the magnetic field is along the $z$--axis.
For simplicity of the expressions, we adopt the following notation
\beq
\bga
\frac{\partial T_0^S}{\partial B_0}(\vec k)\equiv  \delta(k)\frac{\partial G}{\partial B_0}({\bf{\widehat k}},B_0=0),\\
\frac{\partial G_0}{\partial B_0}({\bf{\widehat k}})\equiv\frac{\partial G}{\partial B_0}({\bf{\widehat k}},B_0=0),
\ega
\label{eq:dTdB_dGdB}
\eeq
where $\frac{\partial G_0}{\partial B_0}=\frac{\partial G_0}{\partial B} (1+z)^2$ for adiabatic evolution of the magnetic field. 

The signal power spectrum in the absence of a magnetic field (null case) is given by
\beq
\bga
\left<T_0(\vec k)T_0^*(\vec k')\right> \equiv (2\pi)^3 \delta_D(\vec k-\vec k') P_0^S(\vec k)\\
= (2\pi)^3 \delta_D(\vec k-\vec k')G^2_0({\bf{\widehat k}})P_\delta(k),
\ega
\eeq
where 
\beq
\bga
\left<\delta(\vec k)\delta^*(\vec k')\right> \equiv (2\pi)^3 \delta_D(\vec k-\vec k') P_\delta(k).
\ega
\label{eq:Pdelta_definition}
\eeq
The total measured null--case power spectrum is
\beq
P_\text{null}(\vec k) \equiv P^N(\vec k) + P_0^S(\vec k).
\label{eq:Pnull}
\eeq

In \S\ref{subsec:uniform}, we first consider the case of a field uniform in the entire survey volume; this case is described by a single parameter, $B_0$. In \S\ref{subsec:SI}, we move on to the case of a stochastic magnetic field, with a given power spectrum $P_B(\vec K)$ (where $\vec K$ is the wavevector of a given mode of the field); in this case, the relevant parameter is the amplitude of this power spectrum, $ A_0^2$. In both cases, we assume that there is a valid separation of scales: density--field modes in consideration must have much smaller wavelengths than the coherence scale of the magnetic field (or a given mode wavelength for the case of a stochastic magnetic field), and both length scales must fit within the size of the survey.

\subsection{Uniform field}
\label{subsec:uniform}

We now derive an estimator $\widehat B_0$ for a comoving uniform magnetic field. We adopt the linear--theory approach and start with
\beq
\bga
T^S(\vec k) = T^S_0(\vec k) + B_0\frac{\partial T^S_0}{\partial B_0}(\vec k),
\ega
\label{eq:TS_uniform}
\eeq
where $B_0$ is a small expansion parameter. The observable two--point correlation function in Fourier space is then
\beq
\bga
\langle T(\vec k)T^*(\vec k')\rangle = P_\text{null}(\vec k)(2\pi)^3\delta_D(\vec k-\vec k') \\
+ \langle T^S_0(\vec k)B_0\frac{\partial T_0^{S,*}}{\partial B_0}(\vec k')\rangle + \langle T_0^{S,*}(\vec k')B_0\frac{\partial T_0^S}{\partial B_0}(\vec k)\rangle\\
=\left(P_\text{null}(\vec k)
 + 2B_0P_{\delta}( k)G_0({\bf{\widehat k}})\frac{\partial G_0}{\partial B_0}({\bf{\widehat k}}) \right) \\\times(2\pi)^3\delta_D(\vec k-\vec k'),
\ega
\label{eq:TT_step2}
\eeq
where we use the reality of $G_0$ and $\frac{\partial G_0}{\partial B_0}$, assume that the signal and the noise are uncorrelated, and keep only terms linear in $B_0$. Since we observe only one universe, a proxy for the ensemble average in \eq{\ref{eq:TT_step2}} is measurement of the product $T(\vec k)T^*(\vec k)$. Thus, an estimate of $B_0$ from a single temperature mode $\vec k$ is
\beq
\widehat B_0^{\vec k} = \frac{\frac{1}{V}T(\vec k)T^*(\vec k) - P_\text{null}(\vec k)}{2P_{\delta}( k)G_0({\bf{\widehat k}})\frac{\partial G_0}{\partial B_0}({\bf{\widehat k}})},
\label{eq:hatBk}
\eeq 
where we use the following properties of the Dirac delta function (defined on a finite volume $V$ of the survey)
\beq
\bga
\delta_D(\vec k-\vec k') = \frac{V}{(2\pi)^3},\hspace{0.2in} \text{for }\vec k = \vec k',\\
(2\pi)^3\delta_D(\vec k - \vec k') \equiv \int e^{-i\vec r \cdot (\vec k-\vec k')}d\vec r,
\ega
\label{eq:delta_kk}
\eeq
which is related to the Kronecker delta as
\beq
\delta_{\vec k\vec k'} = \frac{(2\pi)^3}{V}\delta_D(\vec k-\vec k').
\label{eq:deltas}
\eeq

The estimator of \eq{\ref{eq:hatBk}} is unbiased (with a zero mean), $\langle \widehat B_0^{\vec k}\rangle=0$. The covariance $\langle \widehat B_0^{\vec k}\widehat B_0^{{\vec k'},*}\rangle $ of estimators derived from all measured temperature modes involves temperature--field four--point correlation function with three Wick contractions, whose numerator reads
\beq
\bga
\frac{1}{V^2}\langle T(\vec k)T^*(\vec k)T(\vec k')T^*(\vec k') \rangle + P_\text{null}(\vec k)P_\text{null}(\vec k')\\
- \frac{1}{V}P_\text{null}(\vec k)\langle T(\vec k')T^*(\vec k') \rangle
- \frac{1}{V}P_\text{null}(\vec k')\langle T(\vec k)T^*(\vec k) \rangle \\
= P_\text{null}(\vec k)P_\text{null}(\vec k') \left[\frac{(2\pi)^6}{V^2}\right.\delta_D(\vec k-\vec k)\delta_D(\vec k'-\vec k')\\
+\frac{(2\pi)^6}{V^2}\delta_D(\vec k-\vec k')\delta_D(\vec k-\vec k')+
\frac{(2\pi)^6}{V^2}\delta_D(\vec k+\vec k')\delta_D(\vec k+\vec k')\\
-\frac{(2\pi)^3}{V}\delta_D(\vec k'-\vec k')-\left.\frac{(2\pi)^3}{V}\delta_D(\vec k-\vec k)\right]\\
=P_\text{null}(\vec k)P_\text{null}(\vec k')\left(\delta_{\vec k,\vec k'} + \delta_{\vec k,-\vec k'}\right),
\ega
\label{eq:TTTT_expansion}
\eeq
where every ensemble average yielded one factor of volume $V$. Using the final expression in the above Equation, we get  
\beq
\langle \widehat B_0^{\vec k}\widehat B_0^{{\vec k'},*}\rangle = \frac{P_\text{null}^2(\vec k)\left(\delta_{\vec k,\vec k'}  + \delta_{\vec k,-\vec k'} \right)}{4P_{\delta}( k)^2\left[G_0({\bf{\widehat k}})\frac{\partial G_0}{\partial B_0}({\bf{\widehat k}})\right]^2}.
\label{eq:B_covariance}
\eeq
Estimators from all $\vec k$--modes can be combined with inverse--variance weighting as
\beq
\bga
\widehat B_0 = \frac{\sum_{\vec k}\frac{\widehat B_0^{\vec k}}{\langle \widehat B_0^{\vec k}\widehat B_0^{{\vec k},*}\rangle}}{\sum_{\vec k}\frac{1}{\langle \widehat B_0^{\vec k}\widehat B_0^{{\vec k},*}\rangle}}.
\ega
\label{eq:B_mve}
\eeq 
Expanding the above expression, we get the minimum--variance quadratic estimator for $B_0$ obtained from all temperature--fluctuation modes  observed at a given redshift, 
\beq
\bga
\widehat B_0 = \sigma^{2}_{ B_0}\sum_{\vec k}\frac{\frac{1}{V}T(\vec k)T^*(\vec k) - P_\text{null}(\vec k)}{P_\text{null}^2(\vec k)}\\
\times 2P_{\delta}( k)G_0({\bf{\widehat k}})\frac{\partial G_0}{\partial B_0}({\bf{\widehat k}}).
\ega
\label{eq:B_estimator}
\eeq
Its variance $\sigma^{2}_{ B_0}$ is given by
\beq
\bga
\sigma^{-2}_{ B_0} = \frac{1}{2}\sum_{\vec k}\left(\frac{2P_{\delta}(k)G_0({\bf{\widehat k}})\frac{\partial G_0}{\partial B_0}({\bf{\widehat k}})}{P_\text{null}(\vec k)}\right)^{2},
\ega
\label{eq:B_estimator_var}
\eeq
where the sums are unrestricted. Note that $\widehat B_0^{\vec k}=\widehat B_0^{-\vec k}$; this follows from the reality condition on the temperature field, $T(\vec k)=T^*(-\vec k)$, and from the isotropy of space in the null--assumption case, $G_0({\bf{\widehat k}})=G_0(-{\bf{\widehat k}})$. Thus, in order to avoid double counting of modes, a factor of $1/2$ appears at the right--hand side of Eq.~(\ref{eq:B_estimator_var}).

Finally, the total sensitivity of a survey covering a range of redshifts is given by integrating the above Equation as
\beq
\bga
\sigma_{ B_0,\text{tot}}^{-2} = 
\frac{1}{2}\int dV_\mathrm{}(z)
\frac{k^2dk d\phi_k\sin \theta_kd\theta_k}{(2\pi)^3}\\
\times\left( \frac{2P_\delta(k,z)G_0(\theta_k,\phi_k,z)\frac{\partial G_0}{\partial B_0}(\theta_k, \phi_k,z)}{P^N(k,\theta_k,z) + P_\delta(k,z)G_0^2(\theta_k,\phi_k,z)} \right)^2,
\ega
\label{eq:fisher_patch}
\eeq
where we transitioned from a sum over $\vec k$ modes to an integral, using $\sum_{\vec k} \to V\int d\vec k /(2\pi)^3$. 
The integral is performed over the (comoving) volume of the survey of angular size $\Omega_\mathrm{survey}$ (at a given redshift, given in steradians), such that the volume element reads
\beq
dV_\mathrm{} = \frac{c}{H(z)}\chi^2(z)\Omega_\mathrm{survey}dz.
\label{eq:dVpatch}
\eeq

\subsection{Stochastic field}
\label{subsec:SI}

We now examine the case where both the magnitude and the direction of the magnetic field are stochastic random variables, with spatial variation. Note that in this Section we do \textit{not} assume a particular model for their power spectra, but we do assume a separation of scales, in the sense that we are only concerned with the modes $\vec K$ of the magnetic field that correspond to scales much larger than those corresponding to the density and temperature modes used for estimating the field, $|\vec K|\ll|\vec k|,|\vec k'|$. We use $B_0$ to denote a component of the magnetic field along one of the three Cartesian--system axes, and $\vec r$ to denote position vector in physical space, as before, and start with 
\beq
T(\vec r) = T^S_0(\vec r) + B_0(\vec r)\frac{\partial T^S_0}{\partial B_0}(\vec r),
\eeq
where the subscripts and superscripts have the same meaning as before. In Fourier space, we now get
\beq
\bga
T(\vec k) = T^S_0(\vec k) + \int d\vec r e^{-i\vec k \cdot \vec r} B_0(\vec r) \frac{\partial T^S_0}{\partial B_0}(\vec r)\\
= T^S_0(\vec k) + \frac{1}{(2\pi)^3}\int d\vec k_1B_0(\vec k_1) \frac{\partial T^S_0}{\partial B_0}(\vec k-\vec k_1),
\ega
\eeq
where the last step uses the convolution theorem. The observable two--point correlation function in Fourier space then becomes
\beq
\bga
\left < T(\vec k)T^*(\vec k')\right > = (2\pi)^3\delta_D(\vec k-\vec k')P_\text{null}(\vec k)\\
+ \left <T_0^{S,*}(\vec k')\frac{1}{(2\pi)^3}\int d\vec k_1 B_0(\vec k_1) \frac{\partial T^S_0}{\partial B_0}(\vec k-\vec k_1)\right > \\
+ \left <T^S_0(\vec k)\frac{1}{(2\pi)^3}\int d\vec k_1 B_0^*(\vec k_1) \left(\frac{\partial T^S_0}{\partial B_0}(\vec k'-\vec k_1)\right)^*\right >, 
\ega
\eeq
to first order in $B_0$. Note that, in this case, there is cross--mixing of different modes of the temperature field. From Eqs.~(\ref{eq:def_G}), (\ref{eq:dTdB_dGdB}), and (\ref{eq:Pdelta_definition}), we get
\beq
\bga
\left< T(\vec k)T^*(\vec k')\right> = (2\pi)^3\delta_D(\vec k - \vec k')  P_\text{null}(\vec k)+B_0(\vec k - \vec k')\\
\times\left[ P_\delta(k')G_0^*({\bf{\widehat k'}})\frac{\partial G_0}{\partial B_0}({\bf{\widehat k'}}) + P_\delta(k)G_0({\bf{\widehat k}})\frac{\partial G_0^*}{\partial B_0}({\bf{\widehat k}})\right],
\ega
\eeq
where we used the reality condition $B_0^*(-\vec K) = B_0(\vec K)$. In analogy to the procedure of \S\ref{subsec:uniform}, we estimate $B_0(\vec K)$ from $\vec k\vec k'$ pair of modes that satisfy $\vec K=\vec k-\vec k'$ as
\beq
\widehat B_0^{\vec k\vec k'}(\vec K) = \frac{T(\vec k)T^*(\vec k')}{P_\delta(k')G_0^*({\bf{\widehat k'}})\frac{\partial G_0}{\partial B_0}({\bf{\widehat k'}}) + P_\delta(k)G_0({\bf{\widehat k}})\frac{\partial G_0^*}{\partial B_0}({\bf{\widehat k}})},
\label{eq:Bkkp_estimator}
\eeq
where we only focus on terms $\vec K\ne0$ ($\vec k \ne\vec k'$).
The variance $\left< \widehat B_0^{\vec k\vec k'}(\vec K)\left(\widehat B_0^{\vec k\vec k'}(\vec K')\right)^*\right>$ of this estimator (under the null assumption) can be evaluated using the above expression. Furthermore, the full estimator for $B_0(\vec K)$ from all available temperature modes is obtained by combining individual $\widehat B_0^{\vec k\vec k'}(\vec K)$ estimates with inverse--variance weights, and with appropriate normalization, in complete analogy to the uniform--field case. For the purpose of forecasting sensitivities, we are interested in the variance of the minimum--variance estimator, or equivalently, the noise power spectrum $P^N_{B_0}(\vec K)$, given by
\beq
\bga
(2\pi)^3\delta_D(\vec K - \vec K') P^N_{B_0}(\vec K) \equiv \left< \widehat B_0(\vec K)\widehat B_0(\vec K')^*\right>\\
= \left( \sum_{\vec k} \frac{\left(P_\delta(k')G_0^*({\bf{\widehat k'}})\frac{\partial G_0}{\partial B_0}({\bf{\widehat k'}}) + P_\delta(k)G_0({\bf{\widehat k}})\frac{\partial G_0^*}{\partial B_0}({\bf{\widehat k}})\right)^2}{2V^2P_\text{null}(\vec k) P_\text{null}(\vec k')} \right)^{-1},
\ega
\label{eq:NK1}
\eeq
with the restriction $\vec K=\vec k-\vec k'$. The factor of $2$ in the denominator corrects for double counting mode pairs, since $\widehat B_0^{\vec k\vec k'}(\vec K)=\left(\widehat B_0^{-\vec k-\vec k'}(\vec K)\right)^*$, and the sum is unconstrained. If we only consider diagonal terms $\vec K=\vec K'$, then the left--hand side of the above Equation becomes equal to $V P^N_{B_0}(\vec K)$. The explicit expression for the noise power spectrum is then
\beq
\bga
P^N_{B_0}(\vec K) = \\
\left(\sum_{\vec k} \frac{\left(P_\delta(k')G_0^*({\bf{\widehat k'}})\frac{\partial G_0}{\partial B_0}({\bf{\widehat k'}}) + P_\delta(k)G_0({\bf{\widehat k}})\frac{\partial G_0^*}{\partial B_0}({\bf{\widehat k}})\right)^2}{2VP_\text{null}(\vec k) P_\text{null}(\vec k')  } \right)^{-1}.
\label{eq:NK}
\ega
\eeq

Finally, transitioning from a sum to the integral (like in \S\ref{subsec:uniform_fisher}), we get the following expression for the noise power spectrum of one of the components $B_{0,i}$ of the magnetic field in the plane of the sky,
\beq
\bga
\left(P^N_{B_{0,i}}(\vec K)\right)^{-1} = \int k^2d{k}\sin \theta_kd\theta_kd\phi_k \\
\times\frac{\left(P_\delta(k')G_0^*({\bf{\widehat k'}})\frac{\partial G_0}{\partial B_i}({\bf{\widehat k'}}) + P_\delta(k)G_0({\bf{\widehat k}})\frac{\partial G_0^*}{\partial B_i}({\bf{\widehat k}})\right)^2}{2(2\pi)^3P_\text{null}(\vec k) P_\text{null}(\vec k') } ,
\ega
\label{eq:NK2}
\eeq
where $\vec k'=\vec K -\vec k$ and the above expression is evaluated at a particular redshift. Only the components of the magnetic field in the plane of the sky affect the observed brightness--temperature fluctuations, and so Eq.~(\ref{eq:NK2}) can be used to evaluate the noise power spectrum for either one of the two (uncorrelated) components. The noise in the direction along the line of sight can be considered infinite. Finally, note that we can construct a similar estimator for the direction of the magnetic field in a given patch of the sky. However, in this work we focus on the magnitude of the field and ignore considerations with regard to its direction.
\section{Fisher analysis}
\label{sec:fisher}

We now use the key results of \S\ref{sec:estimators} to evaluate sensitivity of future tomographic 21--cm surveys to detecting presence of magnetic fields in high--redshift IGM. In \S\ref{subsec:uniform_fisher}, we derive the expression for sensitivity to a field uniform in the entire survey volume. We start with the unsaturated case where (in the classical picture) hydrogen atoms complete less than a radian of magnetic precession during their lifetime in the triplet state, for all redshifts of interest (weak--field limit), and then move on to considering the saturated case (the fast--precession and stong--field limit). In \S\ref{subsec:SI_fisher}, we derive the expression for sensitivity to detecting a stochastic magnetic field described by a scale--invariant power spectrum.

\subsection{Uniform field}
\label{subsec:uniform_fisher}

Eq.~(\ref{eq:fisher_patch}) provides an expression for evaluating $1\sigma$ sensitivity to reconstructing a uniform magnetic field from measurements of the 21--cm signal at range of redshifts. For our numerical calculations, we take the following integration limits: $\phi_k\in[0,2\pi]$; $\theta_k\in [0,\pi]$; and $k\in[2\pi u_\mathrm{min}/(\chi(z)\sin\theta_k),2\pi u_\mathrm{max}/(\chi(z)\sin\theta_k)]$, where $u_\mathrm{min, max}=\frac{L_\text{min, max}}{\lambda}$ correspond to the maximum and minimum baseline lengths, $L_\text{min}$ and $L_\text{max}$, respectively. If the survey area is big enough that the flat--sky approximation breaks down, $\sigma_{B_0, \text{tot}}^{-2} $ can be evaluated on a small (approximately flat) patch of  size $\Omega_\text{patch}$ centered on the line of sight, and then corrected to account for the total survey volume
\footnote{This accounts for the change in the angle that a uniform magnetic field makes with a line of sight, as the line of sight ``scans'' through the survey area.} as
\beq
\bga
\sigma^{-2}_{ B_0,\text{corr}} = \frac{\sigma^{-2}_{ B_0,\text{tot}}}{\Omega_\text{patch}} \int_0^{\theta_\text{survey}}\int_{0}^{2\pi} \cos^2 \theta d\theta d\phi \\
= \frac{\pi\sigma^{-2}_{ B_0,\text{tot}}}{\Omega_\text{patch}} \left(\theta_\text{survey} + \cos \theta_\text{survey} \sin \theta_\text{survey}\right).
\ega
\label{eq:sigma_sum_survey}
\eeq

So far, we have only focused on the regime of a weak magnetic field. Let us now consider the case where the field is strong enough that the precession period is comparable to (or shorter than) the lifetime of the excited state of the forbidden transition---the saturated regime. In this case, the brightness--temperature signal still captures the presence of the field (as illustrated in Fig.~\ref{fig:hp}), but it loses information about the magnitude of the field, and can only be used to determine the lower limit of the field strength. The ability to distinguish the saturated case from zero magnetic field becomes a relevant measure of survey sensitivity in this scenario. 

To quantify the distinguishability of the two regimes, we write the signal power spectrum as the sum of contributions from both $B_0=0$ and $B_0\to\infty$, 
\beq
P^S(\vec k) = (1-\xi)P^S(\vec k, B=0) + \xi P^S(\vec k, B\to \infty).
\label{eq:saturated_P}
\eeq
We then perform the standard Fisher analysis to evaluate sensitivity to recovering parameter $\xi$,
\beq
\bga
\sigma_{\xi}^{-2} = 
\int dV_\mathrm{}(z)
\frac{d\vec k}{(2\pi)^3}\left(  \frac{\frac{\partial P^S}{\partial \xi}(\vec k)}{P^N (\vec k)+ P_0^S(\vec k,\xi=0) }\right)^2,
\ega
\label{eq:sigma_xi}
\eeq
where
\beq
\frac{\partial P^S}{\partial \xi}(\vec k) = P^S(\vec k, B\to \infty)-P^S(\vec k, B=0),
\eeq
and evaluating $P^S(\vec k, B\to \infty)$ requires the following limit of the transfer function (derived from Eq.~(\ref{eq:G_def}))
\beq
\bga
G({\bf{\widehat k}}, B\to \infty)
=\left( 1 - \frac{T_\gamma}{T_{\rm s}} \right) x_{1{\rm s}} \left( \frac{1+z}{10} \right)^{1/2} \\
\times \biggl[ 26.4 \ {\rm mK}  \left(1 + ({\bf{\widehat k}} \cdot {\bf{\widehat n}})^2 \right)  
- 0.128 \ {\rm mK} \left( \frac{T_\gamma}{T_{\rm s}} \right)\\
\times x_{1{\rm s}} \left( \frac{1+z}{10} \right)^{1/2}  
 \Bigl\{ 2 + 2({\bf{\widehat k}} \cdot {\bf{\widehat n}})^2 
- \frac{1}{60} \frac{1-3\cos ^2\theta_k}{1+x_{\alpha,(2)}+x_{c,(2)} }\Bigr\} \biggr].
\label{eq:G_Binf}
\ega
\eeq
Note that the above Equation is valid in the reference frame where the magnetic field is along the $z$ axis, and the line--of--sight direction is perpendicular to it. When evaluating Eq.~(\ref{eq:sigma_xi}) in \S\ref{sec:results}, we will only be interested in this configuration, since we aim to evaluate the sensitivity to the plane--of--the--sky component of $\vec B$. We interpret $\sigma_\xi^{-1}$ as 1$\sigma$ sensitivity to \textit{detecting} the presence of a strong magnetic field. 

\subsection{Stochastic field}
\label{subsec:SI_fisher}

To compute signal--to--noise ratio (SNR) for measuring the amplitude of a stochastic--field power spectrum, at a given redshift, we start with the general expression  
\beq
\text{SNR}^2 = \frac{1}{2} \text{Tr} \left( N^{-1}SN^{-1}S\right),
\label{eq:snr_general}
\eeq
where Tr denotes a trace of a matrix, and $S$ and $N$ stand for the signal and noise matrices, respectively. In the case of interest, these are $3N_\text{voxels}\times 3N_\text{voxels}$ matrices (there are 3 components of the magnetic field and $N_\text{voxels}$ voxels in the survey). In the null case, voxels are independent and the noise matrix is diagonal. Voxel--noise  variance for measuring a single mode is given by $P^N_{B_{0,i}}(\vec K, z)/V_\text{voxel} (z)$, where $V_\text{voxel}$ is voxel volume. Summing over all voxels and components of the magnetic field with inverse--variance weights gives
\beq
\bga
\text{SNR}^2 (z)= \frac{1}{2} \sum_{i\alpha, j\beta} \frac{S_{i\alpha , j\beta}^2}{P^N_{B_{0,i}}(\vec K, z)P^N_{B_{0,j}}(\vec K, z)} V_\text{voxel}^2\\=
\frac{1}{2} \sum_{ij} \int d\vec r_\alpha \int d\vec r_\beta \frac{\left< B_{0,i}(\vec r_\alpha) B_{0,j}(\vec r_\beta)\right>^2}{P^N_{B_{0,i}}(\vec K, z)P^N_{B_{0,j}}(\vec K, z)},
\ega
\label{eq:snr_z_step1}
\eeq
at a given redshift, where the Greek indices label individual voxels and, as before, Roman indices denote field components; $\vec r_{\alpha/\beta}$ represents spatial position of a given voxel. 

To simplify further calculations, we now focus on a particular class of magnetic--field models where most of the power is on largest scales (small $\vec K$). In this (squeezed) limit, $\vec K \ll \vec k$ and thus $\vec k \approx \vec k'$, such that \eq{\ref{eq:NK2}} reduces to white noise (independent of $\vec K$). A model for the power spectrum is defined through
\beq
(2\pi)^3\delta_D(\vec K - \vec K') P_{B_{0,i}B_{0,j}}(\vec K) \equiv \left<B_{0,i}^*(\vec K) B_{0,j}(\vec K')\right>,
\label{eq:Pbb}
\eeq
which relates to the variance in the transverse component $P_{B_\bot}(\vec K)$ as
\beq
P_{B_{0,i}B_{0,j}}(\vec K) = (\delta_{ij} - \widehat K_i \widehat K_j) P_{B_\bot}(\vec K),
\label{eq:Pbb_Pb}
\eeq
where $\widehat K_{i/j}$ is a unit vector along the direction of the ${i/j}$ component of the wavevector.
In the rest of this discussion, for concreteness, we consider a scale--invariant (SI) power spectrum, 
\beq
P_{{B_\bot}}(\vec K) = A_0^2/K^3.
\label{eq:SI}
\eeq
Here, the amplitude $A_0$ is a free parameter of the model (in units of Gauss).

If homogeneity and isotropy are satisfied, the integrand in Eq.~(\ref{eq:snr_z_step1}) only depends on the separation vector $\vec s \equiv \vec r_\beta -\vec r_\alpha$. Using this and the squeezed limit assumption gives\footnote{In the last step, we used $\int d\vec s |f(\vec s)|^2 = \int \frac{d\vec K}{(2\pi)^3}|\widetilde f(\vec K)|^2$, which holds for an arbitrary function $f$ and its Fourier transform $\widetilde f$.}
\beq  
\bga
\text{SNR}^2 (z) = 
\frac{1}{2} \sum_{ij}  \frac{dV_\text{patch}}{{(P^N_{B_{0,i}}(z))^2}}\int d\vec s \left< B_{0,}i(\vec r_\beta - \vec s) B_{0,j}(\vec r_\beta)\right>^2
\\=
\frac{1}{2(2\pi)^3} \sum_{ij}  \frac{dV_\text{patch}}{{(P^N_{B_{0,i}}(z))^2}} \int d\vec K\left(P_{B_{0,i}B_{0,j}}(\vec K)\right)^2,
\ega
\label{eq:snr_z}
\eeq
where $dV_\text{patch}$ is the volume of a redshift--slice patch defined in \eq{\ref{eq:dVpatch}}. After substituting \eq{\ref{eq:SI}} and integrating over redshifts the total SNR is given by
\beq
\bga
\text{SNR}^2 =  \frac{A_0^4}{2(2\pi)^3}  \int_{z_\text{min}}^{z_\text{max}}\frac{dV_\text{patch}}{{(P^N_{B_{0,i}}(z))^2}}
\int_0^{\pi} \sin\theta d\theta \\
\int_0^{2\pi} d\phi\int_{K_\text{min}(z,\theta,\phi)}^{K_\text{max}(z,\theta,\phi)} \frac{d K}{K^4}\sum_{ij\in \{xx, xy, yx, yy\}}(\delta_{ij} - \widehat K_i\widehat K_j)^2,
\ega
\label{eq:snr_intK}
\eeq
where $x$ and $y$ denote components in the plane of the sky, and
\beq
\widehat K_x = \sin\theta\sin\phi, \text{     }
\widehat K_y = \sin\theta\cos\phi.
\label{eq:hat_K_xy}
\eeq
The sum in the above expression reduces to
\beq
\sum_{ij\in \{xx, xy, yx, yy\}}(\delta_{ij} - \widehat K_i\widehat K_j)^2 = 2\cos^2\theta+\sin^4\theta.
\label{eq:sumij}
\eeq
Substituting  Eq.~(\ref{eq:sumij}) into Eq.~(\ref{eq:snr_intK}) and integrating over $K,\theta,\phi$ gives
\beq
\text{SNR}^2 =  \frac{A_0^4}{10\pi^2} \int_{z_\text{min}}^{z_\text{max}}\frac{dV_\text{patch}}{{(P^N_{B_{0,i}}(z))^2}} \left(\frac{1}{K_\text{min}^3}-\frac{1}{K_\text{max}^3}\right).
\label{eq:snr_ints}
\eeq
Finally, from the above expression, $1\sigma$ sensitivity to measuring $A_0^2$ is given by
\beq
\sigma^{-2}_{A_0^2} =  \frac{1}{10\pi^2} \int_{z_\text{min}}^{z_\text{max}}\frac{dV_\text{patch}}{{(P^N_{B_{0,i}}(z))^2}} \left(\frac{1}{K_\text{min}^3}-\frac{1}{K_\text{max}^3}\right).
\label{eq:sigma_A0}
\eeq
Note at the end that, for our choice of the SI power spectrum, the choice of $K_\text{max}$ does not matter (contribution to sensitivity rapidly decreases at small scales), while we take $K_\text{min}=2\pi /(\chi(z)\sin\theta_k)$ to match the survey size at a given redshift, for the compact--array configuration considered throughout this work. 
\section{Results}
\label{sec:results}
\begin{figure}
\centering
\includegraphics[width=.4\textwidth,keepaspectratio=true]{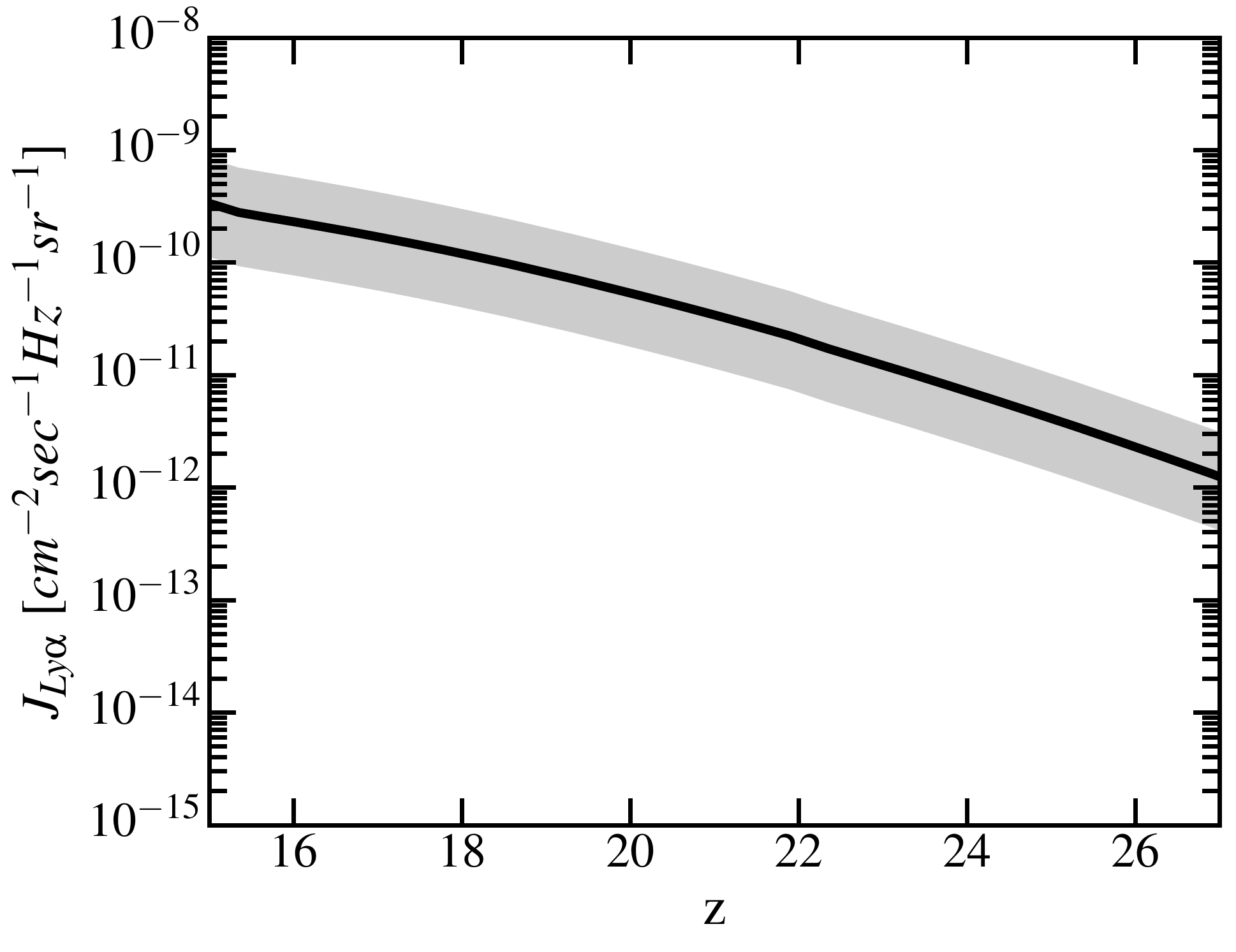}
\includegraphics[width=.4\textwidth,keepaspectratio=true]{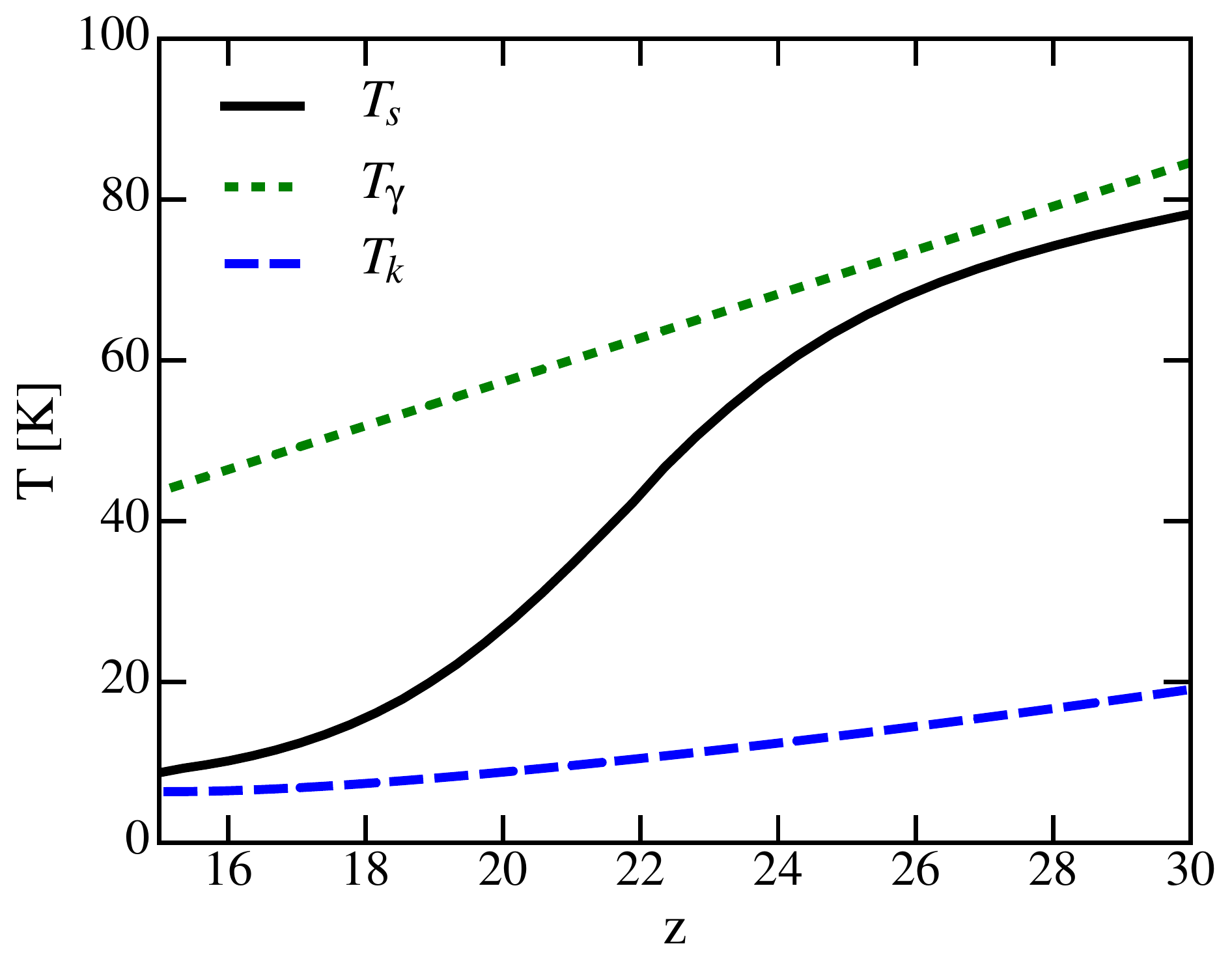}
\includegraphics[width=.4\textwidth,keepaspectratio=true]{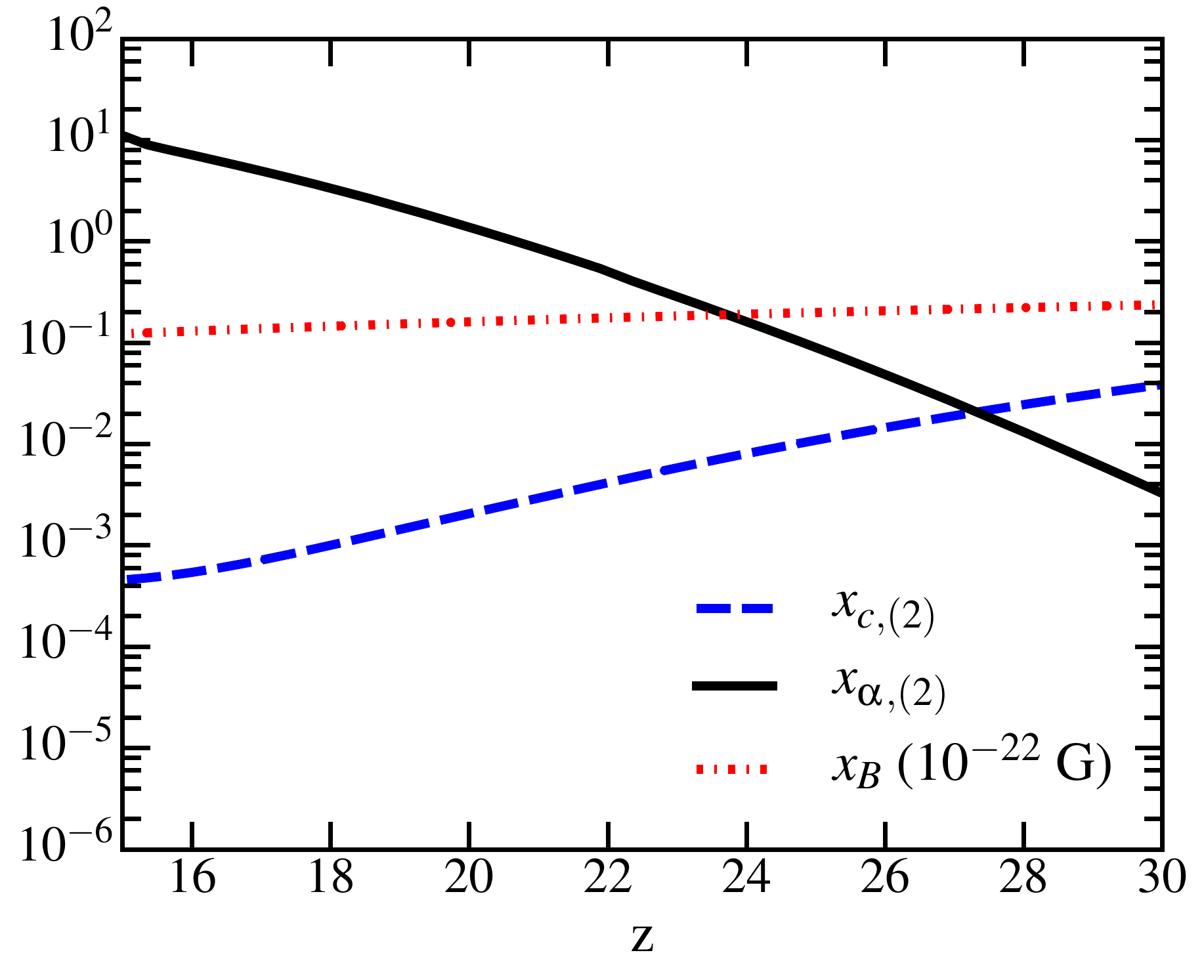}
\caption{Inputs used for the sensitivity calculation, computed for standard cosmology using the \texttt{21CMFAST} code. Top panel: Lyman--$\alpha$ flux model; fiducial choice used for sensitivity calculations is shown with a solid line, while the extrema of the gray band are used to test the effects of the uncertainty in the Lyman--$\alpha$ flux at high redshift (as discussed in the text). Middle panel: fiducial models for spin, kinetic, and CMB temperatures. Bottom panel: fiducial models for quantities that parametrize the rate of depolarization of the ground state by optical pumping and atomic collisions, and the rate of magnetic precession for a representative value of the magnetic field ($10^{-22}$ Gauss comoving). \label{fig:cosmo}}
\end{figure}
We now proceed to numerically evaluate the detection threshold of 21--cm tomography for magnetic fields in the pre--reionization epoch, using the formalism from the previous two Sections. For this purpose, we only focus on one type of experimental setup---an array of dipole antennas arranged in a compact grid. The motivation for this choice is that such a configuration maximizes sensitivity to recovering the power spectrum of the cosmological 21--cm signal \cite{2009PhRvD..79h3530T,2015AAS...22532803D}. We consider an array with a collecting area of $(\Delta L\text{ km})^2$, where $\Delta L$ is taken to be the maximal baseline separation. In this case, the observation time $t_1$ entering the expression for the noise of Eq.~(\ref{eq:Pnoise_K}) is the same as the total survey duration\footnote{Calculation of the observation time $t_1$, given total survey duration $t_\text{obs}$, depends on the type of the experiment. For a radio dish with a beam of solid angle $\Omega_\text{beam}=\lambda^2/A_e$ (smaller than the survey size $\Omega_\text{survey}$), where the telescope scans the sky one beamwidth at a time, $t_1$ is the total time spent observing one $(u,v)$ element, and thus $t_1=t_\text{obs}\Omega_\text{survey}/\Omega_\text{beam}$.}, $t_1=t_\text{obs}$. We do not explicitly account for the fact that any given portion of the sky is above the horizon of a given location only for a part of a day. Therefore, $t_\text{obs}$ we substitute in the noise calculation is shorter than the corresponding wall--clock duration of the survey (by a factor equal to the fraction of the day that a given survey region is above the horizon). To derive numerical results, we assume $\Omega_\text{survey}=1$sr and $t_\text{obs}=1$ year (corresponding to the wall--clock observing time on the order of three years). To compute sky temperature, we assume a simple model for Galactic synchrotron emission from Ref.~\cite{2008PhRvD..78b3529M}, 
\beq
T_\text{sky}  = 60\left(\frac{21}{100} (1+z)\right)^{2.55}\text{   [K]}.
\label{eq:tsys}
\eeq
We take the observed redshift range to be $z\in[15,30]$. 

Other inputs to the sensitivity calculation are shown in Fig.~\ref{fig:cosmo}: the mean Lyman--$\alpha$ flux as a function of redshift (top panel); the spin and kinetic temperatures of the IGM, along with the CMB temperature, also as functions of redshift (middle panel); and the quantities that parametrize the rate of depolarization of the ground state by optical pumping and atomic collisions, and the rate of magnetic precession, for a representative value of the magnetic field (bottom panel). We obtain the quantities from the top two panels from the \texttt{21CMFAST} code \cite{2011MNRAS.411..955M}, and the matter power spectra from the \texttt{CAMB} code \cite{2000ApJ...538..473L}. As inputs to \texttt{21CMFAST} and \texttt{CAMB}, we use standard cosmological parameters ($H_0=67$ km s$^{-1}$ Mpc$^{-1}$, $\Omega_\text{m}=0.32$, $\Omega_K=0$, $n_s=0.96$, $\sigma_8=0.83$, $w=-1$) consistent with Planck measurements \cite{2015arXiv150201589P}. For the \texttt{21CMFAST} runs, we set the sources responsible for early heating to Population III stars by setting \verb|Pop|$=3$, and keep all other input parameters at their default values, with the exception of the star formation efficiency, \verb|F_STAR|. For our fiducial calculation (denoted with solid curves in Fig.~\ref{fig:cosmo}), we choose \verb|F_STAR|=$0.0075$, but we also explore two other reionization models, as discussed below. The fiducial model is chosen to match the models from Ref.~\cite{2012ApJ...746..125H} at $z=15$ (which were computed by extrapolation of the flux measurements from observations at much lower redshifts). We tested that this fiducial model is physically reasonable, in the sense that it produces a sufficient number of ionizing photons to reionize the universe; we detail these tests in Appendix \ref{app:fesc}. 

Since the evolution of the Lyman--$\alpha$ flux prior to reionization is unconstrained by observations, we vary our input flux model (and, correspondingly, the models for the temperaures and depolarization rates) in order to capture the effect of this uncertainty on the key results of our sensitivity calculation. Specifically, we consider two ``extreme'' models for the Lyman--$\alpha$ flux, shown in the top panel of Fig.~\ref{fig:cosmo} as the extrema of the gray band of uncertainty around the fiducial $J_{\text{Ly}\alpha}(z)$ curve. They are obtained from \texttt{21CMFAST} runs with \verb|F_STAR|$=0.01875$  (for the top edge of the gray band), and \verb|F_STAR|$=0.0025$ (bottom edge).  Note that the rest of the panels in this Figure only show the fiducial model in order to avoid clutter, but the corresponding variation in all quantities is consistently included in the calculations. 

\begin{figure}
\centering
\includegraphics[width=.4\textwidth,keepaspectratio=true]{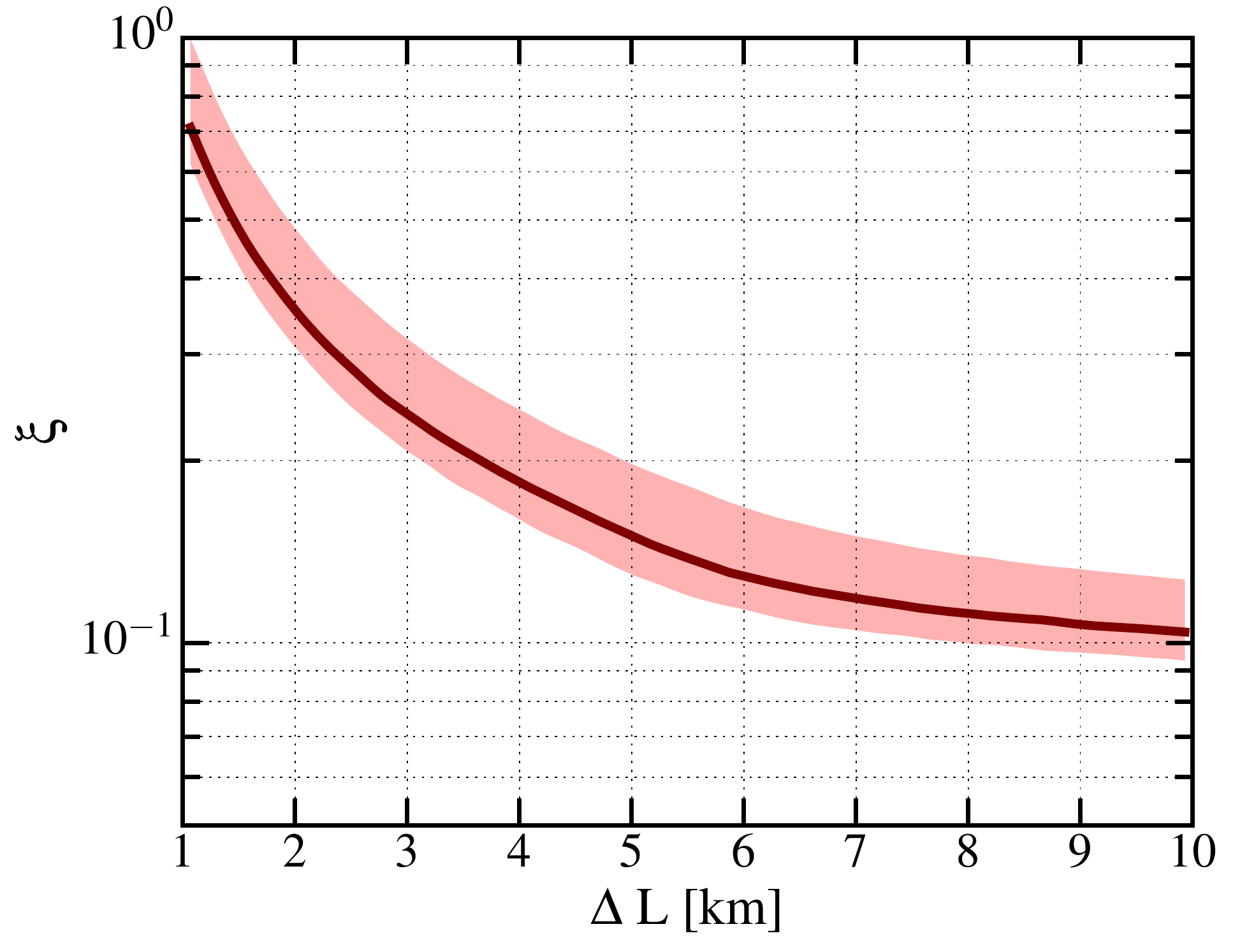}
\caption{Projected $1\sigma$ threshold of an array of dipoles in a compact--grid configuration for detecting a cosmological magnetic field in the saturated regime, as a function of the maximum array baseline. We assume a survey size of 1 sr, a total observation time of three years, and a collecting area of ($\Delta L)^2$. The parameter on the $y$ axis quantifies distinguishability of the case of no magnetic field ($\xi=0$) from a strong magnetic field ($\xi=1$). Smaller thresholds (for larger maximum--baseline values shown on the $x$ axis) correspond to a higher sensitivity for recovering $\xi$, and thus to a better prospect for distinguishing between the two regimes. The light--colored band around the solid line corresponds to the Lyman--$\alpha$ model flux variation represented with a gray band in Fig.~\ref{fig:cosmo}.\label{fig:xi_vs_deltas}}
\end{figure}
\begin{figure}
\centering
\includegraphics[width=.4\textwidth,keepaspectratio=true]{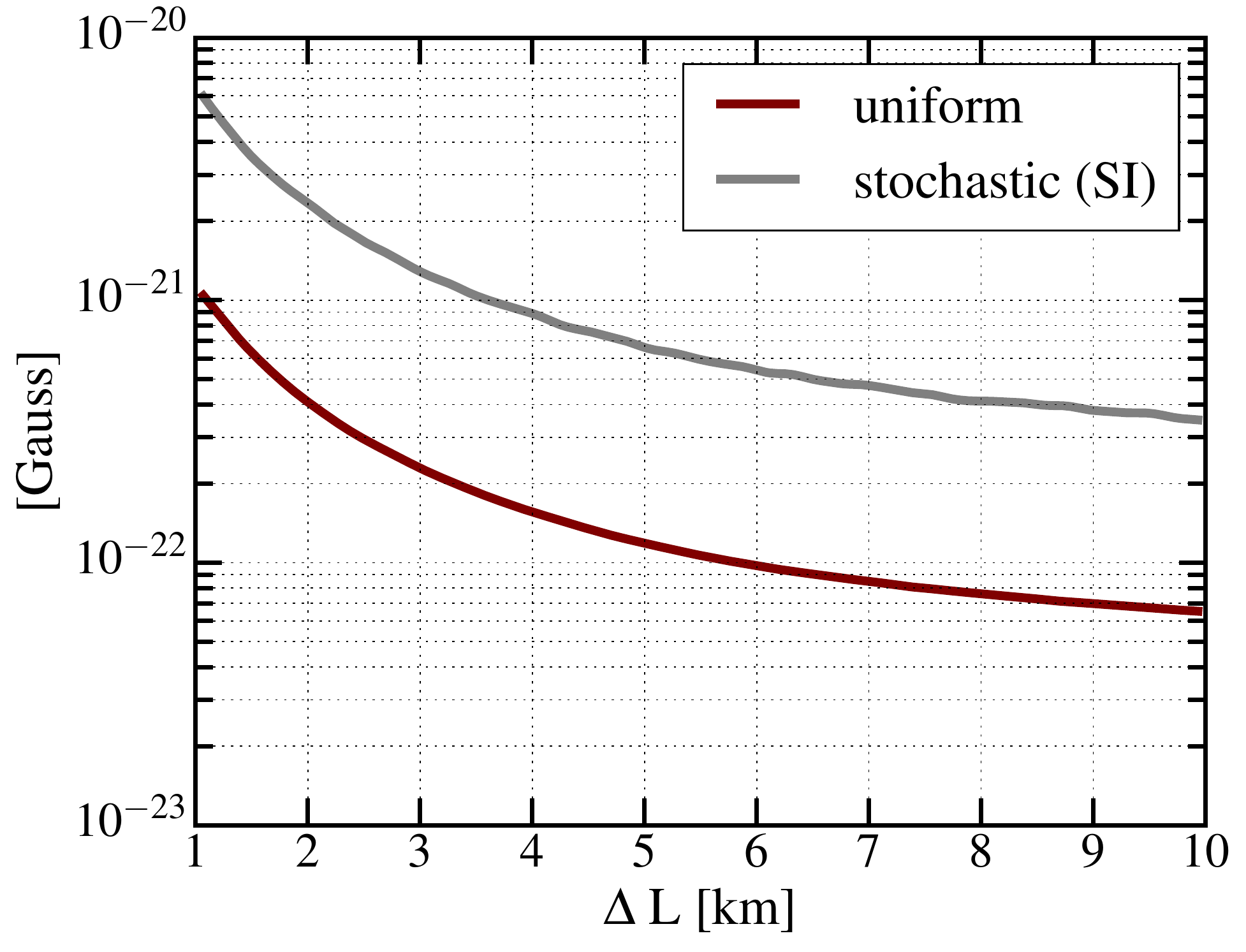}
\caption{Projected $1\sigma$ threshold of an array of dipoles in a compact--grid configuration for detecting cosmological magnetic fields. We assume a survey size of 1 sr, a total observing time of three years, and a collecting area of $(\Delta L)^2$. Thresholds for a uniform (lower red line) and a stochastic (upper gray line) magnetic field are shown as a function of maximum array baseline $\Delta L$.  For the stochastic field, a scale--invariant (SI) power spectrum is assumed; we plot the $1\sigma$ error for measuring the root--mean--square variation of the field magnitude per $\log K$, or $A_0/\pi$ (with $A_0$ defined in the text).\label{fig:B_vs_deltas}}
\end{figure}
\begin{figure}
\centering
\includegraphics[width=.4\textwidth,keepaspectratio=true]{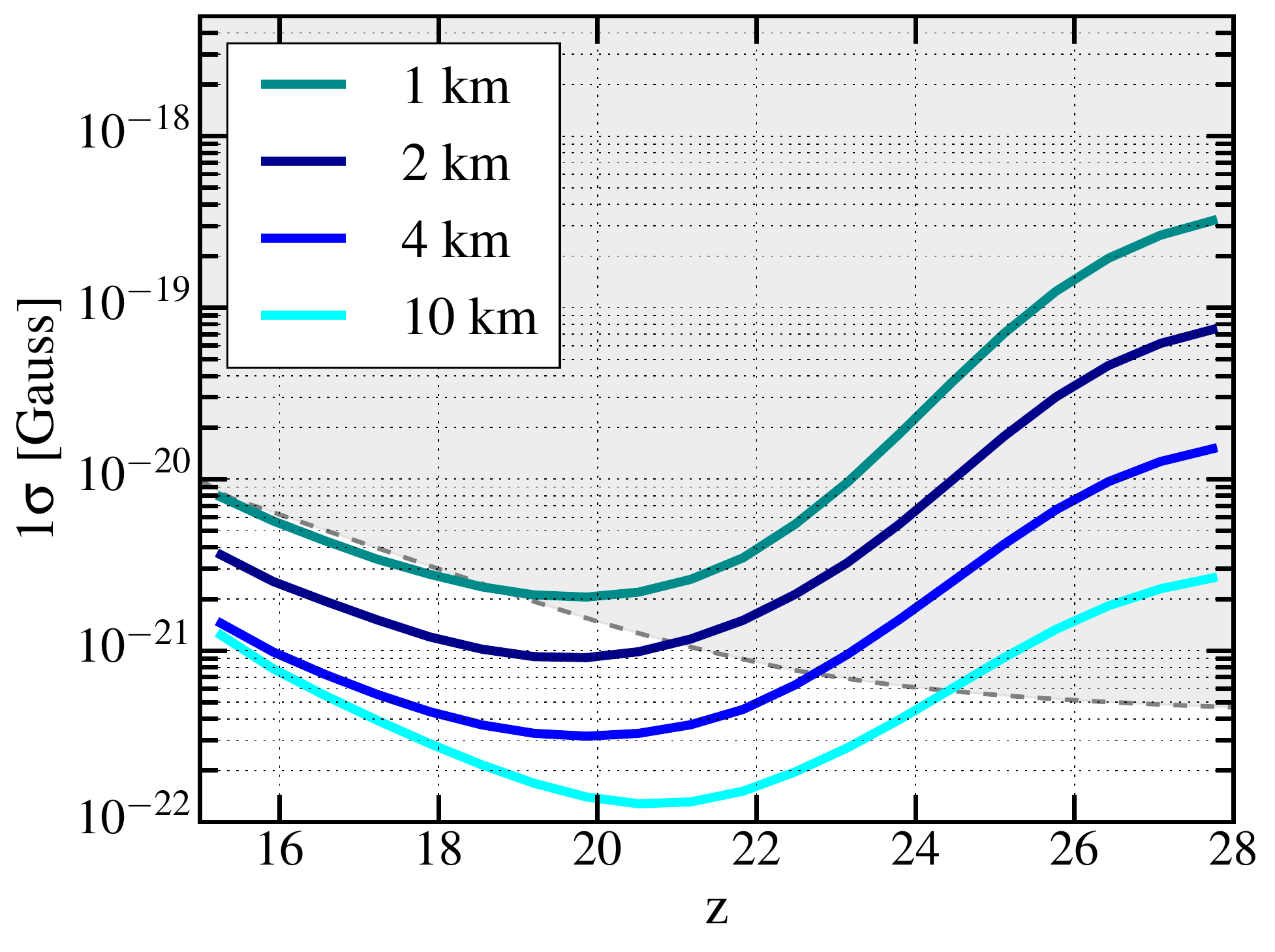}
\caption{Saturation regime is shown as a shaded gray area above the dashed curve (saturation ceiling). Integrand of \eq{\ref{eq:fisher_patch}} (inverse sqare root of it) is shown as a function of redshift, for several maximum--baseline lengths.  When the integrand values are below the saturation ceiling, the analysis assuming the unsaturated regime is valid. For the baseline lengths considered here, this is indeed the case for integrand values around their minima (corresponding to redshifts of maximal signal--to--noise for magnetic--field detection; note that the saturation ceiling is conservatively calculated for the purposes of this illustration), for arrays with collecting areas slightly above a square kilometer. This implies that the projections for the uniform--field case of Fig.~\ref{fig:B_vs_deltas} are valid. \label{fig:Bsat}}
\end{figure}

Figs.~\ref{fig:xi_vs_deltas} and \ref{fig:B_vs_deltas} show our key results: the projected $1\sigma$ detection thresholds for tomographic surveys, as a function of the maximum baseline $\Delta L$ (where different values of $\Delta L$ may correspond to different stages of a single experiment). Fig.~\ref{fig:xi_vs_deltas} shows the $1\sigma$ threshold to measuring the parameter $\xi$ of \eq{\ref{eq:saturated_P}} which quantifies the inferred preference for zero magnetic field versus the case where the field is strong and the signal is in the saturated regime. The value of this parameter is, by definition, bounded between 0 and 1 (representing the case of no magnetic field and the saturated case, respectively). In this Figure, the solid line corresponds to our fiducial calculation, while the light--colored band around it corresponds to the level of variation in the input Lyman--$\alpha$ flux shown as a grey band in Fig.~\ref{fig:cosmo}. The fiducial result implies that an array of dipoles with one square kilometer of collecting area can achieve enough sensitivity to detect a magnetic field in the saturation regime. Such detection of a non--vanishing value of $\xi$ can then be interpreted as a lower bound on a uniform magnetic field, at a $1\sigma$ confidence level (assuming the field is uniform in the entire survey volume). The value of the lower bound as a function of redshift  corresponds, in this case, to the saturation ``ceiling'' at that redshift, which can be roughly evaluated by requiring that the depolarization rates through standard channels equal the rate of magnetic precession, $x_B = 1+x_{\alpha ,(2)} +x_{c,(2)}$. The ceiling is depicted with a dashed line in Fig.~\ref{fig:Bsat}, and it corresponds to $|\vec B|\approx10^{-21}$ Gauss (comoving) at $z=21$, for example.  On the other hand, if a survey were to report a null result, it would rule out such a magnetic field, at the same confidence level. In this case, the result would imply an upper bound on the strength of the magnetic field components in the plane of the sky, as discussed in the following. 

We obtain results in Fig.~\ref{fig:B_vs_deltas} by evaluating Eqs.~(\ref{eq:fisher_patch}) and (\ref{eq:snr_ints}). This Figure shows a projected $1\sigma$ upper bound that can be placed on the value of the magnetic field, in case of no detection with an array of a given size. The result is shown for both the uniform field (lower solid red line), and for the amplitude of a stochastic field with a scale--independent power spectrum (upper gray line). It implies that an array with one square kilometer collecting area may reach a $1\sigma$ detection threshold of $10^{-21}$ Gauss comoving, after three years of observing 1 sr of the sky. 

While the numerical calculation behind this result assumes that the brightness--temperature signal is a linear function of the field strength, this assumption is not guaranteed to hold---it breaks down in the limit of a strong field, as discussed above and in \S\ref{sec:method}. So, the results of Fig.~\ref{fig:B_vs_deltas} are only valid if the value of the $\xi$ parameter is measured to be small at high confidence. In order to demonstrate how these projected constraints compare to the saturation ceiling, Fig.~\ref{fig:Bsat} shows the saturation ceiling and the values of the integrand of \eq{\ref{eq:fisher_patch}} (as a function of redshift, plotted for several array sizes). We see that the sensitivity of arrays with collecting areas slightly above one square kilometer is sufficient to reach below the saturation ceiling for redshifts contributing most of the signal--to--noise, $z$$\sim$$21$ (the minima of these curves). This gives us confidence that the results for the uniform field in Fig.~\ref{fig:B_vs_deltas} are indeed valid, and the linear theory holds in a given regime (the transfer function is a linear function of the field strength). For the stochastic case, however, it is likely that collecting areas larger by a factor of $\sim$10 will be needed to achieve detection thresholds below saturation at relevant redshifts. It is important to note two things here. First, the saturation ceiling presented in this Figure is quite conservatively calculated, and the linear approximation may hold for field strengths a few times above this level (for illustration, see also Fig.~\ref{fig:hp}). Second, a downwards variation of the Lyman--$\alpha$ flux by a factor of a few from our fiducial model at redshift of $\sim$21 can easily change relative values of the ceiling and detection thresholds of a one--square--kilometer array, placing the result into the unsaturated regime and enabling detection of a uniform field on the order of $10^{-21}$ Gauss comoving with such an array; however, the converse is also true. 

\section{Summary and Discussion}
\label{sec:conclusions}

In Paper I of this series, we proposed a new method to detect extremely weak magnetic fields in the IGM during the cosmic Dark Ages, using 21--cm tomography. In this Paper, we forecast the sensitivity of this method for future 21--cm tomography surveys. For this purpose, we developed a minimum--variance estimator for the magnetic field, which can be applied to measurements of the 21--cm brightness--temperature fluctuations prior to the epoch of reionization. The main numerical results are shown in Figs.~\ref{fig:xi_vs_deltas} and \ref{fig:B_vs_deltas}. They imply that a radio array in a compact--grid configuration with a collecting area slightly larger than one square kilometer can achieve $1\sigma$ sensitivity to a uniform magnetic field of strength $\sim$$10^{-21}$ Gauss comoving, after three years of observation. The case of a stochastic field is more challenging (by a factor of a few in sensitivity for a field with a scale--invariant power spectrum), and detection in that case would require $\sim$10 times larger collecting area.

We have only considered an array of dipole antennas in a compact--grid configuration, such as the proposed Fast Fourier Transform Telescope (FFTT) \cite{2009PhRvD..79h3530T}. Our calculations are, however, also applicable to compact arrays of dishes, with the caveat that they have a smaller instantaneous field of view than FFTT and hence have to observe for longer in order to reach the same sensitivity threshold (for a fixed collecting area). Such a design will soon be implemented in the Hydrogen Epoch of Reionization Array (HERA) [29], and our forecasts can be easily rescaled for the next generation of this experiment.

The prospect for measuring cosmological magnetic fields using this method depends on the rate of depolarization of the ground state of hydrogen through Lyman--$\alpha$ pumping, which is proportional to the mean Lyman--$\alpha$ flux prior to reionization. As shown in Fig.~\ref{fig:Bsat}, most of the sensitivity to magnetic fields (for the setup considered in this work) comes from $z\sim 21$, where the Lyman--$\alpha$ flux sufficiently decreases, while the kinetic temperature of the IGM is still low enough. However, the value of the mean Lyman--$\alpha$ flux at these redshifts is completely unconstrained by observation. While the fiducial model we used in our calculations represents one that satisfies modeling constraints and can be extrapolated to match low--redshift observations, it does not capture the full range of possibilities. It is thus important to keep in mind that the projected sensitivity can vary depending on this quantity. We qualitatively capture the variation in projected sensitivity by exploring Lyman--$\alpha$ flux models that vary within a factor of a few from the fiducial model, as shown in Fig.~\ref{fig:cosmo}.

In our analysis, we took into account the noise component arising from Galactic synchrotron emission, but we ignored more subtle effects (such as the frequency dependence of the beams, control of systematic errors from foreground--cleaning residuals, etc.) which may further complicate reconstruction of the magnetic--field signal and should be taken into account when obtaining detailed figures of merit for future experiments.
Finally, we note that the effect of cosmic shear on the 21--cm signal (from weak lensing of the signal by the intervening large scale structure) can produce a noise bias for the magnetic--field measurements. In Appendix \ref{app:lensing}, we examine the level of lensing contamination and show that it is small even for futuristic array sizes of a hundred square kilometers of collecting area. 

It is worth emphasizing again that the main limitation of this method is that it relies on effects that require two--scattering processes. As soon as the quality of cosmological 21--cm statistics reaches the level necessary to probe second--order processes, the effect of magnetic precession we discussed here will lend unprecedented precision to a new \textit{in situ} probe of minuscule, possibly primordial, magnetic fields at high redshifts.

\acknowledgements

VG gratefully acknowledges the support from the W. M. Keck Foundation Fund at the Institute for Advanced Study. TV gratefully acknowledges support from the Schmidt Fellowship and the Fund for Memberships in Natural Sciences at the Institute for Advanced Study. XF is supported by the Simons Foundation and is grateful to Joseph McEwen for useful discussions. AM, CH, and AO are supported by the U.S. Department of Energy, the David \& Lucile Packard Foundation, and the Simons Foundation. The authors thank Juna Kollmeier and Francesco Haardt for useful conversations about the Lyman--$\alpha$ flux evolution. Illustrations in Fig.~\ref{fig:hp} made use of HEALPix \cite{2005ApJ...622..759G} software package\footnote{\url{ http://healpix.sf.net}; \url{https://github.com/healpy/healpy}}. 

The complete code implementing all the calculations presented in this work, along with the \texttt{21CMFAST} reionization histories used as input, is available at \url{https://github.com/veragluscevic/pmfs}.  
\appendix 
\section{Visibility variance}
\label{app:Vrms}

Here we derive the variance of the visibility for an interferometric array of two antennas separated by a baseline $\vec{b}=(b_x,b_y)$, each with an effective collecting area $A_e$, observing a single element in the $uv$ plane for time duration $t_1$, with total bandwidth $\Delta \nu = \nu_\text{max}-\nu_\text{min}$. We choose notation that is consistent with the rest of this Paper, and adapted to the purpose of discussing measurement of a cosmological signal (as opposed to the traditional context of radio imaging). However, similar derivation can be found in the radio astronomy literature (see, e.g., Refs.~\cite{2001isra.book.....T,1986sicn.book.....P}), and in the literature discussing forecasts for 21--cm experiments (see, e.g., Refs.~\cite{2008PhRvD..78b3529M,2009astro2010S..82F,2014ApJ...782...66P,2007ApJ...661....1B,2008PhRvL.100i1302K,2008PhRvD..78b3529M}).

\begin{figure}
\centering
\includegraphics[width=.5\textwidth,keepaspectratio=true]{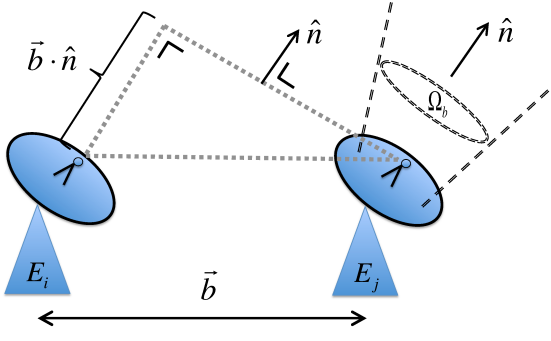}
\caption{Schematic of a two--antenna interferometer.\label{fig:2antennae}}
\end{figure}

A schematic of the experimental setup considered here is shown in Fig.~\ref{fig:2antennae}. Modes with frequencies that differ by less than $1/t_1$ cannot be distinguished, and modes with frequencies in each interval $1/t_1$ are collapsed into a discrete mode with frequency $\nu_n = n/t_1$, where $n\in Z$. Thus, the number of measured (discrete) frequencies is $N_\nu=t_1\Delta \nu$. Electric field induced in a single antenna is
\beq
E(t) = \sum_{n}^{N_\nu}\widetilde{E}(\nu_n)e^{2\pi i\nu_nt},
\eeq
while the quantity an interferometer measures is the correlation coefficient between the electric field $E_i$ in one and the electric field $E_j$ in the other antenna, as a function of frequency,
\beq
\rho_{ij}(\nu) \equiv \frac{\langle \widetilde{E}^*_i(\nu)\widetilde E_j(\nu)\rangle}{\sqrt{\langle |\widetilde{E}_i(\nu)|^2\rangle\langle|\widetilde E_j(\nu)|^2\rangle}}.
\label{eq:rho_ij}
\eeq

Let us now assume that 
\beq
\bga
\langle \widetilde{E}^*_i(\nu_n)\widetilde E_j(\nu_m)\rangle=\sigma(\nu)^2\delta_{mn}.
\ega
\label{eq:var_ReE}
\eeq
In the following, for clarity, we omit the dependence on $\nu$.  The real (or imaginary) part of $\rho$ has the following variance
\beq
\bga
\text{var}(Re[\rho_{ij}]) 
\frac{1}{2N_\nu} = \frac{1}{2t_1\Delta \nu}.
\ega
\label{eq:var_Rerho}
\eeq

Before continuing, let us take a brief digression to show that the above formula implicitly assumes that the electric fields in the two antennas have a very weak correlation, $\rho\ll 1$. Consider two random Gaussian variables, $x$ and $y$, both with zero mean values, where $\text{var(x)}\equiv\langle(x-\langle x\rangle)^2\rangle = \langle x^2\rangle - \langle x \rangle^2=\langle x^2\rangle$, and similarly for $y$. Their correlation coefficient is $\rho\equiv \frac{\langle xy\rangle}{\sqrt{\langle x^2\rangle \langle y^2\rangle}}$. In this case, the following is true
\beq
\bga
\text{var}(xy) = \langle x^2y^2\rangle -  \langle xy \rangle^2 = 
\langle x^2\rangle \langle y^2\rangle + \langle xy\rangle^2\\
=\langle x^2\rangle \langle y^2\rangle+\rho^2\langle x^2\rangle\langle y^2\rangle=\text{var}(x)\text{var}(y)(1+\rho^2),
\ega
\eeq
so that when $\rho$ is small, $\text{var}(xy)=\text{var}(x)\text{var}(y)$, which was assumed in the first equality of \eq{\ref{eq:var_Rerho}}.

Resuming the derivation, if different frequencies are uncorrelated, the result of \eq{\ref{eq:var_Rerho}} implies
\beq
\langle|\rho_{ij}(\nu)|^2\rangle = \frac{1}{t_1\Delta \nu}.
\label{eq:var_rho}
\eeq
The final step requires a relation between intensity on the sky $\mathcal{I}(\theta_x,\theta_y, \nu)$ (within the beam solid angle $\Omega_\text{beam}$, centered on the direction ${\bf{\widehat n}}=(\theta_x,\theta_y)$) and the electric fields measured in the two antennas,
\beq
\bga
\langle \widetilde{E}_i^*(\nu)\widetilde{E}_j(\nu)\rangle \propto \int_{\Omega_\text{beam}} d\theta_xd\theta_y\mathcal{I}(\theta_x,\theta_y,\theta_\nu)\\
\times e^{ i\frac{2\pi\nu}{c}(b_x\theta_x + b_y\theta_y)  }R(\theta_x,\theta_y),
\ega
\label{eq:E_vs_mathcalI}
\eeq
where $R(\theta_x,\theta_y)$ is the antenna response function (the shape of the beam in the sky), which we will assume to be unity. Furthermore, $\frac{2\pi\nu}{c}(b_x\theta_x + b_y\theta_y)\equiv {2\pi}(u\theta_x + v\theta_y)$ is the phase delay between the two antennas (position in the $uv$ plane measures the phase lag between the two dishes in wavelengths). The coefficient of proportionality in the above Equation is set by various instrumental parameters and is not relevant for our purposes. From \eq{\ref{eq:rho_ij}}, it follows that
\beq
\rho_{ij}(\nu) = \frac{\int_{\Omega_\text{beam}}d\theta_xd\theta_y\mathcal{I}(\theta_x,\theta_y,\theta_\nu)e^{2\pi i(u\theta_x+v\theta_y)}}{\int_{\Omega_\text{beam}}d\theta_xd\theta_y\mathcal{I}(\theta_x,\theta_y,\theta_\nu)},
\label{eq:rho_mathcalI}
\eeq
where the denominator in the above formula approximately integrates to (for a small beam)
\beq
\int_{\Omega_\text{beam}}d\theta_xd\theta_y\mathcal{I}(\theta_x,\theta_y,\theta_\nu) \approx
\Omega_\text{beam} \mathcal{I}(\theta_x,\theta_y,\theta_\nu).
\label{eq:rho_denominator}
\eeq
We can now use the approximate expression for the resolution of a single dish,
\beq
\Omega_\text{beam} = \frac{\lambda^2}{A_e},
\label{eq:Omegab}
\eeq
the Rayleigh--Jeans law (or the definition of the brightness temperature),
\beq
\mathcal{I}(\theta_x,\theta_y,\theta_\nu) = \frac{2k_BT_\text{sky}}{\lambda^2},
\label{eq:I_Tsky}
\eeq
and note that the numerator in \eq{\ref{eq:rho_mathcalI}} matches the definition of visibility from \eq{\ref{eq:visibility}}, to get 
\beq
\rho_{ij}(\nu) = \frac{A_e}{2k_BT_\text{sky}}\mathcal{V}(u,v,\theta_\nu).
\label{eq:rho_V}
\eeq

Combining Eq.~(\ref{eq:rho_V}) and Eq.~(\ref{eq:var_rho}), we get the final result of this derivation,
\beq
\bga
\langle|\mathcal{V}(u,v,\theta_\nu)|^2\rangle 
=\frac{1}{\Omega_\text{beam}}\left(\frac{2k_BT_\text{sky}}{A_e\sqrt{t_1\Delta \nu}}\right)^2\\
\times\delta_D(u-u')\delta_D(v-v')\delta_{\theta_\nu\theta_{\nu'}},
\ega
\label{eq:Vrms_final}
\eeq
where the visibility $\mathcal{V}$ is a complex Gaussian variable, centered at zero, and uncorrelated for different values of its arguments, and the factor of $\Omega_\text{beam}$ came from converting from Kronecker delta to a  Dirac delta function. Note finally that we considered the contribution to the visibility from the noise only (the system temperature + the foreground sky temperature, in the absence of a signal); in the presence of a signal, $T_\text{sky}$ should be the sum of the signal and the noise temperatures.

\label{app:Vrms} 
\section{Lensing noise bias}

We now consider weak gravitational lensing of the 21--cm signal by the large scale structure, as a source of noise in searches for magnetic fields using the method proposed in this work. We first compute the transverse shear power spectrum and then evaluate the noise bias it produces for the magnetic--field estimator. We demonstrate that this bias is very small, even for arrays with futuristic collecting areas of one hundred square kilometers.

To follow standard lensing notation, we no longer label cartesian coordinate axes with $x$, $y$, and $z$, but rather with numbers, using the convention where directions $1$ and $2$ lie in the plane of the sky, while $3$ lies along the line of sight. Specifically, we use angular coordinates $(\theta_1, \theta_2)$ to denote direction in the sky ${\bf{\widehat n}}$, and $\theta_3$ to denote a comoving interval $r_z/\chi(z)$ along the line of sight, located at redshift $z$, and corresponding to $\Delta z$ interval. As before, we denote variables in Fourier space with tilde. We use $\vec{\ell}\equiv(\ell_1,\ell_2)$ for a conjugate variable of ${\bf{\widehat n}}$. 
 
We start by generalizing the formalism for two--dimensional weak lensing \cite{Weinberg201387} to the three--dimensional case.
In the presence of lensing, a source coordinate $\theta_i^S$, where $i\in\{1,2,3\}$, maps onto the observed coordinate $\theta_i$ as follows 
\beq
\theta_k^S=\theta_k+\frac{\partial\psi}{\partial\theta_k},\ k=1,2,\ \ \ \theta_3^S=\theta_3,
\label{eq:lensingmapping}
\eeq
where $\psi$ is the lensing potential. The full Jacobian of this coordinate transformation is
\beq
\bga
\mathcal{J}_{ij}\equiv\frac{\partial\theta_i^S}{\partial\theta_j}=\left(\begin{array}{ccc}
1+\psi_{,11} & \psi_{,12} & \psi_{,13}\\
\psi_{,21} & 1+\psi_{,22} & \psi_{,23}\\
0&0&1
\end{array}\right) \\
= \left(\begin{array}{ccc}
1+\kappa+\gamma_{11} & \gamma_{12} & \gamma_{13}\\
\gamma_{12} & 1+\kappa-\gamma_{11} & \gamma_{23}\\
0&0&1
\end{array}\right),
\ega
\label{eq:lensingtsf}
\eeq
where $i,j\in\{1,2,3\}$, and the commas stand for partial derivatives with respect to the corresponding coordinates, as usual. In the above Equation, $\kappa$ and $\gamma$ components represent the magnification and shear, respectively. Fourier transform of the lensing potential is
\begin{equation}
\widetilde{\psi}(\vec{\ell},z)\equiv\int\psi({\bf\widehat{n}},z)e^{-i\vec{\ell}\cdot{\bf{\widehat n}}}\ d\theta_1 d\theta_2,
\label{eq:potentialFourier}
\end{equation}
where the relation between $\psi({\bf\widehat{n}},z)$ and the Newtonian potential $\Phi$ in a flat universe reads
\begin{equation}
\psi({\bf{\widehat n}},z)=
-2\int_0^{\chi(z)}d\chi_1\left[\frac{1}{\chi_1}-\frac{1}{\chi}\right]\Phi({\bf{\widehat n}},\chi_1).
\label{eq:lensingpotential}
\end{equation}
Combining Eqs.~(\ref{eq:potentialFourier}) and (\ref{eq:lensingpotential}), we get 
\begin{equation}
\frac{\partial\widetilde{\psi}(\vec{\ell},z)}{\partial\theta_3}=-\frac{2}{\chi(z)}\int_0^{\chi(z)} d\chi_1\widetilde{\Phi}(\vec{\ell},\chi_1).
\label{eq:dpsi_dtheta3}
\end{equation}
From Eqs.~(\ref{eq:dpsi_dtheta3}) and (\ref{eq:lensingtsf}), it follows 
\beq
\bga
\langle\widetilde{\gamma}_{13}^*(\vec{\ell},z)\widetilde{\gamma}_{13}(\vec{\ell}',z')\rangle=\left\langle \ell_1\ell_1'\frac{\widetilde{\psi}^*(\vec{\ell},z)}{\partial\theta_3}\frac{\widetilde{\psi}(\vec{\ell}',z')}{\partial\theta_3}\right\rangle\\
=\frac{4\ell_1\ell_1'}{\chi(z)\chi(z')}\int_0^{\chi(z)}d\chi_1\int_0^{\chi(z')}d\chi_1'\langle\widetilde{\Phi}^*(\vec{\ell},\chi_1)\widetilde{\Phi}(\vec{\ell}',\chi_1')\rangle.
\ega
\eeq

We now define the three--dimensional Fourier transform $\widetilde{\widetilde\Phi}$ of the Newtonian potential,
\beq
\widetilde{\Phi}(\vec{\ell},\chi)\equiv\int\widetilde{\widetilde{\Phi}}(\vec{\ell},\ell_3)e^{i\ell_3\chi}\frac{d\ell_3}{2\pi}.
\eeq
Using this definition, we get
\beq
\bga
\langle\widetilde{\Phi}^*(\vec{\ell},\chi)\widetilde{\Phi}(\vec{\ell}',\chi')\rangle=\int\int\frac{d\ell_3}{2\pi}\frac{d\ell_3'}{2\pi}\langle\widetilde{\widetilde{\Phi}}^*(\vec{\ell},\ell_3)\widetilde{\widetilde{\Phi}}(\vec{\ell}',\ell_3')\rangle\\
\times e^{i(\ell_3'\chi'-\ell_3\chi)}. 
\ega
\label{eq:doubletilde}
\eeq
Assuming different modes are uncorrelated, we arrive at
\beq
\bga
\langle\widetilde{\widetilde{\Phi}}^*(\vec{\ell},\ell_3)\widetilde{\widetilde{\Phi}}(\vec{\ell}',\ell_3')\rangle\\
=(2\pi)^3\delta(\ell_3-\ell_3')\delta^2(\vec{\ell}-\vec{\ell}')P_{\Phi}(\sqrt{\ell_3^2+\ell^2}),
\ega
\label{eq:expectation_tildetildephi}
\eeq
where
\beq
\bga
P_{\Phi}(\ell)=\frac{P_{\Phi}(k=\ell/\chi(z))}{\chi(z)^2}\\
=\left[\frac{3}{2}\Omega_mH_0^2(1+z)\right]^2\frac{P_{\delta}(k,z)}{k^4\chi(z)^2}.
\ega
\eeq
Substituting Eq.~(\ref{eq:expectation_tildetildephi}) into (\ref{eq:doubletilde}) and applying Limber approximation $\ell_3\ll\ell$, we obtain
\beq
\bga
\langle\widetilde{\Phi}^*(\vec{\ell},\chi)\widetilde{\Phi}(\vec{\ell}',\chi')\rangle\\
=(2\pi)^2\delta(\vec{\ell}-\vec{\ell}')P_{\Phi}(\ell)\delta(\chi'-\chi).
\ega
\eeq
Thus, for $z\leq z'$,
\beq
\bga
\langle\widetilde{\gamma}_{13}^*(\vec{\ell},z)\widetilde{\gamma}_{13}(\vec{\ell}',z')\rangle\\
=\frac{4}{\chi(z)\chi(z')}\ell_1\ell_1'(2\pi)^2\delta^2(\vec{\ell}-\vec{\ell}')\int_0^{\chi(z)}d\chi_1P_{\Phi}(\ell).
\ega
\label{eq:exp_gamma13}
\eeq

We are interested in calculating the power spectrum $P_{13}(\vec{\ell},z,z')$ of $\gamma_{13}$ components, defined as
\beq
\bga
\langle\widetilde{\gamma}_{13}^*(\vec{\ell},z)\widetilde{\gamma}_{13}(\vec{\ell}',z')\rangle\\
\equiv(2\pi)^2P_{13}(\vec{\ell},z,z')\delta(\vec{\ell}-\vec{\ell}').
\ega
\eeq
From Eq.~(\ref{eq:exp_gamma13}), we can express
\beq
P_{13}(\vec{\ell},z,z')=\frac{4\ell_1^2}{\chi(z)\chi(z')}\int_0^{\chi(z)}d\chi_1P_{\Phi}(\ell).
\eeq
A similar result holds for the power spectrum $P_{23}$ of $\gamma_{23}$ component. The transverse power spectrum $P_t$ reads
\beq
\bga
P_t(\ell,z,z')\equiv P_{13}+P_{23}\\
=\frac{4\ell^2}{\chi(z)\chi(z')}\int_0^{\chi(z)}d\chi_1P_{\Phi}(\ell).
\ega
\eeq
If $z=z'$, the above expression simplifies to
\beq
P_t(\ell,z)=\frac{4\ell^2}{\chi(z)^2}\int_0^{\chi(z)}d\chi_1P_{\Phi}(\ell).
\label{eq:Pt}
\eeq

Now that we have computed the transverse power spectrum, we move on to evaluating the contamination it produces for the measurement of the magnetic field. Denoting a vector transpose with ``T'', let us set ${\bf{\widehat k}}=(\sin\theta\cos\phi,\sin\theta\sin\phi,\cos\theta)^{\rm T}$, and consider the line of sight along the direction 3, ${\bf{\widehat n}}=(0,0,1)^{\rm T}$, in the three--dimensional Cartesian reference frame where $x$, $y$, and $z$ axes correspond to 1, 2, and 3, respectively; $\theta$ is the angle between the direction 3 and ${\bf{\widehat k}}$. Lensing distorts ${\vec{k}}$ into
\beq
{\vec{k}}'=[\mathcal{J}^{-1}]^{\rm T}\cdot{\vec{k}}=\left(1-\frac{2\kappa}{3}\right){\vec{k}}+{\bm{\sigma}}\cdot{\vec{k}}+{\bf{\Omega}}\times{\vec{k}},
\label{eq:distorted_kn}
\eeq
where $\mathcal{J}$ is given by Eq.~(\ref{eq:lensingtsf}) and
\beq
\bga
\bm{\sigma}\equiv\left(\begin{array}{ccc}
-\kappa/3-\gamma_{11} & -\gamma_{12} & -\gamma_{13}/2 \\
-\gamma_{12} & -\kappa/3+\gamma_{11} & -\gamma_{23}/2 \\
-\gamma_{23}/2 & -\gamma_{23}/2 & 2\kappa/3
\end{array}\right),\\
\bm{\Omega}\equiv(-\gamma_{23}/2,\gamma_{13}/2,0)^{\rm T},
\ega
\eeq
where $\bm{\sigma}$ is a tensor quantity. The first term in Eq.~(\ref{eq:distorted_kn}) only changes the magnitude of ${\vec{k}}$, the third term only changes its direction, and the second term contributes to both changes. To leading order, the fractional magnitude change is
$(k'-k)/k=-2\kappa/3+\widehat{\bf{k}}\cdot\bm{\sigma}\cdot\widehat{\bf{k}}$.
We now define
\beq
C\equiv 26.4\ {\rm mK}\ \left(1-\frac{T_\gamma}{T_{\rm s}}\right)x_{\rm 1s}\left(\frac{1+z}{10}\right)^{1/2},
\eeq
and use Eqs.~(\ref{eq:distorted_kn}) and (\ref{eq:tbsoln}) to arrive at the expression for the brightness--temperature fluctuation in the presence of lensing (keeping only the leading--order terms and assuming no magnetic fields),
\beq
\bga
{T}_{\rm (lens)}(\widehat{\bf{n}},\vec{k})=\frac{1}{\det(\mathcal{J})} T\left(\widehat{\bf{n}},{\vec{k}}'\right)  \\
=T\left(\widehat{\bf{n}},{\vec{k}}\right)(1-2\kappa) +C\left\lbrace {\delta}({\vec{k}})2(\widehat{\bf{k}}\cdot\widehat{{\bf{n}}})\left[\widehat{{\bf{n}}}\cdot\bm{\sigma}\cdot\widehat{{\bf{k}}}\right.\right. \\
\left.\left.-(\widehat{\bf{k}}\cdot\widehat{{\bf{n}}})(\widehat{\bf{k}}\cdot\bm{\sigma}\cdot\widehat{{\bf{k}}})+(\bm{\Omega}\times\widehat{{\bf{k}}})\cdot\widehat{{\bf{n}}}\right]\right.\\
\left.+\left(-\frac{2\kappa}{3}{\vec{k}}+\bm{\sigma}\cdot{\vec{k}}
+\bm{\Omega}\times{\vec{k}}\right)\cdot\bm{\nabla}_{\vec{k}}{\delta}({\vec{k}})\left[1+(\widehat{\bf{k}}\cdot\widehat{{\bf{n}}})^2\right]\right\rbrace,
\ega
\eeq
where $\det(\mathcal{J})$ corresponds to the determinant of $\mathcal{J}$. The lensed signal power spectrum is then given by
\beq
\bga
P_{\rm (lens)}^S({\vec{k}})=C^2 P_{\delta}(k)\left(1+(\widehat{\bf{k}}\cdot\widehat{{\bf{n}}})^2\right) \\
\times\left\lbrace \left(1+(\widehat{\bf{k}}\cdot\widehat{{\bf{n}}})^2\right) \left[1-2\kappa\left(1+\frac{1}{3}\frac{\partial\ln P_\delta(k)}{\partial\ln k}\right)\right.\right.\\
\left.
+\frac{\partial\ln P_\delta(k)}{\partial\ln k}(\widehat{\bf{k}}\cdot\bm{\sigma}\cdot\widehat{{\bf{k}}})\right]
+4(\widehat{\bf{k}}\cdot\widehat{{\bf{n}}})\\
\left.\times\left(\left[\widehat{\bf{n}}-
(\widehat{\bf{k}}\cdot\widehat{{\bf{n}}})
\widehat{\bf{k}}\right]\cdot\bm{\sigma}\cdot\widehat{\bf{k}}
+(\bm{\Omega}\times\widehat{\bf{k}})\cdot\widehat{\bf{n}}\right)\right\rbrace,
\label{eq:Tb_power}
\ega
\eeq
where we use $\partial\ln P_\delta(k)/\partial\ln k\sim -2.15$ (the slope of the density--fluctuation power spectrum, evaluated at redshift and $k$ values that contribute most to the SNR for magnetic--field measurement). On the other hand, from Eq.~(\ref{eq:tbsoln}), a magnetic field contributes to the signal as
\beq
\bga
P_B^S({\vec{k}})=C^2 P_{\delta}(k)\left(1+(\widehat{\bf{k}}\cdot\widehat{{\bf{n}}})^2\right) \times\\
\left\lbrace \left(1+(\widehat{\bf{k}}\cdot\widehat{{\bf{n}}})^2\right) + 1.353\times 10^{16}\left(\frac{1+z}{10}\right)^{-1/2}\right. \\
\left.\times \frac{T_{\gamma}}{T_{\rm s}}\frac{x_{\rm 1s}}{(1+x_{\alpha,(2)}+x_{c,(2)})^2}\left[{\vec{B}}\cdot(\widehat{\bf{k}}\times\widehat{{\bf{n}}})\right](\widehat{\bf{k}}\cdot\widehat{{\bf{n}}})\right\rbrace,
\label{eq:Tbmag_power}
\ega
\eeq
where ${\vec{B}}$ is given in units of Gauss (physical, rather than comoving). Let us now consider a magnetic field in the $(1,2)$ plane, such that ${\vec{B}}=(B_x,B_y,0)$; the results will be valid for any field orientation. If we explicitly expand both Eq.~(\ref{eq:Tb_power}) and Eq.~(\ref{eq:Tbmag_power}) in terms of spherical harmonics, and consider only $Y_{2\pm 1}$ terms (which dominate the terms that are asymmetric around the line--of--sight direction; contribution from the higher--order harmonics is subdominant), we can match the coefficient of Eq.~(\ref{eq:Tb_power}) that corresponds to the multiplier to the magnetic--field strength of Eq.~(\ref{eq:Tbmag_power}). With this procedure, we arrive at the expression for the comoving value of the lensing--induced spurious magnetic field given by 
\beq
\bga
{\vec{B}}_{\rm (lens)}=1.577\times 10^{-18}{\left[\rm Gauss\right]}\ \times \frac{1}{x_{\rm 1s}}\left(\frac{T_s}{T_{\gamma}}\right)\left(\frac{1+z}{10}\right)^{-3/2}\\
\times(1+x_{\alpha,(2)}+x_{c,(2)})^2\left(1+\frac{11}{16}\frac{\partial\ln P_\delta(k)}{\partial\ln k}\right)\\
\times(-\gamma_{23},\gamma_{13},0)^{\rm T}
\equiv\alpha(-\gamma_{23},\gamma_{13},0)^{\rm T},
\label{eq:alpha_def}
\ega
\eeq
in units of comoving Gauss. The lensing noise bias for magnetic--field reconstruction reads
\beq
P_{(\text{lens})}^{\rm noise}(\ell)=P_{(\text{lens})}^{\rm noise, B_x}+P_{(\text{lens})}^{\rm noise, B_y}=\alpha^2 P_t(\ell),
\eeq
where $\alpha$ is given by Eq.~(\ref{eq:alpha_def}) and $P_t(\ell)$ is given by Eq.~(\ref{eq:Pt}). Finally, the root--mean--square of the contamination is given by
\beq
\Delta_{(\text{lens})}(\ell)=\sqrt{\frac{\ell(\ell+1)}{2\pi}P_{(\text{lens})}^{\rm noise}(\ell)}.
\label{eq:delta_lens}
\eeq

\begin{figure}[h]
\centering
\includegraphics[scale=0.4]{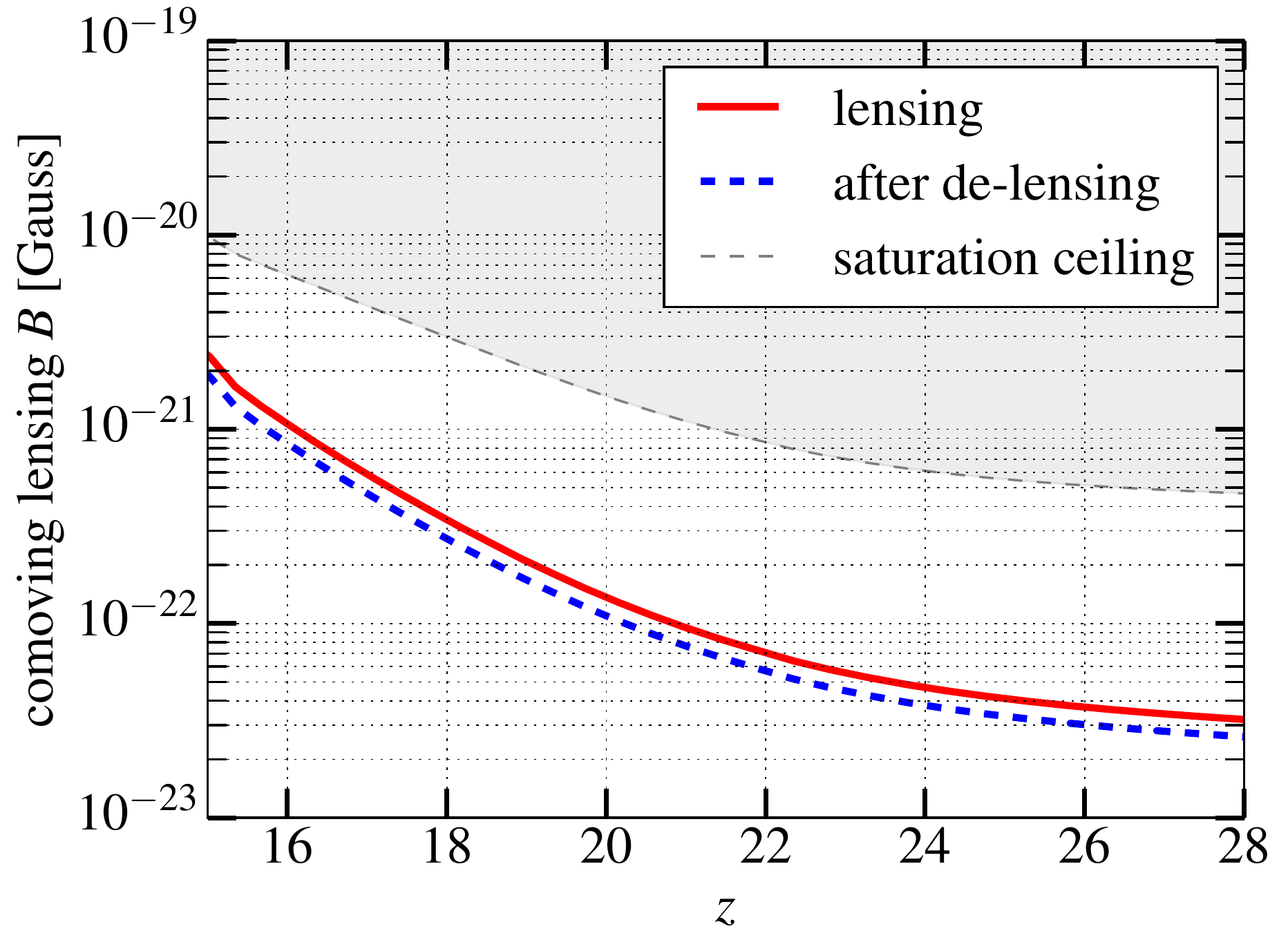}
\caption{The lensing--shear noise bias for the measurement of the magnetic field is shown before (solid red line) and after the de--lensing procedure is applied (dashed blue line). The saturation ceiling is denoted by the shaded region above the thin dashed line. Comparison with Fig.~\ref{fig:Bsat} reveals that lensing noise is below the projected sensitivity even for futuristic array sizes.}
\label{fig:lensing_B}
\end{figure}

A survey of size 1 sr, considered in this work, corresponds to $\ell\sim 6$, which relates to the lensing--potential fluctuations on comoving scale $\ell/D(z)\sim 5\times 10^{-4}{\rm Mpc}^{-1}$ at $z\sim 20$. We evaluate the contamination of Eq.~(\ref{eq:delta_lens}) at this multipole, which has a dominant contribution to the noise bias.\footnote{Note that the derivations shown in this Appendix hold only if the scale of matter fluctuations that contribute most to the lensing contamination are much larger than than those that contribute the most SNR for magnetic--field measurements, which is indeed the case here.}, and show the results in Fig.~\ref{fig:lensing_B}. Comparing this to Fig.~\ref{fig:Bsat}, we see that the contamination due to lensing shear remains below the projected sensitivities even for the case of futuristic array sizes. It may further be possible to distinguish lensing contribution from that of a magnetic field using difference in shapes of the inferred signal power spectra, but such detailed considerations are beyond the scope of this work.

\label{app:lensing}
\section{Estimating the escape fraction of ionizing photons}

This Appendix describes our method for estimating the escape fraction of ionizing photons in semi--numerical simulations of the high--redshift 21--cm signal. We use this estimate to perform a sanity check of the fiducial model of the Lyman--$\alpha$ flux evolution (shown in Fig.~\ref{fig:cosmo}) used for the sensitivity calculations shown in \S\ref{sec:results}. We computed this model using \texttt{21CMFAST}. In order to match the calculations of Ref.~\cite{2012ApJ...746..125H} at the lower end of the relevant redshift range ($z\sim 15$), we changed two of the default input parameters, setting the star--formation efficiency to $0.0075$, and the population of ionizing sources to Population III stars. We then checked that these parameters satisfy the constraint that the escape fraction of ionizing photons is bound to be less than one, at all redshifts of interest. 

\texttt{21CMFAST} sidesteps the computationally expensive tasks of tracking individual radiation sources and performing the radiative transfer of ionizing photons (needed to simulate HII regions in the early universe). It uses an approximate relation between the statistics of HII regions and those of collapsed structures, the latter of which can be efficiently computed in pure large--scale--structure simulations \cite{2004ApJ...613....1F}. Thus, the escape fraction of ionizing photons is not a direct input to these simulations, but can be estimated using the procedure we describe below. 

The number of ionizing photons emitted in a given ionized region, integrated up to a fixed redshift, should equal the number of absorbed ionized photons. These read, respectively, 
\beq    
\bga
N_{\rm em} = \langle f_{\rm esc} \rangle f_\ast N_{\gamma/{\rm b}} f_{\rm coll} N_{\rm b}\\
N_{\rm abs} = f_{\rm H} (1 + \langle n_{\rm rec} \rangle) N_{\rm b},
\ega
\label{eq:pbalance}
\eeq 
where $f_{\rm H} = 0.924$ is the hydrogen number fraction; $f_\ast$ is the star--formation efficiency (the fraction of galactic baryonic mass in stars; this is an input parameter to \texttt{21CMFAST}); $N_{\gamma/{\rm b}}$ is the number of ionizing photons produced by stars per nucleus; $N_{\rm b}$ is the total number of nuclei within a given ionized region; $\langle f_{\rm esc} \rangle$ is the average escape fraction associated with a given region; $\langle n_{\rm rec} \rangle$ is the average number of recombinations per hydrogen atom inside that region; and $f_{\rm coll}$ is the collapse fraction therein. We assume that once regions are ionized, they stay ionized. 

Integrating the number of absorbed photons of Eq.~(\ref{eq:pbalance}) over the set $\mathcal{R}(z)$ of all ionized regions at a given redshift, we get the total number of absorbed ionizing photons,
\beq
\bga
N_{\rm abs, tot}(z) = f_{\rm H} \int_{\mathcal{R}(z)} n_{\rm b} dV  \\
  + f_{\rm H}^2 \int_z^\infty dz^\prime \biggr\vert \frac{dt}{dz^\prime} \biggr\vert \int_{\mathcal{R}(z^\prime)} \mathcal{C} n_{\rm b}^2 \alpha_{\rm B} \ dV , 
\ega
\label{eq:netabs}
\eeq
where $n_{\rm b}$ is the baryon number density; the Jacobian $|dt/dz|$ maps between redshift and proper time; $\mathcal{C} \equiv \langle n_\text{b}^2 \rangle/\langle n_\text{b} \rangle^2$ is the clumping factor; and $\alpha_{\rm B}$ is the case--B recombination coefficient (varies from ionized region to ionized region). On the other hand, using the \texttt{21CMFAST} ansatz that $f_{\rm coll} = 1/\zeta$, where $\zeta$ is an efficiency factor (also given as an input to the code), the total number of emitted ionizing photons reads
\beq
\bga
N_{\rm em, tot}(z)  = \frac{ \overline{f_{\rm esc}}(z) f_\ast N_{\gamma/{\rm b}} }{\zeta} \int_{\mathcal{R}(z)}  n_{\rm b} \ dV,
\ega 
\label{eq:netem}
\eeq
where $\overline{f_{\rm esc}}(z)$ is the overall averaged escape fraction up to redshift $z$---the quantity we aim to estimate. Combining Eqs.~(\ref{eq:netabs}) and (\ref{eq:netem}), we get
\beq
\bga
  \overline{f_{\rm esc}}(z) = \frac{f_{\rm H} \zeta}{ f_\ast N_{\gamma/{\rm b} } } \\
\times\left[ 1 + f_{\rm H} \frac{ \int_z^\infty dz^\prime \biggr\vert \frac{dt}{dz^\prime} \biggr\vert \int_{\mathcal{R}(z^\prime)} \mathcal{C} n_{\rm b}^2 \alpha_{\rm B} \ dV  }{ \int_{\mathcal{R}(z)} n_{\rm b} \ dV} \right].
\ega
\eeq
Rewriting the above integrals in terms of comoving coordinates $\vec r$ and the overdensity $\delta(\vec r, z)$, we finally get
\vspace{-5pt}
\beq
\bga
\overline{f_{\rm esc}}(z) = \frac{f_{\rm H} \zeta}{ f_\ast N_{\gamma/{\rm b} } } \\
\times  \left[ 1 + \frac{ f_{\rm H} n_{\rm b, \text{today}} }{ \int_{ \mathcal{R}(z)} d\vec r[1 + \delta(\vec r, z)] } \int_z^\infty dz^\prime \biggr\vert \frac{dt}{dz^\prime} \biggr\vert \right.\\
\times \left. (1 + z^\prime)^3 \int_{\mathcal{R}(z^\prime)} d\vec r \ \mathcal{C} [1 + \delta(\vec r, z^\prime)]^2 \alpha_{\rm B} \right] .
\ega
\eeq 
where $n_{\rm b, \text{today}}$ is the number density of baryons today. An additional subtlety is that \texttt{21CMFAST} follows the kinetic temperature in the IGM outside the ionized regions, while the recombination coefficient $\alpha_{\rm B}$ depends on the temperature inside these regions. In general, the latter differs from the former due to the energy deposited by the free--electrons released during photoionization. We simplify our calculation by setting the temperature inside the bubbles to $10^4$ K (corresponding to the mean kinetic energy of the particles of a few eV). 

\label{app:fesc}

\bibliographystyle{apsrev4-1}
\bibliography{detectability}

\begin{thebibliography}{47}%
\makeatletter
\providecommand \@ifxundefined [1]{%
 \@ifx{#1\undefined}
}%
\providecommand \@ifnum [1]{%
 \ifnum #1\expandafter \@firstoftwo
 \else \expandafter \@secondoftwo
 \fi
}%
\providecommand \@ifx [1]{%
 \ifx #1\expandafter \@firstoftwo
 \else \expandafter \@secondoftwo
 \fi
}%
\providecommand \natexlab [1]{#1}%
\providecommand \enquote  [1]{``#1''}%
\providecommand \bibnamefont  [1]{#1}%
\providecommand \bibfnamefont [1]{#1}%
\providecommand \citenamefont [1]{#1}%
\providecommand \href@noop [0]{\@secondoftwo}%
\providecommand \href [0]{\begingroup \@sanitize@url \@href}%
\providecommand \@href[1]{\@@startlink{#1}\@@href}%
\providecommand \@@href[1]{\endgroup#1\@@endlink}%
\providecommand \@sanitize@url [0]{\catcode `\\12\catcode `\$12\catcode
  `\&12\catcode `\#12\catcode `\^12\catcode `\_12\catcode `\%12\relax}%
\providecommand \@@startlink[1]{}%
\providecommand \@@endlink[0]{}%
\providecommand \url  [0]{\begingroup\@sanitize@url \@url }%
\providecommand \@url [1]{\endgroup\@href {#1}{\urlprefix }}%
\providecommand \urlprefix  [0]{URL }%
\providecommand \Eprint [0]{\href }%
\providecommand \doibase [0]{http://dx.doi.org/}%
\providecommand \selectlanguage [0]{\@gobble}%
\providecommand \bibinfo  [0]{\@secondoftwo}%
\providecommand \bibfield  [0]{\@secondoftwo}%
\providecommand \translation [1]{[#1]}%
\providecommand \BibitemOpen [0]{}%
\providecommand \bibitemStop [0]{}%
\providecommand \bibitemNoStop [0]{.\EOS\space}%
\providecommand \EOS [0]{\spacefactor3000\relax}%
\providecommand \BibitemShut  [1]{\csname bibitem#1\endcsname}%
\let\auto@bib@innerbib\@empty
\bibitem [{\citenamefont {{Durrer}}\ and\ \citenamefont
  {{Neronov}}(2013)}]{2013A&ARv..21...62D}%
  \BibitemOpen
  \bibfield  {author} {\bibinfo {author} {\bibfnamefont {R.}~\bibnamefont
  {{Durrer}}}\ and\ \bibinfo {author} {\bibfnamefont {A.}~\bibnamefont
  {{Neronov}}},\ }\href {\doibase 10.1007/s00159-013-0062-7} {\bibfield
  {journal} {\bibinfo  {journal} {\aapr}\ }\textbf {\bibinfo {volume} {21}},\
  \bibinfo {eid} {62} (\bibinfo {year} {2013})},\ \Eprint
  {http://arxiv.org/abs/1303.7121} {arXiv:1303.7121 [astro-ph.CO]} \BibitemShut
  {NoStop}%
\bibitem [{\citenamefont {Vallee}(2004)}]{Vallee04}%
  \BibitemOpen
  \bibfield  {author} {\bibinfo {author} {\bibfnamefont {J.~P.}\ \bibnamefont
  {Vallee}},\ }\href {\doibase http://dx.doi.org/10.1016/j.newar.2004.03.017}
  {\bibfield  {journal} {\bibinfo  {journal} {New Astronomy Reviews}\ }\textbf
  {\bibinfo {volume} {48}},\ \bibinfo {pages} {763 } (\bibinfo {year}
  {2004})}\BibitemShut {NoStop}%
\bibitem [{\citenamefont {{Neronov}}\ and\ \citenamefont
  {{Vovk}}(2010)}]{Neronov10}%
  \BibitemOpen
  \bibfield  {author} {\bibinfo {author} {\bibfnamefont {A.}~\bibnamefont
  {{Neronov}}}\ and\ \bibinfo {author} {\bibfnamefont {I.}~\bibnamefont
  {{Vovk}}},\ }\href {\doibase 10.1126/science.1184192} {\bibfield  {journal}
  {\bibinfo  {journal} {Science}\ }\textbf {\bibinfo {volume} {328}},\ \bibinfo
  {pages} {73} (\bibinfo {year} {2010})},\ \Eprint
  {http://arxiv.org/abs/1006.3504} {arXiv:1006.3504 [astro-ph.HE]} \BibitemShut
  {NoStop}%
\bibitem [{\citenamefont {{Wielebinski}}(2005)}]{2005LNP...664...89W}%
  \BibitemOpen
  \bibfield  {author} {\bibinfo {author} {\bibfnamefont {R.}~\bibnamefont
  {{Wielebinski}}},\ }in\ \href {\doibase 10.1007/11369875_5} {\emph {\bibinfo
  {booktitle} {Cosmic Magnetic Fields}}},\ \bibinfo {series} {Lecture Notes in
  Physics, Berlin Springer Verlag}, Vol.\ \bibinfo {volume} {664},\ \bibinfo
  {editor} {edited by\ \bibinfo {editor} {\bibfnamefont {R.}~\bibnamefont
  {{Wielebinski}}}\ and\ \bibinfo {editor} {\bibfnamefont {R.}~\bibnamefont
  {{Beck}}}}\ (\bibinfo {year} {2005})\ p.~\bibinfo {pages} {89}\BibitemShut
  {NoStop}%
\bibitem [{\citenamefont {{Beck}}(2012)}]{2012SSRv..166..215B}%
  \BibitemOpen
  \bibfield  {author} {\bibinfo {author} {\bibfnamefont {R.}~\bibnamefont
  {{Beck}}},\ }\href {\doibase 10.1007/s11214-011-9782-z} {\bibfield  {journal}
  {\bibinfo  {journal} {\ssr}\ }\textbf {\bibinfo {volume} {166}},\ \bibinfo
  {pages} {215} (\bibinfo {year} {2012})}\BibitemShut {NoStop}%
\bibitem [{\citenamefont {{Park}}\ \emph {et~al.}(2013)\citenamefont {{Park}},
  \citenamefont {{Blackman}},\ and\ \citenamefont
  {{Subramanian}}}]{2013PhRvE..87e3110P}%
  \BibitemOpen
  \bibfield  {author} {\bibinfo {author} {\bibfnamefont {K.}~\bibnamefont
  {{Park}}}, \bibinfo {author} {\bibfnamefont {E.~G.}\ \bibnamefont
  {{Blackman}}}, \ and\ \bibinfo {author} {\bibfnamefont {K.}~\bibnamefont
  {{Subramanian}}},\ }\href {\doibase 10.1103/PhysRevE.87.053110} {\bibfield
  {journal} {\bibinfo  {journal} {\pre}\ }\textbf {\bibinfo {volume} {87}},\
  \bibinfo {eid} {053110} (\bibinfo {year} {2013})},\ \Eprint
  {http://arxiv.org/abs/1305.2080} {arXiv:1305.2080 [physics.plasm-ph]}
  \BibitemShut {NoStop}%
\bibitem [{\citenamefont {{Naoz}}\ and\ \citenamefont
  {{Narayan}}(2013{\natexlab{a}})}]{Naoz13}%
  \BibitemOpen
  \bibfield  {author} {\bibinfo {author} {\bibfnamefont {S.}~\bibnamefont
  {{Naoz}}}\ and\ \bibinfo {author} {\bibfnamefont {R.}~\bibnamefont
  {{Narayan}}},\ }\href {\doibase 10.1103/PhysRevLett.111.051303} {\bibfield
  {journal} {\bibinfo  {journal} {Physical Review Letters}\ }\textbf {\bibinfo
  {volume} {111}},\ \bibinfo {eid} {051303} (\bibinfo {year}
  {2013}{\natexlab{a}})},\ \Eprint {http://arxiv.org/abs/1304.5792}
  {arXiv:1304.5792 [astro-ph.CO]} \BibitemShut {NoStop}%
\bibitem [{\citenamefont {{Naoz}}\ and\ \citenamefont
  {{Narayan}}(2013{\natexlab{b}})}]{2013PhRvL.111e1303N}%
  \BibitemOpen
  \bibfield  {author} {\bibinfo {author} {\bibfnamefont {S.}~\bibnamefont
  {{Naoz}}}\ and\ \bibinfo {author} {\bibfnamefont {R.}~\bibnamefont
  {{Narayan}}},\ }\href {\doibase 10.1103/PhysRevLett.111.051303} {\bibfield
  {journal} {\bibinfo  {journal} {Physical Review Letters}\ }\textbf {\bibinfo
  {volume} {111}},\ \bibinfo {eid} {051303} (\bibinfo {year}
  {2013}{\natexlab{b}})},\ \Eprint {http://arxiv.org/abs/1304.5792}
  {arXiv:1304.5792 [astro-ph.CO]} \BibitemShut {NoStop}%
\bibitem [{\citenamefont {{Widrow}}\ \emph {et~al.}(2012)\citenamefont
  {{Widrow}}, \citenamefont {{Ryu}}, \citenamefont {{Schleicher}},
  \citenamefont {{Subramanian}}, \citenamefont {{Tsagas}},\ and\ \citenamefont
  {{Treumann}}}]{2012SSRv..166...37W}%
  \BibitemOpen
  \bibfield  {author} {\bibinfo {author} {\bibfnamefont {L.~M.}\ \bibnamefont
  {{Widrow}}}, \bibinfo {author} {\bibfnamefont {D.}~\bibnamefont {{Ryu}}},
  \bibinfo {author} {\bibfnamefont {D.~R.~G.}\ \bibnamefont {{Schleicher}}},
  \bibinfo {author} {\bibfnamefont {K.}~\bibnamefont {{Subramanian}}}, \bibinfo
  {author} {\bibfnamefont {C.~G.}\ \bibnamefont {{Tsagas}}}, \ and\ \bibinfo
  {author} {\bibfnamefont {R.~A.}\ \bibnamefont {{Treumann}}},\ }\href
  {\doibase 10.1007/s11214-011-9833-5} {\bibfield  {journal} {\bibinfo
  {journal} {\ssr}\ }\textbf {\bibinfo {volume} {166}},\ \bibinfo {pages} {37}
  (\bibinfo {year} {2012})},\ \Eprint {http://arxiv.org/abs/1109.4052}
  {arXiv:1109.4052 [astro-ph.CO]} \BibitemShut {NoStop}%
\bibitem [{\citenamefont {{Kobayashi}}(2014)}]{2014JCAP...05..040K}%
  \BibitemOpen
  \bibfield  {author} {\bibinfo {author} {\bibfnamefont {T.}~\bibnamefont
  {{Kobayashi}}},\ }\href {\doibase 10.1088/1475-7516/2014/05/040} {\bibfield
  {journal} {\bibinfo  {journal} {\jcap}\ }\textbf {\bibinfo {volume} {5}},\
  \bibinfo {eid} {040} (\bibinfo {year} {2014})},\ \Eprint
  {http://arxiv.org/abs/1403.5168} {arXiv:1403.5168} \BibitemShut {NoStop}%
\bibitem [{\citenamefont {{Yamazaki}}\ \emph {et~al.}(2010)\citenamefont
  {{Yamazaki}}, \citenamefont {{Ichiki}}, \citenamefont {{Kajino}},\ and\
  \citenamefont {{Mathews}}}]{Yamazaki10}%
  \BibitemOpen
  \bibfield  {author} {\bibinfo {author} {\bibfnamefont {D.~G.}\ \bibnamefont
  {{Yamazaki}}}, \bibinfo {author} {\bibfnamefont {K.}~\bibnamefont
  {{Ichiki}}}, \bibinfo {author} {\bibfnamefont {T.}~\bibnamefont {{Kajino}}},
  \ and\ \bibinfo {author} {\bibfnamefont {G.~J.}\ \bibnamefont {{Mathews}}},\
  }\href@noop {} {\bibfield  {journal} {\bibinfo  {journal} {Advances in
  Astronomy}\ }\textbf {\bibinfo {volume} {2010}} (\bibinfo {year} {2010})},\
  \Eprint {http://arxiv.org/abs/1112.4922} {arXiv:1112.4922 [astro-ph.CO]}
  \BibitemShut {NoStop}%
\bibitem [{\citenamefont {{Blasi}}\ \emph {et~al.}(1999)\citenamefont
  {{Blasi}}, \citenamefont {{Burles}},\ and\ \citenamefont
  {{Olinto}}}]{Blasi99}%
  \BibitemOpen
  \bibfield  {author} {\bibinfo {author} {\bibfnamefont {P.}~\bibnamefont
  {{Blasi}}}, \bibinfo {author} {\bibfnamefont {S.}~\bibnamefont {{Burles}}}, \
  and\ \bibinfo {author} {\bibfnamefont {A.~V.}\ \bibnamefont {{Olinto}}},\
  }\href@noop {} {\bibfield  {journal} {\bibinfo  {journal} {Astrophysical
  Journal, Letters}\ }\textbf {\bibinfo {volume} {514}},\ \bibinfo {pages}
  {L79} (\bibinfo {year} {1999})},\ \Eprint
  {http://arxiv.org/abs/astro-ph/9812487} {astro-ph/9812487} \BibitemShut
  {NoStop}%
\bibitem [{\citenamefont {{Tavecchio}}\ \emph {et~al.}(2010)\citenamefont
  {{Tavecchio}}, \citenamefont {{Ghisellini}}, \citenamefont {{Foschini}},
  \citenamefont {{Bonnoli}}, \citenamefont {{Ghirlanda}},\ and\ \citenamefont
  {{Coppi}}}]{Tavecchio10}%
  \BibitemOpen
  \bibfield  {author} {\bibinfo {author} {\bibfnamefont {F.}~\bibnamefont
  {{Tavecchio}}}, \bibinfo {author} {\bibfnamefont {G.}~\bibnamefont
  {{Ghisellini}}}, \bibinfo {author} {\bibfnamefont {L.}~\bibnamefont
  {{Foschini}}}, \bibinfo {author} {\bibfnamefont {G.}~\bibnamefont
  {{Bonnoli}}}, \bibinfo {author} {\bibfnamefont {G.}~\bibnamefont
  {{Ghirlanda}}}, \ and\ \bibinfo {author} {\bibfnamefont {P.}~\bibnamefont
  {{Coppi}}},\ }\href {\doibase 10.1111/j.1745-3933.2010.00884.x} {\bibfield
  {journal} {\bibinfo  {journal} {MNRAS}\ }\textbf {\bibinfo {volume} {406}},\
  \bibinfo {pages} {L70} (\bibinfo {year} {2010})},\ \Eprint
  {http://arxiv.org/abs/1004.1329} {arXiv:1004.1329 [astro-ph.CO]} \BibitemShut
  {NoStop}%
\bibitem [{\citenamefont {{Dolag}}\ \emph {et~al.}(2011)\citenamefont
  {{Dolag}}, \citenamefont {{Kachelriess}}, \citenamefont {{Ostapchenko}},\
  and\ \citenamefont {{Tom{\`a}s}}}]{Dolag11}%
  \BibitemOpen
  \bibfield  {author} {\bibinfo {author} {\bibfnamefont {K.}~\bibnamefont
  {{Dolag}}}, \bibinfo {author} {\bibfnamefont {M.}~\bibnamefont
  {{Kachelriess}}}, \bibinfo {author} {\bibfnamefont {S.}~\bibnamefont
  {{Ostapchenko}}}, \ and\ \bibinfo {author} {\bibfnamefont {R.}~\bibnamefont
  {{Tom{\`a}s}}},\ }\href {\doibase 10.1088/2041-8205/727/1/L4} {\bibfield
  {journal} {\bibinfo  {journal} {Astrophysical Journal, Letters}\ }\textbf
  {\bibinfo {volume} {727}},\ \bibinfo {eid} {L4} (\bibinfo {year} {2011})},\
  \Eprint {http://arxiv.org/abs/1009.1782} {arXiv:1009.1782 [astro-ph.HE]}
  \BibitemShut {NoStop}%
\bibitem [{\citenamefont {{Kunze}}\ and\ \citenamefont
  {{Komatsu}}(2014)}]{2014JCAP...01..009K}%
  \BibitemOpen
  \bibfield  {author} {\bibinfo {author} {\bibfnamefont {K.~E.}\ \bibnamefont
  {{Kunze}}}\ and\ \bibinfo {author} {\bibfnamefont {E.}~\bibnamefont
  {{Komatsu}}},\ }\href {\doibase 10.1088/1475-7516/2014/01/009} {\bibfield
  {journal} {\bibinfo  {journal} {\jcap}\ }\textbf {\bibinfo {volume} {1}},\
  \bibinfo {eid} {009} (\bibinfo {year} {2014})},\ \Eprint
  {http://arxiv.org/abs/1309.7994} {arXiv:1309.7994 [astro-ph.CO]} \BibitemShut
  {NoStop}%
\bibitem [{\citenamefont {{Kahniashvili}}\ \emph {et~al.}(2013)\citenamefont
  {{Kahniashvili}}, \citenamefont {{Maravin}}, \citenamefont {{Natarajan}},
  \citenamefont {{Battaglia}},\ and\ \citenamefont
  {{Tevzadze}}}]{2013ApJ...770...47K}%
  \BibitemOpen
  \bibfield  {author} {\bibinfo {author} {\bibfnamefont {T.}~\bibnamefont
  {{Kahniashvili}}}, \bibinfo {author} {\bibfnamefont {Y.}~\bibnamefont
  {{Maravin}}}, \bibinfo {author} {\bibfnamefont {A.}~\bibnamefont
  {{Natarajan}}}, \bibinfo {author} {\bibfnamefont {N.}~\bibnamefont
  {{Battaglia}}}, \ and\ \bibinfo {author} {\bibfnamefont {A.~G.}\ \bibnamefont
  {{Tevzadze}}},\ }\href {\doibase 10.1088/0004-637X/770/1/47} {\bibfield
  {journal} {\bibinfo  {journal} {\apj}\ }\textbf {\bibinfo {volume} {770}},\
  \bibinfo {eid} {47} (\bibinfo {year} {2013})},\ \Eprint
  {http://arxiv.org/abs/1211.2769} {arXiv:1211.2769 [astro-ph.CO]} \BibitemShut
  {NoStop}%
\bibitem [{\citenamefont {{Shiraishi}}\ \emph {et~al.}(2014)\citenamefont
  {{Shiraishi}}, \citenamefont {{Tashiro}},\ and\ \citenamefont
  {{Ichiki}}}]{2014PhRvD..89j3522S}%
  \BibitemOpen
  \bibfield  {author} {\bibinfo {author} {\bibfnamefont {M.}~\bibnamefont
  {{Shiraishi}}}, \bibinfo {author} {\bibfnamefont {H.}~\bibnamefont
  {{Tashiro}}}, \ and\ \bibinfo {author} {\bibfnamefont {K.}~\bibnamefont
  {{Ichiki}}},\ }\href {\doibase 10.1103/PhysRevD.89.103522} {\bibfield
  {journal} {\bibinfo  {journal} {\prd}\ }\textbf {\bibinfo {volume} {89}},\
  \bibinfo {eid} {103522} (\bibinfo {year} {2014})},\ \Eprint
  {http://arxiv.org/abs/1403.2608} {arXiv:1403.2608} \BibitemShut {NoStop}%
\bibitem [{\citenamefont {{Tashiro}}\ and\ \citenamefont
  {{Sugiyama}}(2006)}]{2006MNRAS.372.1060T}%
  \BibitemOpen
  \bibfield  {author} {\bibinfo {author} {\bibfnamefont {H.}~\bibnamefont
  {{Tashiro}}}\ and\ \bibinfo {author} {\bibfnamefont {N.}~\bibnamefont
  {{Sugiyama}}},\ }\href {\doibase 10.1111/j.1365-2966.2006.10901.x} {\bibfield
   {journal} {\bibinfo  {journal} {\mnras}\ }\textbf {\bibinfo {volume}
  {372}},\ \bibinfo {pages} {1060} (\bibinfo {year} {2006})},\ \Eprint
  {http://arxiv.org/abs/astro-ph/0607169} {astro-ph/0607169} \BibitemShut
  {NoStop}%
\bibitem [{\citenamefont {{Schleicher}}\ \emph {et~al.}(2009)\citenamefont
  {{Schleicher}}, \citenamefont {{Banerjee}},\ and\ \citenamefont
  {{Klessen}}}]{2009ApJ...692..236S}%
  \BibitemOpen
  \bibfield  {author} {\bibinfo {author} {\bibfnamefont {D.~R.~G.}\
  \bibnamefont {{Schleicher}}}, \bibinfo {author} {\bibfnamefont
  {R.}~\bibnamefont {{Banerjee}}}, \ and\ \bibinfo {author} {\bibfnamefont
  {R.~S.}\ \bibnamefont {{Klessen}}},\ }\href {\doibase
  10.1088/0004-637X/692/1/236} {\bibfield  {journal} {\bibinfo  {journal}
  {\apj}\ }\textbf {\bibinfo {volume} {692}},\ \bibinfo {pages} {236} (\bibinfo
  {year} {2009})},\ \Eprint {http://arxiv.org/abs/0808.1461} {arXiv:0808.1461}
  \BibitemShut {NoStop}%
\bibitem [{\citenamefont {{Planck Collaboration}}\ \emph
  {et~al.}(2015{\natexlab{a}})\citenamefont {{Planck Collaboration}},
  \citenamefont {{Ade}}, \citenamefont {{Aghanim}}, \citenamefont {{Arnaud}},
  \citenamefont {{Arroja}}, \citenamefont {{Ashdown}}, \citenamefont
  {{Aumont}}, \citenamefont {{Baccigalupi}}, \citenamefont {{Ballardini}},
  \citenamefont {{Banday}},\ and\ \citenamefont
  {et~al.}}]{2015arXiv150201594P}%
  \BibitemOpen
  \bibfield  {author} {\bibinfo {author} {\bibnamefont {{Planck
  Collaboration}}}, \bibinfo {author} {\bibfnamefont {P.~A.~R.}\ \bibnamefont
  {{Ade}}}, \bibinfo {author} {\bibfnamefont {N.}~\bibnamefont {{Aghanim}}},
  \bibinfo {author} {\bibfnamefont {M.}~\bibnamefont {{Arnaud}}}, \bibinfo
  {author} {\bibfnamefont {F.}~\bibnamefont {{Arroja}}}, \bibinfo {author}
  {\bibfnamefont {M.}~\bibnamefont {{Ashdown}}}, \bibinfo {author}
  {\bibfnamefont {J.}~\bibnamefont {{Aumont}}}, \bibinfo {author}
  {\bibfnamefont {C.}~\bibnamefont {{Baccigalupi}}}, \bibinfo {author}
  {\bibfnamefont {M.}~\bibnamefont {{Ballardini}}}, \bibinfo {author}
  {\bibfnamefont {A.~J.}\ \bibnamefont {{Banday}}}, \ and\ \bibinfo {author}
  {\bibnamefont {et~al.}},\ }\href@noop {} {\bibfield  {journal} {\bibinfo
  {journal} {ArXiv e-prints}\ } (\bibinfo {year} {2015}{\natexlab{a}})},\
  \Eprint {http://arxiv.org/abs/1502.01594} {arXiv:1502.01594} \BibitemShut
  {NoStop}%
\bibitem [{\citenamefont {{Venumadhav}}\ \emph {et~al.}(2014)\citenamefont
  {{Venumadhav}}, \citenamefont {{Oklopcic}}, \citenamefont {{Gluscevic}},
  \citenamefont {{Mishra}},\ and\ \citenamefont
  {{Hirata}}}]{2014arXiv1410.2250V}%
  \BibitemOpen
  \bibfield  {author} {\bibinfo {author} {\bibfnamefont {T.}~\bibnamefont
  {{Venumadhav}}}, \bibinfo {author} {\bibfnamefont {A.}~\bibnamefont
  {{Oklopcic}}}, \bibinfo {author} {\bibfnamefont {V.}~\bibnamefont
  {{Gluscevic}}}, \bibinfo {author} {\bibfnamefont {A.}~\bibnamefont
  {{Mishra}}}, \ and\ \bibinfo {author} {\bibfnamefont {C.~M.}\ \bibnamefont
  {{Hirata}}},\ }\href@noop {} {\bibfield  {journal} {\bibinfo  {journal}
  {ArXiv e-prints}\ } (\bibinfo {year} {2014})},\ \Eprint
  {http://arxiv.org/abs/1410.2250} {arXiv:1410.2250} \BibitemShut {NoStop}%
\bibitem [{\citenamefont {{Madau}}\ \emph {et~al.}(1997)\citenamefont
  {{Madau}}, \citenamefont {{Meiksin}},\ and\ \citenamefont
  {{Rees}}}]{1997ApJ...475..429M}%
  \BibitemOpen
  \bibfield  {author} {\bibinfo {author} {\bibfnamefont {P.}~\bibnamefont
  {{Madau}}}, \bibinfo {author} {\bibfnamefont {A.}~\bibnamefont {{Meiksin}}},
  \ and\ \bibinfo {author} {\bibfnamefont {M.~J.}\ \bibnamefont {{Rees}}},\
  }\href@noop {} {\bibfield  {journal} {\bibinfo  {journal} {\apj}\ }\textbf
  {\bibinfo {volume} {475}},\ \bibinfo {pages} {429} (\bibinfo {year}
  {1997})},\ \Eprint {http://arxiv.org/abs/astro-ph/9608010} {astro-ph/9608010}
  \BibitemShut {NoStop}%
\bibitem [{\citenamefont {{Loeb}}\ and\ \citenamefont
  {{Zaldarriaga}}(2004)}]{2004PhRvL..92u1301L}%
  \BibitemOpen
  \bibfield  {author} {\bibinfo {author} {\bibfnamefont {A.}~\bibnamefont
  {{Loeb}}}\ and\ \bibinfo {author} {\bibfnamefont {M.}~\bibnamefont
  {{Zaldarriaga}}},\ }\href {\doibase 10.1103/PhysRevLett.92.211301} {\bibfield
   {journal} {\bibinfo  {journal} {Physical Review Letters}\ }\textbf {\bibinfo
  {volume} {92}},\ \bibinfo {eid} {211301} (\bibinfo {year} {2004})},\ \Eprint
  {http://arxiv.org/abs/astro-ph/0312134} {astro-ph/0312134} \BibitemShut
  {NoStop}%
\bibitem [{\citenamefont {{Greenhill}}\ and\ \citenamefont
  {{Bernardi}}(2012)}]{2012arXiv1201.1700G}%
  \BibitemOpen
  \bibfield  {author} {\bibinfo {author} {\bibfnamefont {L.~J.}\ \bibnamefont
  {{Greenhill}}}\ and\ \bibinfo {author} {\bibfnamefont {G.}~\bibnamefont
  {{Bernardi}}},\ }\href@noop {} {\bibfield  {journal} {\bibinfo  {journal}
  {ArXiv e-prints}\ } (\bibinfo {year} {2012})},\ \Eprint
  {http://arxiv.org/abs/1201.1700} {arXiv:1201.1700 [astro-ph.CO]} \BibitemShut
  {NoStop}%
\bibitem [{\citenamefont {{Bowman}}\ \emph {et~al.}(2011)\citenamefont
  {{Bowman}}, \citenamefont {{Morales}}, \citenamefont {{Hewitt}},\ and\
  \citenamefont {{MWA Collaboration}}}]{2011AAS...21813206B}%
  \BibitemOpen
  \bibfield  {author} {\bibinfo {author} {\bibfnamefont {J.~D.}\ \bibnamefont
  {{Bowman}}}, \bibinfo {author} {\bibfnamefont {M.~F.}\ \bibnamefont
  {{Morales}}}, \bibinfo {author} {\bibfnamefont {J.~N.}\ \bibnamefont
  {{Hewitt}}}, \ and\ \bibinfo {author} {\bibnamefont {{MWA Collaboration}}},\
  }in\ \href@noop {} {\emph {\bibinfo {booktitle} {American Astronomical
  Society Meeting Abstracts \#218}}}\ (\bibinfo {year} {2011})\ p.\ \bibinfo
  {pages} {132.06}\BibitemShut {NoStop}%
\bibitem [{\citenamefont {{Parsons}}\ \emph {et~al.}(2014)\citenamefont
  {{Parsons}}, \citenamefont {{Liu}}, \citenamefont {{Aguirre}}, \citenamefont
  {{Ali}}, \citenamefont {{Bradley}}, \citenamefont {{Carilli}}, \citenamefont
  {{DeBoer}}, \citenamefont {{Dexter}}, \citenamefont {{Gugliucci}},
  \citenamefont {{Jacobs}}, \citenamefont {{Klima}}, \citenamefont
  {{MacMahon}}, \citenamefont {{Manley}}, \citenamefont {{Moore}},
  \citenamefont {{Pober}}, \citenamefont {{Stefan}},\ and\ \citenamefont
  {{Walbrugh}}}]{2014ApJ...788..106P}%
  \BibitemOpen
  \bibfield  {author} {\bibinfo {author} {\bibfnamefont {A.~R.}\ \bibnamefont
  {{Parsons}}}, \bibinfo {author} {\bibfnamefont {A.}~\bibnamefont {{Liu}}},
  \bibinfo {author} {\bibfnamefont {J.~E.}\ \bibnamefont {{Aguirre}}}, \bibinfo
  {author} {\bibfnamefont {Z.~S.}\ \bibnamefont {{Ali}}}, \bibinfo {author}
  {\bibfnamefont {R.~F.}\ \bibnamefont {{Bradley}}}, \bibinfo {author}
  {\bibfnamefont {C.~L.}\ \bibnamefont {{Carilli}}}, \bibinfo {author}
  {\bibfnamefont {D.~R.}\ \bibnamefont {{DeBoer}}}, \bibinfo {author}
  {\bibfnamefont {M.~R.}\ \bibnamefont {{Dexter}}}, \bibinfo {author}
  {\bibfnamefont {N.~E.}\ \bibnamefont {{Gugliucci}}}, \bibinfo {author}
  {\bibfnamefont {D.~C.}\ \bibnamefont {{Jacobs}}}, \bibinfo {author}
  {\bibfnamefont {P.}~\bibnamefont {{Klima}}}, \bibinfo {author} {\bibfnamefont
  {D.~H.~E.}\ \bibnamefont {{MacMahon}}}, \bibinfo {author} {\bibfnamefont
  {J.~R.}\ \bibnamefont {{Manley}}}, \bibinfo {author} {\bibfnamefont {D.~F.}\
  \bibnamefont {{Moore}}}, \bibinfo {author} {\bibfnamefont {J.~C.}\
  \bibnamefont {{Pober}}}, \bibinfo {author} {\bibfnamefont {I.~I.}\
  \bibnamefont {{Stefan}}}, \ and\ \bibinfo {author} {\bibfnamefont {W.~P.}\
  \bibnamefont {{Walbrugh}}},\ }\href {\doibase 10.1088/0004-637X/788/2/106}
  {\bibfield  {journal} {\bibinfo  {journal} {\apj}\ }\textbf {\bibinfo
  {volume} {788}},\ \bibinfo {eid} {106} (\bibinfo {year} {2014})},\ \Eprint
  {http://arxiv.org/abs/1304.4991} {arXiv:1304.4991} \BibitemShut {NoStop}%
\bibitem [{\citenamefont {{Carilli}}(2008)}]{2008arXiv0802.1727C}%
  \BibitemOpen
  \bibfield  {author} {\bibinfo {author} {\bibfnamefont {C.~L.}\ \bibnamefont
  {{Carilli}}},\ }\href@noop {} {\bibfield  {journal} {\bibinfo  {journal}
  {ArXiv e-prints}\ } (\bibinfo {year} {2008})},\ \Eprint
  {http://arxiv.org/abs/0802.1727} {arXiv:0802.1727} \BibitemShut {NoStop}%
\bibitem [{\citenamefont {{Vanderlinde}}\ and\ \citenamefont {{Chime
  Collaboration}}(2014)}]{Vanderlinde14}%
  \BibitemOpen
  \bibfield  {author} {\bibinfo {author} {\bibfnamefont {K.}~\bibnamefont
  {{Vanderlinde}}}\ and\ \bibinfo {author} {\bibnamefont {{Chime
  Collaboration}}},\ }in\ \href@noop {} {\emph {\bibinfo {booktitle} {Exascale
  Radio Astronomy}}}\ (\bibinfo {year} {2014})\ p.\ \bibinfo {pages}
  {10102}\BibitemShut {NoStop}%
\bibitem [{\citenamefont {{DeBoer}}\ and\ \citenamefont
  {{HERA}}(2015)}]{2015AAS...22532803D}%
  \BibitemOpen
  \bibfield  {author} {\bibinfo {author} {\bibfnamefont {D.~R.}\ \bibnamefont
  {{DeBoer}}}\ and\ \bibinfo {author} {\bibnamefont {{HERA}}},\ }in\ \href@noop
  {} {\emph {\bibinfo {booktitle} {American Astronomical Society Meeting
  Abstracts}}},\ \bibinfo {series} {American Astronomical Society Meeting
  Abstracts}, Vol.\ \bibinfo {volume} {225}\ (\bibinfo {year} {2015})\ p.\
  \bibinfo {pages} {328.03}\BibitemShut {NoStop}%
\bibitem [{\citenamefont {{Yan}}\ and\ \citenamefont
  {{Lazarian}}(2008)}]{Yan08}%
  \BibitemOpen
  \bibfield  {author} {\bibinfo {author} {\bibfnamefont {H.}~\bibnamefont
  {{Yan}}}\ and\ \bibinfo {author} {\bibfnamefont {A.}~\bibnamefont
  {{Lazarian}}},\ }\href {\doibase 10.1086/533410} {\bibfield  {journal}
  {\bibinfo  {journal} {\apj}\ }\textbf {\bibinfo {volume} {677}},\ \bibinfo
  {pages} {1401} (\bibinfo {year} {2008})},\ \Eprint
  {http://arxiv.org/abs/0711.0926} {arXiv:0711.0926} \BibitemShut {NoStop}%
\bibitem [{\citenamefont {{Yan}}\ and\ \citenamefont
  {{Lazarian}}(2012)}]{Yan12}%
  \BibitemOpen
  \bibfield  {author} {\bibinfo {author} {\bibfnamefont {H.}~\bibnamefont
  {{Yan}}}\ and\ \bibinfo {author} {\bibfnamefont {A.}~\bibnamefont
  {{Lazarian}}},\ }\href {\doibase 10.1016/j.jqsrt.2012.03.027} {\bibfield
  {journal} {\bibinfo  {journal} {\jqsrt}\ }\textbf {\bibinfo {volume} {113}},\
  \bibinfo {pages} {1409} (\bibinfo {year} {2012})},\ \Eprint
  {http://arxiv.org/abs/1203.5571} {arXiv:1203.5571 [astro-ph.GA]} \BibitemShut
  {NoStop}%
\bibitem [{\citenamefont {{Okamoto}}\ and\ \citenamefont
  {{Hu}}(2003)}]{2003PhRvD..67h3002O}%
  \BibitemOpen
  \bibfield  {author} {\bibinfo {author} {\bibfnamefont {T.}~\bibnamefont
  {{Okamoto}}}\ and\ \bibinfo {author} {\bibfnamefont {W.}~\bibnamefont
  {{Hu}}},\ }\href {\doibase 10.1103/PhysRevD.67.083002} {\bibfield  {journal}
  {\bibinfo  {journal} {\prd}\ }\textbf {\bibinfo {volume} {67}},\ \bibinfo
  {eid} {083002} (\bibinfo {year} {2003})},\ \Eprint
  {http://arxiv.org/abs/astro-ph/0301031} {astro-ph/0301031} \BibitemShut
  {NoStop}%
\bibitem [{\citenamefont {{Tegmark}}\ and\ \citenamefont
  {{Zaldarriaga}}(2009)}]{2009PhRvD..79h3530T}%
  \BibitemOpen
  \bibfield  {author} {\bibinfo {author} {\bibfnamefont {M.}~\bibnamefont
  {{Tegmark}}}\ and\ \bibinfo {author} {\bibfnamefont {M.}~\bibnamefont
  {{Zaldarriaga}}},\ }\href {\doibase 10.1103/PhysRevD.79.083530} {\bibfield
  {journal} {\bibinfo  {journal} {\prd}\ }\textbf {\bibinfo {volume} {79}},\
  \bibinfo {eid} {083530} (\bibinfo {year} {2009})},\ \Eprint
  {http://arxiv.org/abs/0805.4414} {arXiv:0805.4414} \BibitemShut {NoStop}%
\bibitem [{\citenamefont {{Mao}}\ \emph {et~al.}(2008)\citenamefont {{Mao}},
  \citenamefont {{Tegmark}}, \citenamefont {{McQuinn}}, \citenamefont
  {{Zaldarriaga}},\ and\ \citenamefont {{Zahn}}}]{2008PhRvD..78b3529M}%
  \BibitemOpen
  \bibfield  {author} {\bibinfo {author} {\bibfnamefont {Y.}~\bibnamefont
  {{Mao}}}, \bibinfo {author} {\bibfnamefont {M.}~\bibnamefont {{Tegmark}}},
  \bibinfo {author} {\bibfnamefont {M.}~\bibnamefont {{McQuinn}}}, \bibinfo
  {author} {\bibfnamefont {M.}~\bibnamefont {{Zaldarriaga}}}, \ and\ \bibinfo
  {author} {\bibfnamefont {O.}~\bibnamefont {{Zahn}}},\ }\href {\doibase
  10.1103/PhysRevD.78.023529} {\bibfield  {journal} {\bibinfo  {journal}
  {\prd}\ }\textbf {\bibinfo {volume} {78}},\ \bibinfo {eid} {023529} (\bibinfo
  {year} {2008})},\ \Eprint {http://arxiv.org/abs/0802.1710} {arXiv:0802.1710}
  \BibitemShut {NoStop}%
\bibitem [{\citenamefont {{Mesinger}}\ \emph {et~al.}(2011)\citenamefont
  {{Mesinger}}, \citenamefont {{Furlanetto}},\ and\ \citenamefont
  {{Cen}}}]{2011MNRAS.411..955M}%
  \BibitemOpen
  \bibfield  {author} {\bibinfo {author} {\bibfnamefont {A.}~\bibnamefont
  {{Mesinger}}}, \bibinfo {author} {\bibfnamefont {S.}~\bibnamefont
  {{Furlanetto}}}, \ and\ \bibinfo {author} {\bibfnamefont {R.}~\bibnamefont
  {{Cen}}},\ }\href {\doibase 10.1111/j.1365-2966.2010.17731.x} {\bibfield
  {journal} {\bibinfo  {journal} {\mnras}\ }\textbf {\bibinfo {volume} {411}},\
  \bibinfo {pages} {955} (\bibinfo {year} {2011})},\ \Eprint
  {http://arxiv.org/abs/1003.3878} {arXiv:1003.3878} \BibitemShut {NoStop}%
\bibitem [{\citenamefont {{Lewis}}\ \emph {et~al.}(2000)\citenamefont
  {{Lewis}}, \citenamefont {{Challinor}},\ and\ \citenamefont
  {{Lasenby}}}]{2000ApJ...538..473L}%
  \BibitemOpen
  \bibfield  {author} {\bibinfo {author} {\bibfnamefont {A.}~\bibnamefont
  {{Lewis}}}, \bibinfo {author} {\bibfnamefont {A.}~\bibnamefont
  {{Challinor}}}, \ and\ \bibinfo {author} {\bibfnamefont {A.}~\bibnamefont
  {{Lasenby}}},\ }\href {\doibase 10.1086/309179} {\bibfield  {journal}
  {\bibinfo  {journal} {\apj}\ }\textbf {\bibinfo {volume} {538}},\ \bibinfo
  {pages} {473} (\bibinfo {year} {2000})},\ \Eprint
  {http://arxiv.org/abs/astro-ph/9911177} {astro-ph/9911177} \BibitemShut
  {NoStop}%
\bibitem [{\citenamefont {{Planck Collaboration}}\ \emph
  {et~al.}(2015{\natexlab{b}})\citenamefont {{Planck Collaboration}},
  \citenamefont {{Ade}}, \citenamefont {{Aghanim}}, \citenamefont {{Arnaud}},
  \citenamefont {{Ashdown}}, \citenamefont {{Aumont}}, \citenamefont
  {{Baccigalupi}}, \citenamefont {{Banday}}, \citenamefont {{Barreiro}},
  \citenamefont {{Bartlett}},\ and\ \citenamefont
  {et~al.}}]{2015arXiv150201589P}%
  \BibitemOpen
  \bibfield  {author} {\bibinfo {author} {\bibnamefont {{Planck
  Collaboration}}}, \bibinfo {author} {\bibfnamefont {P.~A.~R.}\ \bibnamefont
  {{Ade}}}, \bibinfo {author} {\bibfnamefont {N.}~\bibnamefont {{Aghanim}}},
  \bibinfo {author} {\bibfnamefont {M.}~\bibnamefont {{Arnaud}}}, \bibinfo
  {author} {\bibfnamefont {M.}~\bibnamefont {{Ashdown}}}, \bibinfo {author}
  {\bibfnamefont {J.}~\bibnamefont {{Aumont}}}, \bibinfo {author}
  {\bibfnamefont {C.}~\bibnamefont {{Baccigalupi}}}, \bibinfo {author}
  {\bibfnamefont {A.~J.}\ \bibnamefont {{Banday}}}, \bibinfo {author}
  {\bibfnamefont {R.~B.}\ \bibnamefont {{Barreiro}}}, \bibinfo {author}
  {\bibfnamefont {J.~G.}\ \bibnamefont {{Bartlett}}}, \ and\ \bibinfo {author}
  {\bibnamefont {et~al.}},\ }\href@noop {} {\bibfield  {journal} {\bibinfo
  {journal} {ArXiv e-prints}\ } (\bibinfo {year} {2015}{\natexlab{b}})},\
  \Eprint {http://arxiv.org/abs/1502.01589} {arXiv:1502.01589} \BibitemShut
  {NoStop}%
\bibitem [{\citenamefont {{Haardt}}\ and\ \citenamefont
  {{Madau}}(2012)}]{2012ApJ...746..125H}%
  \BibitemOpen
  \bibfield  {author} {\bibinfo {author} {\bibfnamefont {F.}~\bibnamefont
  {{Haardt}}}\ and\ \bibinfo {author} {\bibfnamefont {P.}~\bibnamefont
  {{Madau}}},\ }\href {\doibase 10.1088/0004-637X/746/2/125} {\bibfield
  {journal} {\bibinfo  {journal} {\apj}\ }\textbf {\bibinfo {volume} {746}},\
  \bibinfo {eid} {125} (\bibinfo {year} {2012})},\ \Eprint
  {http://arxiv.org/abs/1105.2039} {arXiv:1105.2039} \BibitemShut {NoStop}%
\bibitem [{\citenamefont {{G{\'o}rski}}\ \emph {et~al.}(2005)\citenamefont
  {{G{\'o}rski}}, \citenamefont {{Hivon}}, \citenamefont {{Banday}},
  \citenamefont {{Wandelt}}, \citenamefont {{Hansen}}, \citenamefont
  {{Reinecke}},\ and\ \citenamefont {{Bartelmann}}}]{2005ApJ...622..759G}%
  \BibitemOpen
  \bibfield  {author} {\bibinfo {author} {\bibfnamefont {K.~M.}\ \bibnamefont
  {{G{\'o}rski}}}, \bibinfo {author} {\bibfnamefont {E.}~\bibnamefont
  {{Hivon}}}, \bibinfo {author} {\bibfnamefont {A.~J.}\ \bibnamefont
  {{Banday}}}, \bibinfo {author} {\bibfnamefont {B.~D.}\ \bibnamefont
  {{Wandelt}}}, \bibinfo {author} {\bibfnamefont {F.~K.}\ \bibnamefont
  {{Hansen}}}, \bibinfo {author} {\bibfnamefont {M.}~\bibnamefont
  {{Reinecke}}}, \ and\ \bibinfo {author} {\bibfnamefont {M.}~\bibnamefont
  {{Bartelmann}}},\ }\href {\doibase 10.1086/427976} {\bibfield  {journal}
  {\bibinfo  {journal} {\apj}\ }\textbf {\bibinfo {volume} {622}},\ \bibinfo
  {pages} {759} (\bibinfo {year} {2005})},\ \Eprint
  {http://arxiv.org/abs/astro-ph/0409513} {astro-ph/0409513} \BibitemShut
  {NoStop}%
\bibitem [{\citenamefont {{Thompson}}\ \emph {et~al.}()\citenamefont
  {{Thompson}}, \citenamefont {{Moran}},\ and\ \citenamefont
  {{Swenson}}}]{2001isra.book.....T}%
  \BibitemOpen
  \bibfield  {author} {\bibinfo {author} {\bibfnamefont {A.~R.}\ \bibnamefont
  {{Thompson}}}, \bibinfo {author} {\bibfnamefont {J.~M.}\ \bibnamefont
  {{Moran}}}, \ and\ \bibinfo {author} {\bibfnamefont {G.~W.}\ \bibnamefont
  {{Swenson}}, \bibfnamefont {Jr.}},\ }\href@noop {} {\emph {\bibinfo {title}
  {{Interferometry and Synthesis in Radio Astronomy, 2nd
  Edition}}}}\BibitemShut {NoStop}%
\bibitem [{\citenamefont {{Perley}}\ \emph {et~al.}()\citenamefont {{Perley}},
  \citenamefont {{Schwab}}, \citenamefont {{Bridle}},\ and\ \citenamefont
  {{Ekers}}}]{1986sicn.book.....P}%
  \BibitemOpen
  \bibfield  {author} {\bibinfo {author} {\bibfnamefont {R.~A.}\ \bibnamefont
  {{Perley}}}, \bibinfo {author} {\bibfnamefont {F.~R.}\ \bibnamefont
  {{Schwab}}}, \bibinfo {author} {\bibfnamefont {A.~H.}\ \bibnamefont
  {{Bridle}}}, \ and\ \bibinfo {author} {\bibfnamefont {R.~D.}\ \bibnamefont
  {{Ekers}}},\ }\href@noop {} {\emph {\bibinfo {title} {{Synthesis imaging.
  Course notes from an NRAO summer school, held at Socorro, New Mexico, USA, 5
  - 9 August 1985.}}}}\BibitemShut {Stop}%
\bibitem [{\citenamefont {{Furlanetto}}\ \emph {et~al.}(2009)\citenamefont
  {{Furlanetto}}, \citenamefont {{Lidz}}, \citenamefont {{Loeb}}, \citenamefont
  {{McQuinn}}, \citenamefont {{Pritchard}}, \citenamefont {{Shapiro}},
  \citenamefont {{Alvarez}}, \citenamefont {{Backer}}, \citenamefont
  {{Bowman}}, \citenamefont {{Burns}}, \citenamefont {{Carilli}}, \citenamefont
  {{Cen}}, \citenamefont {{Cooray}}, \citenamefont {{Gnedin}}, \citenamefont
  {{Greenhill}}, \citenamefont {{Haiman}}, \citenamefont {{Hewitt}},
  \citenamefont {{Hirata}}, \citenamefont {{Lazio}}, \citenamefont
  {{Mesinger}}, \citenamefont {{Madau}}, \citenamefont {{Morales}},
  \citenamefont {{Oh}}, \citenamefont {{Peterson}}, \citenamefont
  {{Pihlstr{\"o}m}}, \citenamefont {{Tegmark}}, \citenamefont {{Trac}},
  \citenamefont {{Zahn}},\ and\ \citenamefont
  {{Zaldarriaga}}}]{2009astro2010S..82F}%
  \BibitemOpen
  \bibfield  {author} {\bibinfo {author} {\bibfnamefont {S.~R.}\ \bibnamefont
  {{Furlanetto}}}, \bibinfo {author} {\bibfnamefont {A.}~\bibnamefont
  {{Lidz}}}, \bibinfo {author} {\bibfnamefont {A.}~\bibnamefont {{Loeb}}},
  \bibinfo {author} {\bibfnamefont {M.}~\bibnamefont {{McQuinn}}}, \bibinfo
  {author} {\bibfnamefont {J.~R.}\ \bibnamefont {{Pritchard}}}, \bibinfo
  {author} {\bibfnamefont {P.~R.}\ \bibnamefont {{Shapiro}}}, \bibinfo {author}
  {\bibfnamefont {M.~A.}\ \bibnamefont {{Alvarez}}}, \bibinfo {author}
  {\bibfnamefont {D.~C.}\ \bibnamefont {{Backer}}}, \bibinfo {author}
  {\bibfnamefont {J.~D.}\ \bibnamefont {{Bowman}}}, \bibinfo {author}
  {\bibfnamefont {J.~O.}\ \bibnamefont {{Burns}}}, \bibinfo {author}
  {\bibfnamefont {C.~L.}\ \bibnamefont {{Carilli}}}, \bibinfo {author}
  {\bibfnamefont {R.}~\bibnamefont {{Cen}}}, \bibinfo {author} {\bibfnamefont
  {A.}~\bibnamefont {{Cooray}}}, \bibinfo {author} {\bibfnamefont
  {N.}~\bibnamefont {{Gnedin}}}, \bibinfo {author} {\bibfnamefont {L.~J.}\
  \bibnamefont {{Greenhill}}}, \bibinfo {author} {\bibfnamefont
  {Z.}~\bibnamefont {{Haiman}}}, \bibinfo {author} {\bibfnamefont {J.~N.}\
  \bibnamefont {{Hewitt}}}, \bibinfo {author} {\bibfnamefont {C.~M.}\
  \bibnamefont {{Hirata}}}, \bibinfo {author} {\bibfnamefont {J.}~\bibnamefont
  {{Lazio}}}, \bibinfo {author} {\bibfnamefont {A.}~\bibnamefont {{Mesinger}}},
  \bibinfo {author} {\bibfnamefont {P.}~\bibnamefont {{Madau}}}, \bibinfo
  {author} {\bibfnamefont {M.~F.}\ \bibnamefont {{Morales}}}, \bibinfo {author}
  {\bibfnamefont {S.~P.}\ \bibnamefont {{Oh}}}, \bibinfo {author}
  {\bibfnamefont {J.~B.}\ \bibnamefont {{Peterson}}}, \bibinfo {author}
  {\bibfnamefont {Y.~M.}\ \bibnamefont {{Pihlstr{\"o}m}}}, \bibinfo {author}
  {\bibfnamefont {M.}~\bibnamefont {{Tegmark}}}, \bibinfo {author}
  {\bibfnamefont {H.}~\bibnamefont {{Trac}}}, \bibinfo {author} {\bibfnamefont
  {O.}~\bibnamefont {{Zahn}}}, \ and\ \bibinfo {author} {\bibfnamefont
  {M.}~\bibnamefont {{Zaldarriaga}}},\ }in\ \href@noop {} {\emph {\bibinfo
  {booktitle} {astro2010: The Astronomy and Astrophysics Decadal Survey}}},\
  \bibinfo {series} {Astronomy}, Vol.\ \bibinfo {volume} {2010}\ (\bibinfo
  {year} {2009})\ \Eprint {http://arxiv.org/abs/0902.3259} {arXiv:0902.3259
  [astro-ph.CO]} \BibitemShut {NoStop}%
\bibitem [{\citenamefont {{Pober}}\ \emph {et~al.}(2014)\citenamefont
  {{Pober}}, \citenamefont {{Liu}}, \citenamefont {{Dillon}}, \citenamefont
  {{Aguirre}}, \citenamefont {{Bowman}}, \citenamefont {{Bradley}},
  \citenamefont {{Carilli}}, \citenamefont {{DeBoer}}, \citenamefont
  {{Hewitt}}, \citenamefont {{Jacobs}}, \citenamefont {{McQuinn}},
  \citenamefont {{Morales}}, \citenamefont {{Parsons}}, \citenamefont
  {{Tegmark}},\ and\ \citenamefont {{Werthimer}}}]{2014ApJ...782...66P}%
  \BibitemOpen
  \bibfield  {author} {\bibinfo {author} {\bibfnamefont {J.~C.}\ \bibnamefont
  {{Pober}}}, \bibinfo {author} {\bibfnamefont {A.}~\bibnamefont {{Liu}}},
  \bibinfo {author} {\bibfnamefont {J.~S.}\ \bibnamefont {{Dillon}}}, \bibinfo
  {author} {\bibfnamefont {J.~E.}\ \bibnamefont {{Aguirre}}}, \bibinfo {author}
  {\bibfnamefont {J.~D.}\ \bibnamefont {{Bowman}}}, \bibinfo {author}
  {\bibfnamefont {R.~F.}\ \bibnamefont {{Bradley}}}, \bibinfo {author}
  {\bibfnamefont {C.~L.}\ \bibnamefont {{Carilli}}}, \bibinfo {author}
  {\bibfnamefont {D.~R.}\ \bibnamefont {{DeBoer}}}, \bibinfo {author}
  {\bibfnamefont {J.~N.}\ \bibnamefont {{Hewitt}}}, \bibinfo {author}
  {\bibfnamefont {D.~C.}\ \bibnamefont {{Jacobs}}}, \bibinfo {author}
  {\bibfnamefont {M.}~\bibnamefont {{McQuinn}}}, \bibinfo {author}
  {\bibfnamefont {M.~F.}\ \bibnamefont {{Morales}}}, \bibinfo {author}
  {\bibfnamefont {A.~R.}\ \bibnamefont {{Parsons}}}, \bibinfo {author}
  {\bibfnamefont {M.}~\bibnamefont {{Tegmark}}}, \ and\ \bibinfo {author}
  {\bibfnamefont {D.~J.}\ \bibnamefont {{Werthimer}}},\ }\href {\doibase
  10.1088/0004-637X/782/2/66} {\bibfield  {journal} {\bibinfo  {journal}
  {\apj}\ }\textbf {\bibinfo {volume} {782}},\ \bibinfo {eid} {66} (\bibinfo
  {year} {2014})},\ \Eprint {http://arxiv.org/abs/1310.7031} {arXiv:1310.7031}
  \BibitemShut {NoStop}%
\bibitem [{\citenamefont {{Bowman}}\ \emph {et~al.}(2007)\citenamefont
  {{Bowman}}, \citenamefont {{Morales}},\ and\ \citenamefont
  {{Hewitt}}}]{2007ApJ...661....1B}%
  \BibitemOpen
  \bibfield  {author} {\bibinfo {author} {\bibfnamefont {J.~D.}\ \bibnamefont
  {{Bowman}}}, \bibinfo {author} {\bibfnamefont {M.~F.}\ \bibnamefont
  {{Morales}}}, \ and\ \bibinfo {author} {\bibfnamefont {J.~N.}\ \bibnamefont
  {{Hewitt}}},\ }\href {\doibase 10.1086/516560} {\bibfield  {journal}
  {\bibinfo  {journal} {\apj}\ }\textbf {\bibinfo {volume} {661}},\ \bibinfo
  {pages} {1} (\bibinfo {year} {2007})},\ \Eprint
  {http://arxiv.org/abs/astro-ph/0512262} {astro-ph/0512262} \BibitemShut
  {NoStop}%
\bibitem [{\citenamefont {{Khatri}}\ and\ \citenamefont
  {{Wandelt}}(2008)}]{2008PhRvL.100i1302K}%
  \BibitemOpen
  \bibfield  {author} {\bibinfo {author} {\bibfnamefont {R.}~\bibnamefont
  {{Khatri}}}\ and\ \bibinfo {author} {\bibfnamefont {B.~D.}\ \bibnamefont
  {{Wandelt}}},\ }\href {\doibase 10.1103/PhysRevLett.100.091302} {\bibfield
  {journal} {\bibinfo  {journal} {Physical Review Letters}\ }\textbf {\bibinfo
  {volume} {100}},\ \bibinfo {eid} {091302} (\bibinfo {year} {2008})},\ \Eprint
  {http://arxiv.org/abs/0801.4406} {arXiv:0801.4406} \BibitemShut {NoStop}%
\bibitem [{\citenamefont {Weinberg}\ \emph {et~al.}(2013)\citenamefont
  {Weinberg}, \citenamefont {Mortonson}, \citenamefont {Eisenstein},
  \citenamefont {Hirata}, \citenamefont {Riess},\ and\ \citenamefont
  {Rozo}}]{Weinberg201387}%
  \BibitemOpen
  \bibfield  {author} {\bibinfo {author} {\bibfnamefont {D.~H.}\ \bibnamefont
  {Weinberg}}, \bibinfo {author} {\bibfnamefont {M.~J.}\ \bibnamefont
  {Mortonson}}, \bibinfo {author} {\bibfnamefont {D.~J.}\ \bibnamefont
  {Eisenstein}}, \bibinfo {author} {\bibfnamefont {C.}~\bibnamefont {Hirata}},
  \bibinfo {author} {\bibfnamefont {A.~G.}\ \bibnamefont {Riess}}, \ and\
  \bibinfo {author} {\bibfnamefont {E.}~\bibnamefont {Rozo}},\ }\href {\doibase
  http://dx.doi.org/10.1016/j.physrep.2013.05.001} {\bibfield  {journal}
  {\bibinfo  {journal} {Physics Reports}\ }\textbf {\bibinfo {volume} {530}},\
  \bibinfo {pages} {87 } (\bibinfo {year} {2013})},\ \bibinfo {note}
  {observational Probes of Cosmic Acceleration}\BibitemShut {NoStop}%
\bibitem [{\citenamefont {{Furlanetto}}\ \emph {et~al.}(2004)\citenamefont
  {{Furlanetto}}, \citenamefont {{Zaldarriaga}},\ and\ \citenamefont
  {{Hernquist}}}]{2004ApJ...613....1F}%
  \BibitemOpen
  \bibfield  {author} {\bibinfo {author} {\bibfnamefont {S.~R.}\ \bibnamefont
  {{Furlanetto}}}, \bibinfo {author} {\bibfnamefont {M.}~\bibnamefont
  {{Zaldarriaga}}}, \ and\ \bibinfo {author} {\bibfnamefont {L.}~\bibnamefont
  {{Hernquist}}},\ }\href {\doibase 10.1086/423025} {\bibfield  {journal}
  {\bibinfo  {journal} {\apj}\ }\textbf {\bibinfo {volume} {613}},\ \bibinfo
  {pages} {1} (\bibinfo {year} {2004})},\ \Eprint
  {http://arxiv.org/abs/astro-ph/0403697} {astro-ph/0403697} \BibitemShut
  {NoStop}%
\end{thebibliography}%

\end{document}